\documentclass[a4paper,showpacs,nofootinbib]{revtex4-1}
\usepackage[utf8]{inputenc}
\usepackage{graphicx,bm,color,amssymb,amsmath}
\usepackage{slashed,verbatim}
\usepackage{subfigure}
\usepackage[colorlinks=true,linktocpage=true,linkcolor=blue,citecolor=blue]{hyperref}
\usepackage{placeins}
\usepackage{array}
\usepackage{xcolor} 
\usepackage[toc,page]{appendix}

\DeclareRobustCommand{\rchi}{{\mathpalette\irchi\relax}}
\newcommand{\irchi}[2]{\raisebox{\depth}{$#1\chi$}} % inner command, used by \rchi

\def\c{\centering}

\begin{document}
\hspace{15cm}TIFR/TH/20-36\\
\title{Studying explicit $U(1)_A$ symmetry breaking in hot and magnetised two flavour non-local NJL model constrained using lattice results}
\author{Mahammad Sabir Ali$^{1,a}$, Chowdhury Aminul Islam$^{2,1,b}$, Rishi Sharma$^{1,c}$}
\affiliation{$^1$ Department of theoretical Physics, Tata Institute of Fundamental Research, Homi Bhabha Road, Mumbai 400005, India}
\affiliation{$^2$ School of Nuclear Science and Technology, University of Chinese Academy of Sciences, Beijing 100049, China}
\email{$^a$ sabir@theory.tifr.res.in}
\email{$^b$ chowdhury.aminulislam@gmail.com}
\email{$^c$ rishi@theory.tifr.res.in}

\begin{abstract}
{We study the two flavour non-local Nambu\textemdash Jona-Lasinio (NJL) model in the
presence of a magnetic field and explore the chiral crossover in presence of a
non-local form of the 't Hooft determinant term. Its coupling is governed by a
dimensionless parameter $c$. This term is responsible for the explicit breaking
of $U(1)_A$ symmetry. We have attempted a systematic analysis of the model
parameters by fitting to self-consistent lattice QCD calculations. Three
parameters of the model are fixed by $eB=0$ results from published lattice QCD
on the chiral condensate, the pion decay constant ($F_\pi$), and the pion mass ($m_\pi$).
The difference of the $u$ and $d$ quark condensates in the presence of a
magnetic field ($eB$) is quite sensitive to $c$ and we fix $c$ using published
lattice QCD results for this observable. We see no evidence that $c$ depends on $eB$.
The crossover temperature decreases with increasing $eB$ only for condensate
values at the lower end of the allowed values (as already seen in~\cite{Pagura:2016pwr}) and $F_\pi$ at the upper end of the allowed values. We
further check our model predictions by calculating the topological
susceptibility with the fitted $c$ values and comparing it with lattice results.
Since the topological susceptibility is related to the extent of the $U(1)_A$
symmetry breaking, we find that it is sensitive to the value of $c$.}
\end{abstract}

\maketitle

\section{Introduction} 
The study of quantum chromodynamics (QCD) matter in the presence of strong
magnetic fields is a topic of current interest (see~\cite{Kharzeev:2013jha} for
a broad review). Phenomenologically, it is interesting because a strong
magnetic field is present in the initial stages of Heavy Ion Collisions (HICs)
($\sim 15 m_\pi^2; m_\pi^2\approx10^{18}\, {\mathrm G}
$)~\cite{Skokov:2009qp,Tuchin:2015oka}.  Intense fields in the range
$10^{14}-10^{19}$G may also exist in the core of
magnetars~\cite{Lattimer:2006xb}.

The QCD phase diagram in the presence of magnetic fields, particularly the chiral
crossover, has been heavily studied for the last few decades. From different
effective QCD model
investigations~\cite{Schramm:1991ex,Gusynin:1995nb,Lee:1997zj,Fraga:2008qn,Boomsma:2009yk,Gatto:2010pt,Chatterjee:2011ry,Ferrari:2012yw,Ferreira:2013tba}
and from the earliest lattice QCD study~\cite{DElia:2010abb} it was generally
believed that at zero chemical potential the value of the chiral condensate at any
temperature $T$ in the presence of a magnetic field will be larger than its value
at $eB=0$. This property is termed as magnetic catalysis (MC). 

With more controlled lattice calculations~\cite{Bali:2011qj,Bali:2012zg}
the chiral condensate was found to decrease with an increasing magnetic field near the crossover temperature. This behaviour was termed as inverse magnetic
catalysis (IMC). These effects have been studied for various magnetic fields
and $m_\pi$~\cite{Bali:2011qj,Bali:2012zg,Endrodi:2015oba,DElia:2018xwo,Endrodi:2019zrl}. This sharper reduction in the condensate in the crossover region often leads to a reduction of the crossover temperature with increasing $eB$, which we will also loosely call the IMC effect below. (A reduction in the crossover temperature does not imply the IMC effect for $m_\pi$ significantly larger than the physical $\pi$ mass~\cite{DElia:2018xwo}.)

Since IMC was discovered using lattice QCD, several attempts have been made to
understand it through effective QCD
models~\cite{Chao:2013qpa,Farias:2014eca,Ferreira:2014kpa,Ayala:2014iba,Ayala:2014gwa,Ferrer:2014qka,Andersen:2014oaa,Yu:2014xoa,Providencia:2014txa,Mueller:2015fka,Mao:2016fha,Farias:2016gmy,Pagura:2016pwr,GomezDumm:2017iex}.
A family of local Nambu\textemdash
Jona-Lasinio~\cite{Nambu:1961tp,Nambu:1961fr} (NJL) models explains the IMC by
incorporating the effect of the energy scale $eB$ on the four fermion coupling.
The motivation for this choice is that in QCD the presence of this additional
energy scale is expected to weaken the coupling as $eB$ increases. Modelling this effect by reducing the four Fermi interaction strength with $eB$ 
leads to IMC in the crossover region. 

The NJL interaction is local, which makes the model simple yet powerful. The
price paid for the model's simplicity is the fact that the results of
observables might depend on the regularization procedure used. The most popular
regularization used is the introduction of a three momentum cut-off. In this scheme, the QCD
interaction is assumed to be constant up to the value of the cut-off and modes
above the cut-off are dropped. This procedure misses the important nature of
running of QCD coupling constant with energy and hence the reduction in the coupling constant as a function of $eB$ needs to be put in by hand.

The non-local
~\cite{Bowler:1994ir,Plant:1997jr,General:2000zx,Praszalowicz:2001wy,GomezDumm:2001fz}
version of the NJL model is introduced to overcome some of the above-mentioned
drawbacks of the local NJL model. Technically the resulting expressions for the
observables in the theory are similar to implementing a soft cut-off using form
factors that decrease with increasing momenta but the intuition is more than
that. The reduction of the quark interaction with increasing energy qualitatively mimics the nature of
the running of the QCD coupling constant. The non-local version of the model
also describes spontaneous chiral symmetry breaking. Some aspects of
confinement have also been described in the
model~\cite{Bowler:1994ir,Plant:1997jr}.

The non-local version of the NJL model has been used to describe QCD matter
under strong magnetic field. In contrast to the standard NJL model (where one
needs to use multiple fitted
parameters~\cite{Farias:2014eca,Ferreira:2014kpa}), it naturally leads to the
effect of IMC~\cite{Pagura:2016pwr} without using a four Fermi interaction that
explicitly depends on $eB$. To keep note of the chronology of the actual
development, it should be mentioned here that the first attempt of incorporation
of a magnetic field in non-local NJL model showed MC at all
temperatures~\cite{Kashiwa:2011js}. There the magnetic field was introduced
just like what is usually done in the local NJL model. On the other hand in
Ref.~\cite{Pagura:2016pwr} the analysis was performed following a more rigorous
procedure based on the Ritus eigenfunction method~\cite{Ritus:1978cj} with the
inclusion of a non-local quark model with separable interactions including a
coupling to a uniform magnetic field.  Our choice of working with this model
particularly stems from this important fact that it naturally shows the IMC
effect near crossover, in agreement with lattice results. 

Here we build upon the work of Refs.~\cite{GomezDumm:2006vz,Pagura:2016pwr} and
study the two flavour non-local NJL model in presence of a magnetic field to
explore the chiral phase transition. In this paper, we will focus on the
Gaussian form factor, which is one of the two considered by Refs.~\cite{GomezDumm:2006vz,Pagura:2016pwr}. 

The first addition to Ref.~\cite{GomezDumm:2006vz,Pagura:2016pwr} is that we add to the
non-local form of the standard four Fermi NJL interaction, the 't Hooft
determinant term with an arbitrary coupling constant. The usual NJL interaction
has the well-known form 
\begin{equation}
\frac{G_0}{2}[(\bar{\psi}\psi)^2+(\bar{\psi}i\gamma^5\tau^a\psi)^2] 
~\label{eq:NJL}
\end{equation} 
which can be written as the sum of $U(1)_A$ symmetric term (${\cal{L}}_1$
in Eq.~\ref{eq:lag_1} below) and a $U(1)_A$ breaking  't Hooft determinant term
(${\cal{L}}_2$ in Eq.~\ref{eq:lag_2} below) with equal coupling. The 't Hooft
determinant term arises due to instantons and is included in such effective QCD
models to break the $U(1)_A$ symmetry which mimics the axial anomaly in QCD.
The difference between the two couplings is governed by the dimensionless
parameter $c$ (defined in Eq.~\ref{eq:cdef} below). For $c=1/2$ the strength of
the two couplings are equal and the interaction is of the form Eq.~\ref{eq:NJL}.
Ref.~\cite{GomezDumm:2006vz,Pagura:2016pwr} considered the non-local
generalization of this case. By allowing the two couplings to be independent
($c\neq1/2$), the sum of the $u$ and $d$ quark condensates and the difference
of the $u$ and $d$ quark condensates are governed by two independent coupling
constants. 

In the absence of any isospin symmetry breaking the $u$ and $d$ condensates
are equal (we assume $m_u=m_d$) and so are the respective constituent quark
masses. The value of $c$ does not play any role as only the sum of the $u$ and
$d$ condensates is non-trivial. 

If one considers non-zero isospin chemical potential
($\mu_I$~\cite{Frank:2003ve}) or/and magnetic field
($eB$~\cite{Boomsma:2009yk}), the independent appearance of both the $u$ and $d$
quark condensates in the $u$ and $d$ constituent masses become important.
This effect has been termed as ``flavour mixing''~\cite{Frank:2003ve} in the
literature (although ``flavour coupling'' might be a more appropriate term).
This ``flavour mixing'' depends on $c$.

These facts make the consideration of an arbitrary strength of 't Hooft
interaction in the presence of a magnetic field quite relevant. With these
combined effects of instantons and magnetic field, the exploration becomes more
interesting. On one hand, the ``flavour mixing'' effects coming via the
instantons try to restore the isospin symmetry and on the other hand, the
strength of the magnetic field breaks it further as the different flavours
couple with the magnetic field with different strengths~\cite{Boomsma:2009yk}.

One important result of our analysis is that an $eB$ independent $c$ describes
the lattice results for the $u$, $d$ condensate difference at $T=0$ quite well,
and this allows us to extract the value of $c$ using lattice results on the
$u$, $d$ condensate difference. Results for the thermodynamics of iso-symmetric
matter cannot be used to constrain $c$. To our knowledge, this is the first
attempt to constrain $c$ using lattice results.

The results of~\cite{Pagura:2016pwr,GomezDumm:2017iex} for the $u$, $d$
condensate difference at $T=0$ as a function of $eB$ with $c=1/2$ agree quite
well with the lattice results when the model is fitted to a larger value of the
sum of the $u$, $d$ condensate at $eB=0$. However, the condensate differences
for different values of the magnetic field, have not been contrasted against the lattice
QCD data for finite $T$. Here we aim to compare our findings for both zero and nonzero
temperature with the lattice data with $c$ fitted to $T=0$ results.

Our second addition is a more systematic analysis of the parameters of the
model by fitting to a self-consistent set of lattice results. The pertinent
parameters for $eB=0$ in the non-local NJL model are the overall four Fermi
coupling strength $G_0$, the ``cutoff'' $\Lambda$ (which determines the
momentum beyond which the form factor drops rapidly), and the bare quark mass
$m$.  These are usually~\cite{Klevansky:1992qe,Hatsuda:1994pi} fitted to match $F_\pi$, $m_\pi$ and the condensate $\langle
\bar{\psi}\psi\rangle$ and we follow the same procedure. (Recall
that $c$ does not play a role for $eB=0$.) While $F_\pi$, $m_\pi$ are very well
constrained from experiments and lattice~\cite{Aoki:2019cca}, the condensate
value is often taken from models or phenomenology.

In Ref.~\cite{Pagura:2016pwr} it has already been demonstrated that IMC in the
crossover region is seen only for smaller values of the chiral condensate within a
physically motivated range of values (Ref.~\cite{Pagura:2016pwr} considered a
range of values of $(\langle \bar{\psi}\psi\rangle)^{1/3}$ from $210$ MeV to
$240$ MeV.), especially for the Gaussian form factor.

In this paper, we fit the model to self-consistent calculations of the chiral
condensate and $F_\pi$ for realistic $m_\pi$, on the lattice. We consider the
parameter set by the JLQCD collaboration, with the central value of condensate being
$\langle\bar{\psi}_f\psi_f\rangle^{1/3}=240\ {\rm MeV}$~\cite{Fukaya:2007pn} (at the renormalisation scale $2$ GeV, see Sec.~\ref{ssec:para_set}). 
We denote a specific flavor $f$ ($u$ or $d$) of the $\psi$ as
$\psi_f$. When $f$ is summed over, we denote the bilinear as $\bar{\psi}\psi$.
To increase the exploration range in the space of the chiral condensate we consider
another LQCD calculation~\cite{Brandt:2013dua}, for which the value of the condensate is a bit larger 
($\langle\bar{\psi}_f\psi_f\rangle^{1/3}=261\ {\rm MeV}$). From this inclusion, it becomes evident
that within such effective models the IMC effect can only be observed for smaller values of chiral condensates. This corroborates the findings in Ref.~\cite{Pagura:2016pwr}. We put the whole discussion related to the second lattice data in Appendix~\ref{sec:analysis_B13}.

Like Ref.~\cite{Pagura:2016pwr} we also find that within the error band of
$\langle\bar{\psi}_f\psi_f\rangle^{1/3}$ IMC is obtained for the condensates near
the lower edge of the range. In addition, we find that within the error band of
$F_\pi$ to get a better match with the phase diagram given by
LQCD~\cite{Bali:2011qj} one needs to consider $F_\pi$ towards the upper edge of
the range. Our analysis clearly indicates that the obtainment of the IMC effect in such
a non-local effective QCD model also depends on the values of $F_\pi$.

In addition to the above two improvements, the inclusion of the parameter $c$
presents us with the scope of exploring the temperature evolution of the axial
anomaly breaking term. In particular, it is expected that the $U(1)_A$ symmetry
gets restored at high enough temperature, estimated to be near or above the chiral
crossover temperature~\cite{Bazavov:2012qja}. Here the investigation becomes
more interesting in the presence of a magnetic field, which breaks the iso-spin
symmetry as well. One observable that can be impacted by a finite $c$ is the
topological susceptibility ($\chi_t$) and we also calculate it as a function of
$eB$ and $T$ for the fitted values of the parameters. $\chi_t$ has been
calculated for the local NJL model for physically motivated values of $c$
before~\cite{Lu:2018ukl,Bandyopadhyay:2019pml}. We now calculate this in the
non-local case and compare it with lattice calculations, which helps to provide an extra check on our model. 

We organise the paper as follows: In section~\ref{sec:form} we briefly review
the formalism for the model used in this article. We start the discussion by considering the instanton term at zero temperature and magnetic field in non-local model and subsequently in the subsection~\ref{ssec:form_nz_T_eB} we shift the focus to the scenario for non-zero temperature and magnetic field. In the subsection~\ref{ssec:form_top_sus} we give a brief description for the topological susceptibility. Then in section~\ref{sec:res}, the results of the paper have been outlined. In the subsection~\ref{ssec:para_set} we give details of the fitting of the model parameters and the necessary criteria. We show the results for the fitting of the $U(1)_A$ symmetry breaking parameter $c$ at zero $eB$ and $T$ in the subsection~\ref{ssec:c_fit_eB0_T0}. In the subsection~\ref{ssec:cond_ave_pd_cond_diff} we show the model prediction for condensate average, phase diagram and condensate difference with the fitted $c$ value and compare them with the available LQCD results. We further show the model predictions for $\chi_t$ with the fitted $c$ values along with the comparison with LQCD results in the subsection~\ref{ssec:res_top_sus}. Finally, in section~\ref{sec:con} we conclude.

To avoid breaking the flow of the paper we have moved some material to the appendix. In Appendix~\ref{sec:Metaplot} the dependence of the mass of (2 flavours) $U(1)_A$ fluctuations on $c$ is shown. In ~Appendix~\ref{sec:analysis_B13} results for Brandt13 are discussed. Finally, in Appendix~\ref{Chi_t_JLH_AsC} the dependence of $\chi_t$ on $c$ is discussed.

\section{Formalism}
\label{sec:form}
In this section, we briefly discuss the formalism used in the non-local NJL model.
As mentioned earlier our main goal is to study the interplay between the effect
of the magnetic field and the 't Hooft determinant term, particularly the
interplay of the magnetic field with the strength of the explicit axial symmetry
breaking.

In the standard NJL model the strengths of axial symmetric and axial symmetry
breaking  interaction terms are equal~\cite{Klevansky:1992qe,Hatsuda:1994pi}. We follow the prescription of
Ref.~\cite{Frank:2003ve} where they have considered a general NJL Lagrangian
with arbitrary interaction strengths for $U(1)_A$ symmetric and breaking
interactions,
\begin{eqnarray}
{\cal L}_{\rm NJL}&=&{\cal L}_0+{\cal L}_1+{\cal L}_2,\;~\label{eq:LNJL}
\end{eqnarray}
where the kinetic term is
\begin{eqnarray}
{\cal L}_0&=&\bar{\psi}\left( i\slashed{\partial}-m\right)\psi,\;
\label{eq:lag_0}
\end{eqnarray}
and the interactions are given by,
\begin{eqnarray}
{\cal L}_1&=&G_1\left\{(\bar{\psi}\psi)^2+(\bar{\psi}\vec{\tau}\psi)^2+ (\bar{\psi}i\gamma_5\psi)^2+(\bar{\psi}i\gamma_5\vec{\tau}\psi)^2 \right\}\label{eq:lag_1}\,\, {\mathrm{and}}\\
{\cal L}_2&=&G_2\left\{(\bar{\psi}\psi)^2-(\bar{\psi}\vec{\tau}\psi)^2- (\bar{\psi}i\gamma_5\psi)^2+(\bar{\psi}i\gamma_5\vec{\tau}\psi)^2 \right\}
~\label{eq:lag_2}
\end{eqnarray}
with ${\cal L}_1$ being symmetric under $U(1)_A$ but ${\cal L}_2$ is not.
$\vec\tau$ represents Pauli matrices. 

In the absence of iso-spin chemical potential ($\mu_I$) and magnetic field
($eB$) (we take $m_u=m_d$), the chiral condensate
\begin{equation}
\langle\bar{\psi}(x)\psi(x)\rangle
\end{equation}
spontaneously breaks the (approximate) $SU(2)_A$
symmetry. In mean field theory it depends only on the combination $(G_1+G_2)$.
The state is $SU(2)_V$ symmetric.

In the presence of $\mu_I$ or/and $eB$ as the $SU(2)_V$ symmetry is explicitly
broken one can have $\langle\bar{\psi}\tau_3\psi\rangle$ condensate which
depends also on the combination $(G_1-G_2)$. We can parameterize the coupling
constants as 
\begin{eqnarray}
\nonumber
G_1&=&(1-c)G_0/2\\
G_2&=&cG_0/2\;.~\label{eq:cdef}
\end{eqnarray}
where $c=1/2$ corresponds to the usual NJL model.

The local version of the NJL model lacks some important features of the full QCD
theory. For example, asymptotic freedom, momentum dependent constituent mass
etc. To implement these features in NJL type models qualitatively, one may
consider the non-local version of it. With a non-local form factor, one can
qualitatively incorporate the idea of asymptotic freedom in the NJL model.  

The non-local NJL Lagrangian has the same structure as Eq.~\ref{eq:LNJL} where the
interaction term can be written
as~\cite{Bowler:1994ir,Plant:1997jr,General:2000zx,Praszalowicz:2001wy,GomezDumm:2001fz,GomezDumm:2006vz,Pagura:2016pwr,GomezDumm:2017iex},
\begin{eqnarray}
{\cal L}_1&=&G_1\left\{j_a(x)j_a(x)+\tilde{j}_a(x)\tilde{j}_a(x)\right\}\,\, {\mathrm {and}}\nonumber\\
{\cal L}_2&=&G_2\left\{j_a(x)j_a(x)-\tilde{j}_a(x)\tilde{j}_a(x)\right\}\;.\nonumber
\label{eq:nl_l}
\end{eqnarray}

Here $j_a(x)$ and $\tilde{j}_a(x)$ are the non-local currents, given by (Scheme II)~\cite{GomezDumm:2006vz}
\begin{eqnarray}
j_{a}(x)/\tilde{j}_a(x)=\int d^4z\ {\cal H}(z)\bar{\psi}\left(x+\frac{z}{2}\right)\Gamma_{a}/\tilde{\Gamma}_a\psi(x-\frac{z}{2}),
\label{eq:current}
\end{eqnarray}
where $\Gamma_a=(\mathbb{I},i\gamma_5\vec{\tau})$,
$\tilde{\Gamma}_a=(i\gamma_5,\vec{\tau})$ with $a=\{0,1,2,3\}$ and ${\cal H}(z)$ is the non-local form
factor in position space.

For the rest of this sub-section let us assume that iso-spin is a symmetry of the
system (neither $\mu_I$ nor $eB$ is present). With $G_2\ne 0$ in the chiral
limit the symmetry of the above Lagrangian is
\begin{equation}
SU(2)_V\times SU(2)_A\times U(1)_V\;.
\end{equation}
With $G_2=0$ it has an additional $U_A(1)$ symmetry.

The next step is to integrate out the fermionic degrees of freedom. To do that, moving to the Euclidean space from the Minkowski space will make the calculation easier. This can be achieved with the help of Wick rotation which can be performed through the following transformations,
\begin{equation}
t\rightarrow-ix_4\;\, {\mathrm{and}}\;\, \gamma_0\rightarrow i\gamma_4.
\end{equation}
With the metric $g_{\mu\nu}={\rm diag}(-1,-1,-1,-1)$ all the calculations done below are in Euclidean space.

To integrate out the fermionic degrees of freedom one needs to linearise the theory which can be done with the help of
Hubbard–Stratonovich (HS) transformation. In the HS transformation, one can
introduce $4$ auxiliary fields associated with the $4$ different types of
interactions. For a detailed bosonisation calculation, one can look in Appendix
A of Ref.~\cite{Hell:2008cc}.  We mark the isoscalar and isovector auxiliary
fields by $\sigma$ and $\pi$, respectively, and use `$s$' and `$ps$' in the
subscripts to denote Lorentz scalar and pseudoscalar, respectively. In a
mean-field approximation, some of the auxiliary fields have an equilibrium
expectation value. 

The expectation value of operators in the pseudoscalar channel is $0$ due to
parity conservation. In absence of $\mu_I$ and $eB$, isospin
symmetry ensures 
\begin{equation}
\langle\bar{\psi}(x)\vec{\tau}\psi(x)\rangle=0
\end{equation}
This makes the Free Energy ($\Omega$) and other vacuum
observables independent of $c$. 

The auxiliary field $\sigma_s$ is given by
\begin{equation}
\sigma_s(x)=-\, \frac{G_0}{2}\int d^4z\ {\cal H}(z)\bar{\psi}\left(x+\frac{z}{2}\right)\psi(x-\frac{z}{2})
~\label{eq:sigmasdef}
\end{equation}
With only the non-zero scalar-isoscalar auxiliary field, one gets the mean-field
Lagrangian as
\begin{align}
{\cal L}_{\rm MF}=\bar{\psi}(x)\left(\delta^4(x-y) (-i\slashed{\partial}+m)+{\cal H}(x-y)\sigma_s\left(\frac{x+y}{2}\right)\right)\psi(y)+\frac{1}{2G_0}\sigma_s^2\left(\frac{x+y}{2}\right).
\label{eq:l_mf}
\end{align} 

Assuming that the mean-field is homogeneous and isotropic (for $eB=0$)
throughout space and time we take $\sigma_s$ in Eq.~\ref{eq:sigmasdef}
independent of $x$. With the above assumption one can obtain the formal
expression for the free energy per unit volume as
\begin{align}
\Omega=\frac{S_{\rm MF}}{V^{(4)}}=
-2N_fN_c\int \frac{d^4q}{(2\pi)^4}\ln\left[q^2+M^2(q)\right]
+\frac{\sigma_s^2}{2G_0},~\label{eq:SMFB0}
\end{align}

In the mean field approximation the constituent quark mass is given by, 
\begin{equation}
M(q)=m+h(q,q)\sigma_s\;.~\label{eq:MconstB0}
\end{equation}
$h(p,p')$ is the non-local form factor in momentum space, the Fourier
transformation of ${\cal H}(x-y)$. It is function of only $p+p'$ as one can
see from Eq~\ref{eq:l_mf}. 

We follow the procedure used in Ref.~\cite{GomezDumm:2006vz} and consider the
non-local form factor to be Gaussian. The explicit form is,
\begin{equation}
h(p,p')=e^{-(p+p')^2/(4\Lambda^2)}
\end{equation}

The self-consistent gap equation has the following form,
\begin{eqnarray}
\sigma_s = 8 N_c \ G_0 \int \frac{d^4 q}{(2 \pi)^4}\  h(q,q) \ \frac{
	M(q) } {q^2 + M^2(q)}\;.\label{eq:gapeq}
\end{eqnarray}

With ${\sigma_s}$ given by Eq.~(\ref{eq:gapeq}) one can calculate the formal
expression for the local condensate by differentiating the $\Omega$ with
respect to current quark mass as 
\begin{eqnarray}
\langle\bar{\psi}_f(x)\psi_f(x)\rangle=\frac{\partial\Omega}{\partial m} &=& - \, 4 N_c \int \frac{d^4 q}{(2 \pi)^4}\
\frac{M(q)} {q^2 + M^2(q)} \ \ .
\label{eq:cond0}
\end{eqnarray}
We note that the right hand side in Eq.~\ref{eq:cond0} is not convergent away from
the chiral limit (substituting $m=0$ in Eq.~\ref{eq:MconstB0} implies that
$M(p)\rightarrow 0$ at large $p$) and needs to be regularized. This can be done
by subtracting the identical expression with $M=m$. This prescription can
be understood from Eq.~\ref{eq:SMFB0} if we subtract the analogous term
from the formal expression of the free energy to make it regular.

Now, to fit the model parameters we use pion mass and also pion decay constant.
To get the pion mass we need to calculate the pion propagator which can be
obtained from the quadratic term in the pionic fluctuation from the bosonized
action as
\begin{eqnarray}
G^{\pm}(p^2) = \frac{1}{G_0} - \, 8 \,N_c \int \frac{d^4 q}{(2 \pi)^4}\
h^2(q^+,q^-) \frac{  \left[ (q^+ \cdot q^-) \mp M(q^+)
	M(q^-)\right]}{\left[ (q^+)^2 + M^2(q^+) \right]
	\left[ (q^-)^2 + M^2(q^-)\right]}
\label{eq:pion_prop}
\end{eqnarray}
where $+$ sign in $G^{\pm}$ corresponds to the $\sigma$ mode and $-$ sign to
the pionic mode with $q^\pm = q \pm p/2.$ The pion mass is obtained from
\begin{eqnarray}
G^-(-m_\pi^2) = 0 \ .
\label{eq:pion_mass}
\end{eqnarray}

Following the steps given in Ref.~\cite{GomezDumm:2006vz} we use the
expression for pion decay constant
\begin{equation}
m_\pi^2 \; F_\pi  = m \; Z_\pi^{1/2}\; J(-m_\pi^2) \ ,
\label{eq:exp_pdc}
\end{equation}

where $J(p^2)$ given by
\begin{eqnarray}
J(p^2) = \, 8 \,N_c \int \frac{d^4 q}{(2 \pi)^4}\
h(q^+,q^-) \frac{  \left[ (q^+ \cdot q^-) + M(q^+)
	M(q^-)\right]}{\left[ (q^+)^2 + M^2(q^+) \right]
	\left[ (q^-)^2 + M^2(q^-)\right]},
\label{eq:jp2}
\end{eqnarray}

and $Z_\pi$ is related to the
$\pi\bar{\psi}_f\psi_f$ coupling constant and is given by
\begin{equation}
Z_\pi^{-1} \ = \ \frac{d G^-(p) }{dp^2}
\bigg|_{p^2=-m_\pi^2} \ .
\label{eq:zpi}
\end{equation}

Eqs.~(\ref{eq:cond}),~(\ref{eq:pion_mass}) and~(\ref{eq:exp_pdc}) are used to fit the free parameters of the model, $m$ (current quark mass), $G_0$ and $\Lambda$ with a given form factor. In the case of the Gaussian form factor, $\Lambda$ characterizes the range of non-local interaction. In other words, it controls at what scale the coupling starts becoming small. These parameters are fitted to obtain a phenomenologically allowed quark condensate, physical pion mass and pion decay constant.

\subsection{Non-zero Temperature and Magnetic Field}
\label{ssec:form_nz_T_eB}
To include both the temperature and magnetic field we follow the procedure given in
Ref.~\cite{Pagura:2016pwr}. Since the isospin $SU(2)$  symmetry (both vector
and axial) is broken in the presence of a magnetic field, we
introduce another auxiliary field $\pi_s$~\cite{Boomsma:2009yk}. Pseudoscalar
mean-fields are still not allowed due to the parity symmetry. 

Introducing  these two auxiliary fields ($\sigma_s$, $\pi_s$, introduced in
the previous section) one can obtain the effective Euclidean action using the
mean field Lagrangian,
\begin{equation}
S_{\mathrm{bos}}=-\ln\det\mathcal{D}+\frac{1}{2G_0}
\int d^{4}x\ \sigma_s^2(x)+\frac{1}{2(1-2c)G_0}
\int d^{4}x\ \vec{\pi}_s(x)\cdot\vec{\pi}_s(x),
\end{equation}
where the fermionic determinant is given by
\begin{align}
\mathcal{D}\left(  x+\frac{z}{2}\,,x-\frac{z}{2}\right)=\,\gamma_{0}\;W\left(  x+\frac{z}{2},x\right)  \gamma_{0}
\, \bigg[\,\delta^{(4)}(z)\,\big(-i\rlap/\partial+m\big) + \mathcal{H}(z)\big[\sigma_s(x)+\vec{\tau}\cdot\vec{\pi}_{s}(x) \big]
\bigg]\; W\left(  x,x-\frac{z}{2}\right).
\label{eq:ferm_det}
\end{align}
Here $W(x,y)$ is given as $W(x,y) = \mathrm{P}\exp\left[ -\, i \hat Q\int_{x}^{y}dr_{\mu}\  \mathcal{A}_{\mu
}(r)\right]$. As for the non-magnetic field scenario here also we will assign space-time independent meanfield values to the auxiliary fields. Without loss of generality we can choose $\vec\pi_s$ to be in the $\tau^3$ ($\pi^3_s$) direction. As already mentioned, all other pseudo-scalar auxiliary fields are chosen to have zero mean field values. Then the fermionic determinant and the action become
\begin{eqnarray}
\mathcal{D}^{\mbox{\tiny MFA}} (  x , x') &=& \delta^{(4)}(x-x') \left( - i \rlap/\partial
- \hat Q \, B \, x_1 \, \gamma_2 + m \right)  + \nonumber \\
&&\left( \sigma_s +\tau_3{\pi}^3_s\right) \mathcal{H}(x-x') \; \exp\left[ \frac{i}{2} \, \hat Q \, B \, (x_2 - x_2')\, (x_1 +
x_1')\right]\ {\mathrm{and}}
\end{eqnarray}
\begin{equation}
S_{\mathrm{bos}}=-\ln\det\mathcal{D}^{\mbox{\tiny MFA}} +\frac{1}{2G_0}
\int d^{4}x\ \sigma_s^2+\frac{1}{2(1-2c)G_0}
\int d^{4}x\ \left(\pi^3_s\right)^2,
\end{equation}
respectively. 

Following the Ritus eigenfunction method~\cite{Ritus:1978cj} as employed in
Ref.~\cite{Pagura:2016pwr}, we obtain the constituent mass for a Gaussian
non-locality form factor as,
\begin{equation}
M^{\lambda,f}_{q_\parallel,k} = m + \left(\sigma_s+s_f{\pi}^3_s\right) \
\frac{ \left(1- |q_f B|/\Lambda^2\right)^{k+\frac{\lambda s_{\! f}-1}{2}}}
{ \left(1+ |q_f B|/\Lambda^2\right)^{k+\frac{\lambda s_{\! f}+1}{2}}}
\;\exp\!\big(-{q_{\parallel}}^{2}/\Lambda^2\big)\,
\label{eq:cons_mass_gaus}
\end{equation} 
where $q_{\parallel}=(q_3,q_4)$, $s_f=\rm sign(q_f)$, $k$ is the Landau level index and $\lambda=\pm1$ is the spin. The free energy per unit volume is given by
\begin{eqnarray}
\Omega=\frac{S^{\mbox{\tiny MFA}}_{\mathrm{bos}}}{V^{(4)}} & = & \frac{\sigma_s^2}{2 G_0} +\frac{({\pi}^3_s)^2}{2(1-2c)G_0}- N_c \sum_{f=u,d} \frac{  |q_f B|}{2 \pi} \int \frac{d^2
	q_\parallel}{(2\pi)^2} \ \Bigg\{ \ln\left[q_\parallel^2 + \left({M^{s_{\! f},f}_{q_\parallel,0}\,}\right)^2\right]
+ \nonumber \\
& & \sum_{k=1}^\infty \ \ln\left[ \left( 2 k |q_f B| + q_\parallel^2 +
M^{-1,f}_{q_\parallel,k} M^{+1,f}_{q_\parallel,k}\right)^2 \! \! + q_\parallel^2 \left(
M^{+1,f}_{q_\parallel,k} - M^{-1,f}_{q_\parallel,k} \right)^2\right]\Bigg\}\ .
\label{eq:s_mfa}
\end{eqnarray}

The two gap equations can be obtained by differentiating the above equation with respect to ${\sigma_s}$ and ${\pi}^3_{s}$ as
\begin{eqnarray}
\frac{\partial\Omega}{\partial\sigma_s} & = & \frac{ 
	\sigma_s}{G_0} - N_c \sum_{f=u,d} \frac{2|q_f B|}{2 \pi} \int \frac{d^2q_\parallel}{(2\pi)^2} \ \Bigg\{ \frac{{M^{s_{\! f},f}_{q_\parallel,0}\,}{A^{s_{\! f},f}_{q_\parallel,0}\,}}{q_\parallel^2 + \left({M^{s_{\! f},f}_{q_\parallel,0}\,}\right)^2}+ \nonumber \\
& & \sum_{k=1}^\infty\Bigg[ \frac{ \left( 2 k |q_f B| + q_\parallel^2 +
	M^{-1,f}_{q_\parallel,k} M^{+1,f}_{q_\parallel,k}\right)\left(
	A^{-1,f}_{q_\parallel,k}M^{+1,f}_{q_\parallel,k} + M^{-1,f}_{q_\parallel,k}A^{+1,f}_{q_\parallel,k} \right)}{ \left( 2 k |q_f B| + q_\parallel^2 +
	M^{-1,f}_{q_\parallel,k} M^{+1,f}_{q_\parallel,k}\right)^2 \! \! + q_\parallel^2 \left(
	M^{+1,f}_{q_\parallel,k} - M^{-1,f}_{q_\parallel,k} \right)^2}+\nonumber \\
& & \frac{  q_\parallel^2 \left(
	M^{+1,f}_{q_\parallel,k} - M^{-1,f}_{q_\parallel,k} \right)\left(
	A^{+1,f}_{q_\parallel,k} - A^{-1,f}_{q_\parallel,k} \right)}{ \left( 2 k |q_f B| + q_\parallel^2 +
	M^{-1,f}_{q_\parallel,k} M^{+1,f}_{q_\parallel,k}\right)^2 \! \! + q_\parallel^2 \left(
	M^{+1,f}_{q_\parallel,k} - M^{-1,f}_{q_\parallel,k} \right)^2}\Bigg] \Bigg\}=0\,\, \mathrm{and}
\label{eq:gap_one}
\end{eqnarray}
\begin{eqnarray}
\frac{\partial\Omega}{\partial\pi^3_s} & = & \frac{\pi^3_s}{(1-2c)G_0} - N_c \sum_{f=u,d} s_f\frac{2|q_f B|}{2 \pi} \int \frac{d^2q_\parallel}{(2\pi)^2} \ \Bigg\{ \frac{{M^{s_{\! f},f}_{q_\parallel,0}\,}{A^{s_{\! f},f}_{q_\parallel,0}\,}}{q_\parallel^2 + \left({M^{s_{\! f},f}_{q_\parallel,0}\,}\right)^2}+ \nonumber \\
& & \sum_{k=1}^\infty\Bigg[ \frac{ \left( 2 k |q_f B| + q_\parallel^2 +
	M^{-1,f}_{q_\parallel,k} M^{+1,f}_{q_\parallel,k}\right)\left(
	A^{-1,f}_{q_\parallel,k}M^{+1,f}_{q_\parallel,k} + M^{-1,f}_{q_\parallel,k}A^{+1,f}_{q_\parallel,k} \right)}{ \left( 2 k |q_f B| + q_\parallel^2 +
	M^{-1,f}_{q_\parallel,k} M^{+1,f}_{q_\parallel,k}\right)^2 \! \! + q_\parallel^2 \left(
	M^{+1,f}_{q_\parallel,k} - M^{-1,f}_{q_\parallel,k} \right)^2}+\nonumber \\
& & \frac{  q_\parallel^2 \left(
	M^{+1,f}_{q_\parallel,k} - M^{-1,f}_{q_\parallel,k} \right)\left(
	A^{+1,f}_{q_\parallel,k} - A^{-1,f}_{q_\parallel,k} \right)}{ \left( 2 k |q_f B| + q_\parallel^2 +
	M^{-1,f}_{q_\parallel,k} M^{+1,f}_{q_\parallel,k}\right)^2 \! \! + q_\parallel^2 \left(
	M^{+1,f}_{q_\parallel,k} - M^{-1,f}_{q_\parallel,k} \right)^2}\Bigg] \Bigg\}=0,
\label{eq:gap_two}
\end{eqnarray}
respectively. Here $A^{\lambda,f}_{q_\parallel,k} $ is given by

\begin{equation}
A^{\lambda,f}_{q_\parallel,k} = \frac{ \left(1- |q_f B|/\Lambda^2\right)^{k+\frac{\lambda s_{\! f}-1}{2}}}
{ \left(1+ |q_f B|/\Lambda^2\right)^{k+\frac{\lambda s_{\! f}+1}{2}}}
\;\exp\!\big(-{q_{\parallel}}^{2}/\Lambda^2\big).
\label{eq:nonlocality_form}
\end{equation} 
The finite temperature is introduced using the Matsubara formalism, which connects the Euclidean time component to the temperature as follows
\begin{equation}
    \int\frac{dq_4}{2\pi}f(q_4)=T\sum_{n=-\infty}^{\infty}f(\omega_n),
    \label{eq:matsubara}
\end{equation}
where the Mastubara frequency ($\omega_n$) for fermions are given by $\omega_n=(2n+1)\pi T$.

With the above prescription (Eq.~\ref{eq:matsubara}), $q_\parallel^2$ is replaced by $(q_3^2+\omega_n^2)$ in Eqs.~\ref{eq:cons_mass_gaus},~\ref{eq:s_mfa},~\ref{eq:gap_one} and~\ref{eq:gap_two}; and the integration $\int\frac{d^2q_\parallel}{(2\pi)^2}$ is replaced by $T\sum_{n=-\infty}^{\infty}\int\frac{dq_3}{(2\pi)}$. Here we would like to highlight the major difference with the standard NJL model scenario where one can separate out the vacuum and thermal contribution coming from the fermionic determinant. There the temperature appears only through the momentum component $q_4$ in the propagator, and using a summation identity one can easily separate out the above-mentioned two components. In the non-local case, however, we cannot separate out the vacuum and the thermal contributions. In the presence of a non-local form factor, one gets an explicit temperature dependence in the constituent mass which then prevents us from making any further analytic simplification, and the Matsubara sums needs to be performed numerically.

One can easily notice that the free energy form (Eq.~\ref{eq:s_mfa}) is divergent due to the integration over $q_3$. Our purpose is to minimise the free energy and obtain the associated meanfield values. To remove the divergence from the action we can subtract from Eq.~\ref{eq:s_mfa} the same expression with constituent mass ($M$) being replaced by the current mass ($m$). This replacement does not affect the position of the extremum in the meanfield space but removes the divergence. The subtraction also ensures that the summations over Landau levels and Matsubara frequencies converge.

The values of the meanfields can be obtained by solving the gap equations. Now if we look into the gap equations~(\ref{eq:gap_one} and~\ref{eq:gap_two}) we can see that the integrands are multiplied by $A_{q_\parallel,k}^{\lambda,f}$, which goes to zero as $q_3\rightarrow\infty$ due to the exponential factor. Hence the gap equations are also divergence free. The same exponential factor makes sure that the summation over Matsubara frequencies converges too, whereas the Landau level summation converges as the effective power of $k$ is $-1$ with a suppressing factor in the numerator coming from Eq.~\ref{eq:nonlocality_form}. 

Using the two gap equations (\ref{eq:gap_one},\ref{eq:gap_two}) we obtain the mean field values $\sigma_s$ and
$\pi_s^3$. $\sigma_s$ is proportional to the average of $u$ and $d$ condensates
and $\pi_s^3$ to the difference of them and since $\pi_s^3=0$ for $eB=0$, we
expect that for $eB$ small enough 
\begin{equation}
{\rm abs}(\sigma_s)\ge{\rm abs}(\pi_s^3)\;.
\label{eq:c_boundary}
\end{equation} 

Now looking at the Euclidean action (Eq.~\ref{eq:s_mfa}) in the chiral limit
($m=0$) without the mean-field part (i.e., terms one and two on the right-hand side), one observes
that it is symmetric under the interchange of $\sigma_s$ and $\pi_s^3$. Then it
becomes important to know what guarantees a higher numerical value to
$\sigma_s$ solution compared to $\pi_s^3$ and not vice versa. Where a smaller
solution for $\sigma_s$ compared to $\pi_s^3$ simply implies that the sign of
$u$ and $d$ condensates are different which is unphysical. This symmetry
is broken by the meanfield terms.  $c\neq0$ breaks this symmetry and in the
chiral limit we found out that with $c>0$ we always end up with physically acceptable
solutions, but for $c<0$ we do not. Hence $c=0$ is the boundary line between these
two scenarios. 

The introduction of a small quark mass ($m$) shifts this boundary to a
slightly negative value in $c$. Motivated by this discussion we will restrict ourselves to $c>0$. We will further see that $c>1/2$ can be ruled out of other physical considerations, and hence we shall restrict ourselves to $c\in[0,1/2]$

The formal expression for the quark condensate for individual flavors ($u$ and
$d$) can be obtained by differentiating the action with respect to the
corresponding current quark mass as before,
\begin{eqnarray}
\langle\bar\psi_f\psi_f\rangle & = & - N_c \sum_{f=u,d} \frac{2|q_f B|}{2 \pi} \int \frac{d^2q_\parallel}{(2\pi)^2} \ \Bigg\{ \frac{{M^{s_{\! f},f}_{q_\parallel,0}\,}}{q_\parallel^2 + \left({M^{s_{\! f},f}_{q_\parallel,0}\,}\right)^2}+ \nonumber \\
& & \sum_{k=1}^\infty \frac{ \left( 2 k |q_f B| + q_\parallel^2 +
	M^{-1,f}_{q_\parallel,k} M^{+1,f}_{q_\parallel,k}\right)\left(
	M^{+1,f}_{q_\parallel,k} + M^{-1,f}_{q_\parallel,k} \right)}{ \left( 2 k |q_f B| + q_\parallel^2 +
	M^{-1,f}_{q_\parallel,k} M^{+1,f}_{q_\parallel,k}\right)^2 \! \! + q_\parallel^2 \left(
	M^{+1,f}_{q_\parallel,k} - M^{-1,f}_{q_\parallel,k} \right)^2} \Bigg\}.
\label{eq:cond}
\end{eqnarray}

In the large-$p$ region one easily find out that the above integral is divergent
with non-zero quark masses. To obtain finite condensate one need to regularize
it. Here we have used the same regularization procedure as used in
Ref.~\cite{Pagura:2016pwr},
\begin{equation}
\langle\bar\psi_f\psi_f\rangle^{\rm reg}_{B,T}=\langle\bar\psi_f\psi_f\rangle_{B,T}-\langle\bar\psi_f\psi_f\rangle^{\rm free}_{B,T}+\langle\bar\psi_f\psi_f\rangle^{\rm free,reg}_{B,T},
\label{eq:cond_reg}
\end{equation} 
where "free" implies that there is no self-interaction and $\langle\bar\psi_f\psi_f\rangle^{\rm free,reg}_{B,T}$ is given by
\begin{eqnarray}
\langle\bar\psi_f\psi_f\rangle^{\rm free,reg}_{B,T}&=&\frac{N_cm^3}{4\pi^2}\Bigg[\frac{\ln\Gamma(x_f)}{x_f}-\frac{\ln(2\pi)}{2x_f}+1-\left(1-\frac{1}{2x_f}\ln x_f\right)\Bigg]\nonumber\\
&&+\frac{N_c\left|q_fB\right|}{\pi}\sum_{k=0}^{\infty}\alpha_k\int\frac{dq}{2\pi}\frac{m}{E^f_k\left(1+\exp[E^f_k/T]\right)};
\label{eq:cond_free_reg}
\end{eqnarray}
with $x_f=m^2/(2\left|q_fB\right|)$. It is obvious that the `$\mathrm{free, reg}$' term will be zero in absence of magnetic field.\\

Finally to compare our findings with LQCD results~\cite{Bali:2012zg} we use
their definition of the renormalised condensate, which was used to cancel both additive and multiplicative divergences
that appear in the lattice calculation. The form is given below
\begin{equation}
\Sigma_{B,T}^f=\frac{2m}{{\cal N}^4}\left[\langle\bar\psi_f\psi_f\rangle^{\rm reg}_{B,T}-\langle\bar\psi_f\psi_f\rangle^{\rm reg}_{0,0}\right]+1,
\label{eq:sigma_scaled_lat}
\end{equation} 
where $\cal N$ is given by ${\cal N}=(m_{\pi}F_{\pi,0})^{1/2}$ which follows from
Gell-Mann-Oakes-Renner (GOR) relation, $m_{\pi}$ is the neutral pion mass and
$F_{\pi,0}$ is the pion decay constant in the chiral limit.

\subsection{Topological susceptibility ($\chi_t$)}
\label{ssec:form_top_sus}
The QCD Lagrangian features a CP-violating~\cite{tHooft:1976snw,tHooft:1986ooh} topological term that can be written as~\cite{weinberg_1996},
\begin{equation}
\Delta {\cal{L}} = \epsilon^{\mu\nu\sigma\lambda}\frac{\theta}{64 \pi^2}{\cal F}^a_{\mu\nu}{\cal F}^a_{\sigma\lambda} \;,~\label{eq:FFD}
\end{equation}
where $\theta$ is the QCD vacuum angle and ${\cal F}^a$ stands for the gluonic field strength tensor. For direct comparison to lattice measurements we follow the conventions where $g$ is absorbed in the definition of ${\cal F}$.

An important quantity is the topological susceptibility~\cite{DelDebbio:2004ns} as a function of $T$,
\begin{equation}
    \chi_t = \int d^4 x \langle Q_t(x) Q_t(0)\rangle 
\end{equation}
where
\begin{equation}
    Q_t(x) = -\epsilon^{\mu\nu\sigma\lambda}\frac{1}{64 \pi^2}{\cal F}^a_{\mu\nu}{\cal F}^a_{\sigma\lambda}.
\end{equation}
It can be related to the thermodynamic derivative of the free energy~\cite{Fukushima:2001hr}
\begin{align}
\frac{d^2\Omega}{d \theta^2}\bigg |_{a=0}=\chi_t.
\label{eq:top_sus}
\end{align}

One reason why $\chi_t$ is important is that in QCD $\chi_t$ can be formally related to the mass of the axion field~\cite{Weinberg:1977ma,Wantz:2009mi}. The existence of a dynamical axion is considered to be a possible solution to the strong $CP$ problem (the absence of charge and parity
violation in strong interaction)~\cite{Weinberg:1977ma}. $\theta$ can be related to the QCD axion field ($a$) via the relation $\theta=a/f_a$, with $f_a$ being the axion decay constant.

It is also well known~\cite{Fujikawa:1980eg} that this term (Eq.~\ref{eq:FFD}) can be removed by making a $U(1)_A$ transformation on the fermionic field
\begin{equation}
    \psi \rightarrow e^{-i\gamma^5 \theta/4}\psi,\;\;\bar{\psi} \rightarrow \bar{\psi} e^{-i\gamma^5 \theta/4}\;.~\label{eq:u1Atrans}
\end{equation}
Using the fact that the fermionic measure of the gauge theory is not invariant under the chiral transformations it was shown that $\Delta {\cal{L}}$ (Eq.~\ref{eq:FFD}) cancels out. Now, under above mentioned chiral transformation both the Lagrangian~\eqref{eq:lag_2} and the mass term are not invariant and pick up a phase factor. $\theta$ then appears in the fermionic sector.

For the calculation of $\chi_t$ in effective models without dynamic gluonic degrees of freedom, it is assumed that the same procedure follows~\cite{Boer:2008ct,Boomsma:2009eh,Schwartz:2014sze,Buballa:2003qv}. Namely, $\theta$ can be introduced in the fermionic sector using Eq.~\ref{eq:u1Atrans}.

Under Eq.~\ref{eq:u1Atrans} the currents transform as, 
\begin{eqnarray}
j_{a}(x)/\tilde{j}_a(x) &\rightarrow&\int d^4z\ {\cal H}(z)\bar{\psi}\left(x+\frac{z}{2}\right)e^{-i\gamma_5\theta}\,\Gamma_{a}/\tilde{\Gamma}_a\psi(x-\frac{z}{2})\;.
~\label{eq:u1AtransJ}
\end{eqnarray}
It is clear that the transformations (Eq.~\ref{eq:u1AtransJ}) are the same as those for local currents. This is because the $U(1)_A$ transformations that we are making are global.

Therefore, ${\cal L}_1$ remains unchanged after going through the transformations Eq.~\ref{eq:u1AtransJ}, but ${\cal{L}}_2$ transforms to
\begin{align}
	{\cal L}_2\rightarrow {\cal L}_a =& \,G_2\left[\mathrm{cos}\theta\left\{j_a(x)j_a(x)-\tilde{j}_a(x)\tilde{j}_a(x)\right\}+2\mathrm{sin}\theta\left\{j_{0}(x)\tilde{j}_{0}(x)-j_{i}(x)\tilde{j}_{i}(x)\right\}\right],
	\label{eq:lag_axion_2nl}
\end{align}
where $i$ runs from $1$ to $3$. It can be easily verified that the Lagrangian in Eq.~\ref{eq:lag_axion_2nl} is chirally symmetric but breaks the $U(1)_A$ symmetry as required by QCD. Now with the inclusion of the axion field the new working Lagrangian becomes
\begin{align}
{\cal L}_{\rm NJL}={\cal L}_0+{\cal L}_1+{\cal L}_a.
\end{align}

There is also an additional change in the mass term (Eq.~\ref{eq:lag_0}), but this additional contribution is small because of the smallness of $m$. Therefore, we ignore the explicit breaking of $U(1)_A$ by the mass term and consider only contributions from Eq.~\ref{eq:lag_2}.

Once we have the total Lagrangian, we can get the thermodynamic potential ($\Omega$) using the meanfield approximation~\cite{Boer:2008ct,Boomsma:2009eh,Lu:2018ukl,Bandyopadhyay:2019pml}. To make the above Lagrangian bilinear in quark fields we follow the same procedure used in the previous subsection. As non-zero $\theta$ breaks $P$, $T$ and $CP$ symmetry, we can have non-zero meanfields associated with the axial currents. Thus with the mean-fields $\sigma_s$, $\pi_s^3$ we now, as well, consider the parity violating mean-fields $\sigma_{ps}$ and $\pi_{ps}^3$. The topological susceptibility is calculated from Eq~\ref{eq:top_sus} at $\theta=0$ which makes the newly introduced fields go to zero. The quark contribution to the free energy looks the same as obtained in the previous subsection with modifications in the mass term as given below,
\begin{eqnarray}
    M^{\lambda,f}_{q_\parallel,k} &=& M^{\lambda,f}_{0,q_\parallel,k}+s_f\,M^{\lambda,f}_{3,q_\parallel,k},\\
    M^{\lambda,f}_{0,q_\parallel,k} &=& m + A^{\lambda,f}_{q_\parallel,k}\left\{ A\sigma_s +D\sigma_{ps}+(C\sigma_{s}+B\sigma_{ps})i\gamma_5\right\}\;\;{\mathrm{and}}\nonumber\\
    M^{\lambda,f}_{3,q_\parallel,k} &=&  A^{\lambda,f}_{q_\parallel,k}\left\{B\pi_{s}^3-C\pi_{ps}^3+(A\pi_{ps}^3-D\pi_s^3)i\gamma_5\right\}.\nonumber
\end{eqnarray}
Here, $A=(1-2c\sin^2(\theta/2))$, $B=(1-2c\cos^2(\theta/2))/(1-2c)$, $C=c\sin(\theta)$ and $D=c\sin(\theta)/(1-2c)$. In absence of axion field (i.e. at $\theta=0$) we have $A=B=1$ and $C=D=0$, which then lead us to the expression obtained in previous subsection. The eigenvalues are straightforward to calculate. The pure meanfield contribution to the free energy is given below,
\begin{equation}
    \Omega_{MF}=\frac{A}{2G_0}\left(\sigma_s^2+(\pi_{ps}^3)^2\right)+\frac{B}{2(1-2c)G_0}\left(\sigma_{ps}^2+(\pi_s^3)^2\right)+\frac{C}{(1-2c)G_0}\left(\sigma_s\sigma_{ps}-\pi_s^3\pi_{ps}^3\right).
    \label{eq:ts_mf}
\end{equation}.
With these modifications, one can write down the free energy which looks like Eq.~\ref{eq:s_mfa} with the modified mass eigenvalues and the meanfield part replaced by the above expression (Eq.~\ref{eq:ts_mf}). With this free energy, we have four gap equations associated with four meanfields. Now from $\Omega$ one can obtain different meanfields by solving the gap equations simultaneously. But as the topological susceptibility is calculated at $\theta=0$, which forces $\sigma_{ps}$ and $\pi_{ps}^3$ to acquire zero values. Hence the free energy is the same as what we obtained by considering only two meanfields. But to calculate the topological susceptibility we need the second derivatives of the free energy with respect to all the meanfields and then set $\sigma_{ps}$ and $\pi_{ps}^3$ to zero. For specific temperature and magnetic field, these solutions will also depend on $\theta$ and thus as a whole, the potential will be,
\begin{align}
\Omega=\Omega(T,eB,\theta).
\end{align}
The total derivative of the free energy with respect to $\theta$ can be expressed in terms of the partial derivatives of the same with respect to the meanfields and $\theta$:
\begin{eqnarray}
\frac{d^2\Omega}{d\theta^2} &=& \frac{\partial^2\Omega}{\partial\theta^2}+ 2\frac{\partial^2\Omega}{\partial\pi_{ps}^3\partial\theta}\frac{\partial\pi_{ps}^3}{\partial\theta} + \frac{\partial^2\Omega}{\partial{\pi_{ps}^3}^2}\left(\frac{\partial\pi_{ps}^3}{\partial\theta}\right)^2 + 2\frac{\partial^2\Omega}{\partial\pi_s^3\partial\theta}\frac{\partial\pi_s^3}{\partial\theta} + 2\frac{\partial^2\Omega}{\partial\pi_s^3\partial\pi_{ps}^3}\frac{\partial\pi_s^3}{\partial\theta}\frac{\partial\pi_{ps}^3}{\partial\theta} + \frac{\partial^2\Omega}{\partial{\pi_s^3}^2}\left(\frac{\partial\pi_s^3}{\partial\theta}\right)^2\nonumber\\
&&+ 2\frac{\partial^2\Omega}{\partial\sigma_{ps}\partial\theta}\frac{\partial\sigma_{ps}}{\partial\theta} + 2\frac{\partial^2\Omega}{\partial\sigma_{ps}\partial\pi_{ps}^3}\frac{\partial\sigma_{ps}}{\partial\theta}\frac{\partial\pi_{ps}^3}{\partial\theta} + 2\frac{\partial^2\Omega}{\partial\sigma_{ps}\partial\pi_s^3}\frac{\partial\sigma_{ps}}{\partial\theta}\frac{\partial\pi_s^3}{\partial\theta} + \frac{\partial^2\Omega}{\partial{\sigma_{ps}}^2}\left(\frac{\partial\sigma_{ps}}{\partial\theta}\right)^2 +2\frac{\partial^2\Omega}{\partial\sigma_s\partial\theta}\frac{\partial\sigma_s}{\partial\theta}\nonumber\\
&& + 2\frac{\partial^2\Omega}{\partial\sigma_s\partial\pi_{ps}^3}\frac{\partial\sigma_s}{\partial\theta}\frac{\partial\pi_{ps}^3}{\partial\theta} + 2\frac{\partial^2\Omega}{\partial\sigma_s\partial\pi_{s}^3}\frac{\partial\sigma_s}{\partial\theta}\frac{\partial\pi_s^3}{\partial\theta} + 2\frac{\partial^2\Omega}{\partial\sigma_s\partial\sigma_{ps}}\frac{\partial\sigma_s}{\partial\theta}\frac{\partial\sigma_{ps}}{\partial\theta} + \frac{\partial^2\Omega}{\partial{\sigma_s}^2}\left(\frac{\partial\sigma_s}{\partial\theta}\right)^2 + \frac{\partial\Omega}{\partial\sigma_s}\frac{\partial^2\sigma_s}{\partial\theta^2}\nonumber\\
&& +\frac{\partial\Omega}{\partial\pi_s^3}\frac{\partial^2\pi_s^3}{\partial\theta^2} + \frac{\partial\Omega}{\partial\sigma_{ps}}\frac{\partial^2\sigma_{ps}}{\partial\theta^2} +\frac{\partial\Omega}{\partial\pi_{ps}^3}\frac{\partial^2\pi_{ps}^3}{\partial\theta^2}.
\end{eqnarray}
Each of the last four terms contains gap equations which are zero at the minima of the free energy. 
Using the above expression we can now calculate $\chi_t$ from Eq.~\ref{eq:top_sus}.  For a recent calculation of $\chi_t$ in the local NJL model with $eB=0$ see~\cite{Lu:2018ukl} and for nonzero $eB$, see~\cite{Bandyopadhyay:2019pml}.

\section{Results}
\label{sec:res}

\subsection{Choice of the fitting observables for $eB=0$ and the ranges of parameters}
\label{ssec:para_set}  
As discussed above, the non-local NJL model for $eB=0$ has three parameters
$G_0$, $\Lambda$, and $m$. ($c$ plays a role only for $eB\neq0$.) To fix these,
we would like to fit the model to match three independent observables. In this
paper, we will work only with the two-flavour model. We fit the three parameters to
self-consistently determined data from (two-flavour) lattice calculations for $m_\pi$,
$F_\pi$ and $\langle\bar{\psi}_f\psi_f\rangle$ in the absence of a magnetic
field. With this parameter set, we study the behaviour of $u$ and $d$
condensates in the presence of magnetic field within the non-local NJL model
and compare them with the lattice results of Ref.~\cite{Bali:2012zg} for zero
as well as non-zero temperatures. 

The lattice study in Ref.~\cite{Bali:2012zg} is done for physical pions, and
they also quote the value of $F_\pi$ in the chiral limit ($86\rm\  MeV$).
However, since the value of the condensate for this lattice calculation was not
available in the literature, we use other lattice calculations to fit the
parameters of our model.

We do note that the results of Ref.~\cite{Bali:2012zg} are for $1+1+1$ QCD and
hence the model fitted to two flavour data cannot be expected to capture the
physics of the three flavour model completely. Indeed, as we will see below, the
crossover temperature for $eB=0$ we obtain is lower than the crossover
temperature in Ref.~\cite{Bali:2012zg}. This is a well-known property of the
non-local NJL model~\cite{Pagura:2016pwr}. A direct comparison will require the
generalization of the model to three flavours which we will consider in future
work. However, here, we study to what extent the two flavour model captures the
results found in Ref.~\cite{Bali:2012zg} and hope that once the temperature is
scaled by the scale $T_{CO}$, the dependence on the flavour content is not very
strong. 

To cover a range of self-consistently determined parameter values,  we consider
a 2-flavour LQCD calculation to constrain our model~\cite{Fukaya:2007pn}
(referred in the text as JLQCD), which we discuss in detail in the following
sections. We also consider another LQCD data set~\cite{Brandt:2013dua}
(we refer to it as Brandt13), for which the analysis is briefly discussed in the
appendix. One can look into the Ref.~\cite{Aoki:2019cca} for the available
lattice results and the latest developments. The calculation
in~\cite{Fukaya:2007pn} is performed at the physical $m_\pi$ and the error on
$m_\pi$ is assumed to be negligible. We fit to $F_\pi$ at the physical pion
mass.   

\begin{figure}[!htbp]
	\includegraphics[scale=0.54]{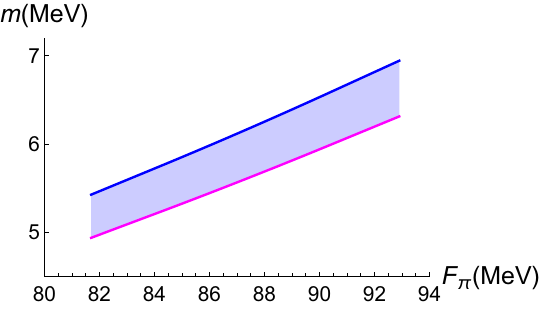}
	\includegraphics[scale=0.54]{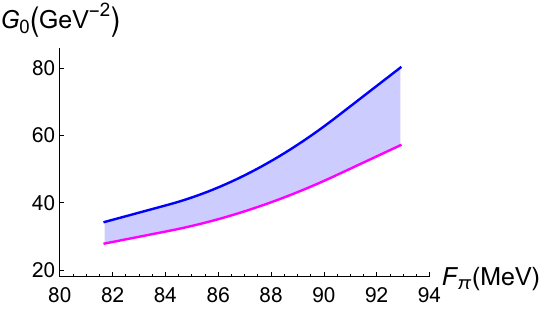}
	\includegraphics[scale=0.54]{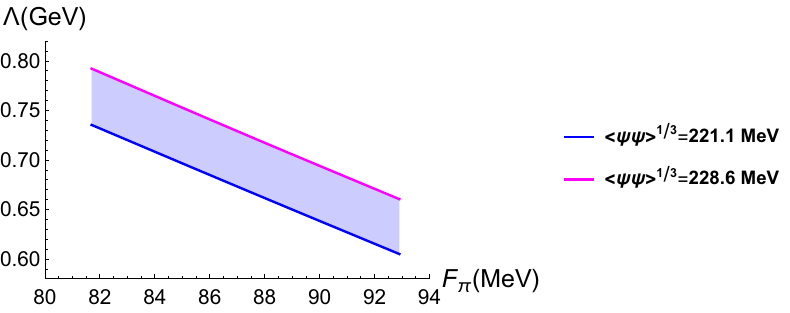}
	\caption{Range of model parameters to access the full allowed range of the
		condensate including the errors as given in LQCD~\cite{Fukaya:2007pn}
		(JLQCD).}
	\label{fig:para_c_fix}
\end{figure}

\begin{figure}[!htbp]
	\includegraphics[scale=0.56]{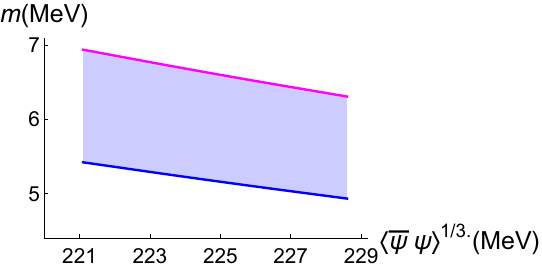}
	\includegraphics[scale=0.56]{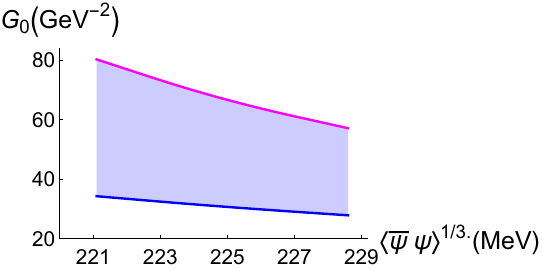}
	\includegraphics[scale=0.56]{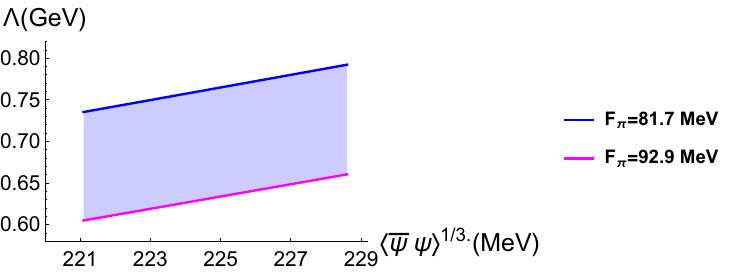}
	\caption{Range of model parameters to access the full allowed range of the
		PDC including the errors as given in LQCD~\cite{Fukaya:2007pn} (JLQCD).}
	\label{fig:para_fpi_fix}
\end{figure}

We begin our search for finding parameter space for the model with the JLQCD
data set. The condensate and the decay constant (with the error estimates) are
given in Ref.~\cite{Aoki:2019cca} and we quote them below,
\begin{eqnarray}
 \langle\bar{\psi}\psi\rangle^{1/3}|_{\mu=2\;\rm GeV}&=&240(4)\;\rm MeV\,and\nonumber\\
 F_{\pi}&=&87.3(5.6)\;\rm MeV.
\end{eqnarray}
One should note that the quoted value of the condensate from the LQCD study
has been obtained at the renormalisation scale $\mu=2$
GeV~\cite{Fukaya:2007pn}. A direct matching of this value to the effective
models is difficult because of two reasons. First, the natural scale for the
kind of effective models that we are considering here is around $1$ GeV (for the
non-local model, $\Lambda$ provides a rough estimate of the cutoff which is
$\sim1$ GeV Figs.~\ref{fig:para_c_fix} and~\ref{fig:para_fpi_fix}). Second, the renormalisation scheme in the
effective models is usually not $\overline{\rm MS}$. 

The correction due to a difference in schemes is a systematic that we can not
presently correct. However, we can consider the effect of the difference in the
energy scales by renormalisation group (RG) techniques. In order to express the
LQCD results at the scales suitable for the model, we exploit the perturbative
renormalisation group running for the condensate calculated in~\cite{Vermaseren:1997fq,vanRitbergen:1997va,Chetyrkin:1997dh}.
This has been previously used for example in Ref.~\cite{Giusti:1998wy}. We
estimate the condensate value at $\mu=1$ GeV and found it to be,
\begin{equation}
 \langle\bar{\psi}\psi\rangle^{1/3}|_{\mu=1\;\rm GeV}=224.8(3.7)\;\rm MeV.
\end{equation}
On the other hand, $F_\pi$ being a scale-independent quantity remains unchanged.

With $m_\pi$ fixed to its physical value $135$ MeV, we explore the allowed region for
the parameters by considering the central value of $F_\pi$ and its $\pm
1\sigma$ variation, and similarly for the condensate. Thus, we obtain a total of $9$
parameter sets to cover the allowed range provided by the quoted data. 

Ref.~\cite{GomezDumm:2006vz} considered condensates in the range
$\langle\bar{\psi}_f\psi_f\rangle^{1/3}=210-240$ MeV and showed that the
appearance of IMC in the crossover region in the non-local NJL model depends on
the values of the condensates. They showed that for condensates larger than
$240$ MeV $T_{CO}$ increases as one increases $eB$ for smaller $eB$ (for the
Gaussian form factor), in contradiction with lattice observations. The
motivation for the considered range of the condensate in
Ref.~\cite{GomezDumm:2006vz}, comes from the cited
articles~\cite{Dosch:1997wb,Berges:1998rc}. One of them~\cite{Dosch:1997wb},
using chiral perturbation theory, calculated the condensate to be in the range
$(200-260\ \rm MeV)^3$ and the other article~\cite{Berges:1998rc}, from a model
calculation with fermions interacting via instanton-induced interaction,
deduces the condensate to be $(270\ \rm MeV)^3$. However, the value of $F_\pi$
was fixed to be the physical value. With the more controlled lattice results
now available on $F_\pi$ and the condensate, one of the motivations for our
paper is to fit the model to these. We find that IMC near $T_{CO}$,
sensitively, also depends on the value of $F_\pi$.

Figs.~\ref{fig:para_c_fix} and~\ref{fig:para_fpi_fix} represent the range of
the model parameters as allowed by the JLQCD observables in the form of bands.
These bands are drawn using the $9$ parameter sets that we have, including the
errors. In Fig.~\ref{fig:para_c_fix} the bands are obtained for different
condensate values as a function of $F_\pi$. On the other hand, they are shown
for different $F_\pi$'s as a function of condensate in
Fig.~\ref{fig:para_fpi_fix}.  The upper and lower lines of the bands are for
two extreme sets of observables ($\langle\bar\psi\psi\rangle^{1/3},F_\pi$)
\textemdash\, the blue and the magenta lines are for lowest and highest values,
respectively. It is further to be noted from Fig.~\ref{fig:para_c_fix}, that as
we increase $F_\pi$ for a given value of condensate the parameter $\Lambda$
decreases whereas $G_0$ increases. On the other hand in
Fig.~\ref{fig:para_fpi_fix} it is the other way around, i.e., as the condensate
is increased $\Lambda$ increases and $G_0$ decreases, for a given value of
$F_\pi$. Now $\Lambda$ controls how the effective coupling runs as a function
of momentum, i.e., how fast the effective coupling decreases with momentum.
This information will be helpful in qualitatively understanding the results for
finite $eB$ from different parameter sets. Out of these $9$ parameter sets, we
will investigate in detail only five \textemdash\, the central value along with
the four corners.

In Table~\ref{tab:ParametersJ240}, for further illustration, we have presented
the central parameter set and the four corner parameter sets associated with
Figs.~\ref{fig:para_c_fix} and~\ref{fig:para_fpi_fix}. The letters C, H and L
stand for central, highest and lowest values, respectively. The first letter
corresponds to the value of condensate and the second one to the value of
$F_\pi$. These are the five parameter sets from JLQCD that we work with. Just
to remind our readers we mention here again that the condensate values shown in
the second column of the Table~\ref{tab:ParametersJ240} are calculated at
$\mu=1$ GeV by using the perturbation RG running~\cite{Giusti:1998wy} and the
values are roughly $15$ MeV smaller than their corresponding LQCD values
estimated at $\mu=2$ GeV.

\begin{figure}[!htbp]
	\includegraphics[scale=0.8]{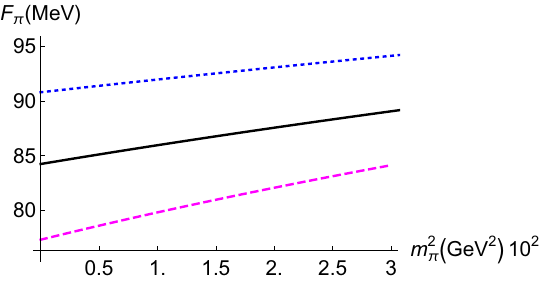}
	\includegraphics[scale=0.86]{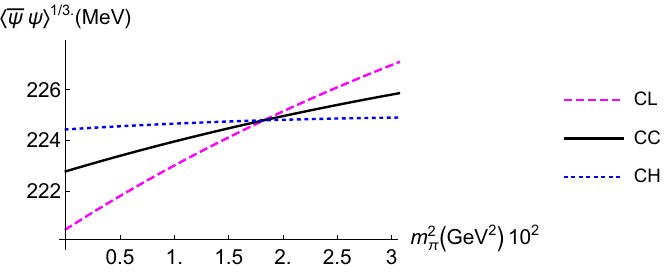}
    \caption{Pion decay constant and condensate in the chiral limit with $G_0$
	and $\Lambda$ kept fixed. The left panel is obtained for the central value
	of condensate ($224.8$ MeV), whereas for the right panel plot three different
	values of $F_\pi$ have been used (see Table~\ref{tab:ParametersJ240}).}
	\label{fig:ch_cond_fpi}
\end{figure}

In the scaled definition of the condensate given by
Eq.~\ref{eq:sigma_scaled_lat} we need to use $F_\pi$ in the chiral limit
(denoted as $F_{\pi,0}$). In Fig.~\ref{fig:ch_cond_fpi} we present the chiral
limit behaviour of the $F_\pi$ and the condensate in the model by keeping $G_0$
and $\Lambda$ fixed and only changing $m$ to $0$. This is a self-consistent way
to obtain the chiral limit within the model.

To give more details about the chiral extrapolation, in the left panel of
Fig.~\ref{fig:ch_cond_fpi}, we show the plot of $F_\pi$ for the central value
of condensate ($224.8$ MeV) and in the right panel the plot for condensate,
for three different values of $F_\pi$ (L, C and H) as allowed by the JLQCD
observables. One should necessarily note here that the mentioned values for the
condensate and $F_\pi$ (CC, CL and CH) are to be interpreted only for the
fitted values of current quark mass ($m$), because eventually in both the plots
neither of them is constant as a function of $m$.  We learn from there that as
we increase the pion mass, $F_\pi$ increases with an almost constant slope. On the
other hand, the condensate also increases with the pion mass but the
slope depends on the values of $F_\pi$ \textemdash\, it decreases as the value
of $F_\pi$ is increased.

In the next subsection, we describe the parameter fitting at zero temperature
but nonzero magnetic field.

\begin{table}[t]
	\centering
	\begin{tabular}{|m{2.8cm} | m{2.8cm} | m{1.8cm} | m{1.6cm} || m{1.8cm} |m{1.6cm}| m{1.8cm} | m{1.6cm}| }
		\hline
		& $\langle\bar{\psi_f}\psi_f\rangle^{1/3}$(MeV) & $m_\pi$(MeV) & $F_\pi$(MeV)& $F_{\pi,0}$(MeV)& $m$(MeV) & $G_0({\rm GeV^{-2}})$ & $\Lambda$(MeV) \\  
		\hline
		Parameter Set CC  & 224.8    & 135  & 87.3   & 84.25 & 5.87  & 43.34   & 697.22   \\ 
		\hline
		Parameter Set HH  & 228.6  & 135  & 92.9   & 90.63 & 6.31  & 57.15   & 660.46   \\ 
		\hline
		Parameter Set HL  & 228.6    & 135  & 81.7   & 77.04 & 4.94  & 27.90   & 792.22   \\ 
		\hline
		Parameter Set LH  & 221.1    & 135  & 92.9   & 91.00 & 6.94  & 80.26   & 605.05   \\ 
		\hline
		Parameter Set LL  & 221.1     & 135  & 81.7   & 77.61 & 5.42  & 34.32   & 735.38   \\ 
		\hline
	\end{tabular}
    \caption{Central and the four corner parameter sets associated with the
	Figs.~\ref{fig:para_c_fix} and~\ref{fig:para_fpi_fix} for LQCD
	data~\cite{Fukaya:2007pn}. The quantities on the right of the
	double bars pertain to the non-local NJL model. The values of the condensate are at $\mu=1$ GeV, obtained following perturbation RG running used in the Ref.~\cite{Giusti:1998wy}.} 
	\label{tab:ParametersJ240}
\end{table}

\subsection{Magnetic field dependence at zero temperature}
\label{ssec:c_fit_eB0_T0}
In this subsection, we fit the explicit $U(1)_A$ symmetry breaking parameter
($c$) with the LQCD data~\cite{Bali:2012zg} at zero $T$ and nonzero $eB$. The
lattice simulation provides us with the average and difference of the
condensates. In the model calculation, the condensate average is independent of
$c$ but the difference depends on it. So we use the values of condensate
differences to fit $c$ at zero $T$ and nonzero $eB$. Then we further use this
fitted value to predict the nonzero $T$ and $eB$ behaviour of the condensate in the
model and compare them with the lattice data in the next subsection. \\

\begin{figure}[!htbp]
	\centering
	\begin{minipage}{0.5\textwidth}
		\centering
		\includegraphics[width=\linewidth]{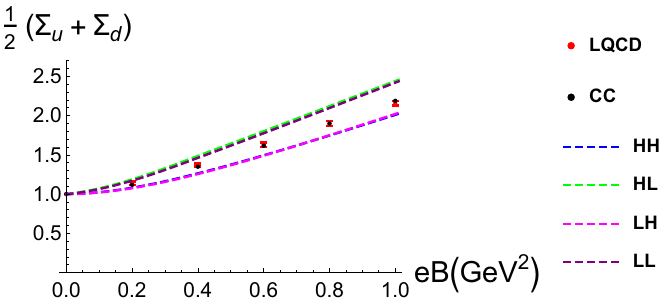}
		\caption{The condensate average as function of magnetic field as compared with LQCD~\cite{Bali:2012zg} data for JLQCD.}
		\label{fig:FinalASB13}
	\end{minipage}\hspace{0.2cm}
	\begin{minipage}{0.47\textwidth}
		\centering
		\includegraphics[width=\linewidth]{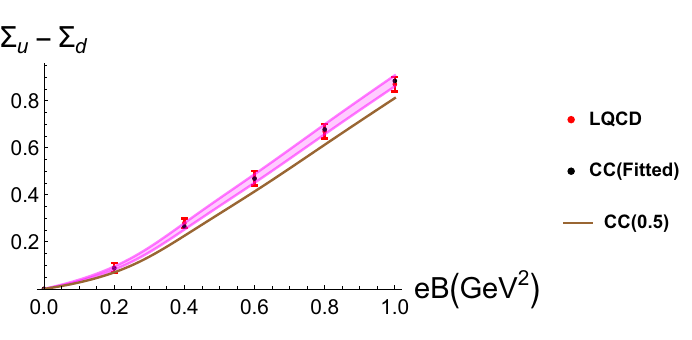}
		\caption{The condensate differences ($u$ and $d$) fitted in c for parameter set CC in JLQCD for LQCD data~\cite{Bali:2012zg}.}
	    \label{fig:FinalDSJ13}
	\end{minipage}
	\vskip\baselineskip % Leave a vertical skip below the figure
\end{figure}

First, let us consider the condensate average. Analyzing the JLQCD set, we
compare the model results with the LQCD data in Fig.~\ref{fig:FinalASB13},
using their normalized definition of condensate as given in
Eq.~\ref{eq:sigma_scaled_lat}. It shows that for the CC parameter set the
matching with the LQCD data is very good even for finite $eB$. Black points
are used to denote the CC parameter and LQCD data
are shown with the red points. Using the corner parameter sets we obtain the 
spread in average condensate which is also shown with the dashed blue (HL) 
and green (LH) lines.   For $eB=0$, the curves meet at unity just from the 
definition (Eq.~\ref{eq:sigma_scaled_lat}).  

What is interesting is that for the central value (CC) of the parameters, the
condensate average agrees very well with results from Ref.~\cite{Bali:2012zg}.
(The black points completely overlap with the red points within error bars.)
However, for the corners of the parameter space (LL, LH, HL, and HH)  the
agreement with the lattice results is not as good: the slope for the $eB$
variation does not match LQCD. We examine this more carefully below.

In Ref.~\cite{Pagura:2016pwr} it was already shown that the scaled condensate given by
Eq.~\ref{eq:sigma_scaled_lat} has a mild dependence on the actual value of
condensate and $F_\pi$ was kept fixed. Here we have explored the $F_\pi$
dependence on all observable quantities and as there is a significant
difference in $F_\pi$, hence in $F_{\pi,0}$ for the corner parameter sets, we
obtain a spread in the (scaled) average condensate. 

The implication for the comparison in Fig.~\ref{fig:FinalASB13} is that the
normalization factor ${\cal{N}}^2=(m_\pi F_{\pi,0})^2$ is very sensitive to
$F_{\pi,0}$ and hence gives rise to the different slopes: the larger values of
$F_{\pi,0}$ lead to a smaller slope and vice versa. In the lattice calculation
in Ref.~\cite{Bali:2012zg}, $F_{\pi,0}$ is taken to be $86$ MeV.
Fig.~\ref{fig:FinalASB13} shows the range of average condensate we obtain when
we vary $F_{\pi,0}$ in the range of values self-consistently determined with
the condensate. 

Now we analyze the effect of $c$. First, we keep $eB=0$ and think about constraints on $c$ due to
general physical considerations. As mentioned above, $F_\pi$, $m_\pi$, and the
condensate average are independent of $c$.  $\Sigma_u-\Sigma_d=0$ for $eB=0$
irrespective of the value of $c$. But certain fluctuations are sensitive to the
value of $c$ even for $eB=0$. More specifically, the mass of the fluctuations
in the isoscalar pseudoscalar channel~\cite{Dmitrasinovic:1996fi} depends on
$c$. In the two flavour problem, this is often called
$\eta^*$~\cite{Dmitrasinovic:1996fi} and can be intuitively thought of as a
fictitious mixture of $\eta$ and $\eta'$ mesons present in the full
three-flavour theory. Since there is no physical particle directly corresponding
to the $\eta^*$, the meson spectrum in the two flavour theory can not be
directly used to find $c$. However, physical considerations do restrict the
allowed values of $c$ in the theory.

At $c=0$, $\eta^*$ becomes degenerate with $\pi^0$. This is a simple
consequence of the restoration of the $U(1)_A$ symmetry in this limit. On the
other hand, considering $\eta^*$ as a mixture of $\eta$ and $\eta'$ one would
expect the mass of $\eta^*$ to be a few times to that of $\pi^0$. This sets a
lower bound on the value of $c$. To be precise, we impose the physically
motivated constraint that $M_{\eta^*}>400$~MeV\footnote{We should remind ourselves that this limit is not of a very strict nature because of the flavour number we are considering here. But we will eventually find out that the kind of $c$ limit that this approximated value of $M_{\eta^*}$ provides us is reasonable for having the appropriate strength of topological susceptibility and is also comparable with other known studies.}. In Figure~\ref{fig:etamass} (in Appendix~\ref{sec:Metaplot}) we have shown the $M_{\eta^*}$, obtained using the expression from Ref.~\cite{Sabir:2021nws}, as a function of $c$ for the CC parameter set and the assumed constraint on $M_{\eta^*}$ allows $c$ to be greater than $0.12$. This should be mentioned here that changing the parameter set will have a negligible effect on the constraint. This gives $c>0.12$.  At the other end, $\eta^*$ becomes tachyonic for $c>1/2$. Therefore, these physical
constraints restrict, $c$ to be in the following region,
\begin{equation}
c\in [0.12,0.5]~\label{eq:cPhysicalConstraints}\;.
\end{equation}

Below we first consider the effect of $c$ on condensate difference for finite
$eB$ without imposing the physically motivated constraints on $c$
(Eq.~\ref{eq:cPhysicalConstraints}). These results are given in
Table~\ref{tab:J240ChiSquared}. 

For finite $eB$, $\Sigma_u-\Sigma_d$ grows as $eB$ increases and the rate of
this increase is sensitive to $c$ and we use this dependence to fit $c$.

As an illustrative example, in Fig.~\ref{fig:FinalDSJ13} we show the best fit for the CC parameters along with the uncertainty in $c$. In principle, there is no reason why $c$ cannot depend on $eB$
but we see from the same figure that a one-parameter fit in $c$ is
quite adequate to describe the data. It is drawn with the fitted value of $c$ along with its uncertainty shown as a magenta coloured band. We see from Table~\ref{tab:J240ChiSquared}
where the $\rchi^2/{\mathrm{DoF}}$ is shown, $c=0.276\pm0.068$ describes the data well
for CC. The conclusion is that the data for $\Sigma_u-\Sigma_d$ as a function
of $eB$ allow us to extract the value of $c$, which can then be used to
compute additional observables (for example topological susceptibility)
everywhere through the range from $eB=0$ to about $1$ GeV$^2$. 

In the pursuit of narrowing down further to find the best possible parameter
set for the model we look at the fitted $c$ values from
Table~\ref{tab:J240ChiSquared}. 
\begin{table}[t]
	\centering
	\begin{tabular}{|m{2.8cm} | m{2.2cm} | m{2.cm} | }
		\hline
		&   c        &  $\rchi^2$ per DoF  \\  
		\hline
		Parameter Set CC   &  $0.276 \pm 0.068$   &   $0.211$  \\ 
		\hline
		Parameter Set HH  &  $0.044 \pm 0.079$   &   $ 0.149 $  \\ 
		\hline
		Parameter Set HL  &  $0.374 \pm 0.051$   &   $ 0.290 $  \\ 
		\hline
		Parameter Set LH  &  $0.149 \pm 0.103$   &   $ 0.634 $  \\ 
		\hline
		Parameter Set LL  &  $0.465 \pm 0.062$   &   $ 0.551 $  \\ 
		\hline
	\end{tabular}
	\caption{$\rchi^2$ fitting of condensate difference in $c$ for all the five parameter sets from JLQCD.}
	\label{tab:J240ChiSquared}
\end{table} 

The fittings for the other four corner parameter sets are summarized in Table~\ref{tab:J240ChiSquared}, which contains both the fitted
$c$ values along with their uncertainty and the corresponding $\rchi^2/{\mathrm{DoF}}$. We see that the quality of fits in $c$ is quite satisfactory. We would like to mention here that the parameter $c$ can be constrained further with reduced errors, once we have better-controlled LQCD data with narrower error bands for the $\Sigma_u-\Sigma_d$.

Now with these fitted $c$ values we will explore the nonzero $T$ behaviour of
both the condensate average and the difference.

\subsection{Results for nonzero temperature}
\label{ssec:cond_ave_pd_cond_diff}
\subsubsection{Condensate average and the phase diagram}
\label{sssec:cond_ave_pd}

\begin{table}[t]
	\centering
	\begin{tabular}{|m{2.8cm} | m{2cm} | m{2cm} | m{2cm} | m{2cm} | m{2cm} | }
		\hline
		Parameter Set   &   CC       &  HH     & HL & LH & LL \\  
		\hline
		JLQCD              &   122.282   &   130.156  &  115.577   & 130.664   & 115.041   \\ 
		\hline
	\end{tabular}
	\caption{Crossover temperature at $eB=0$ (we call this $T_{CO}$) for all parameter sets of JLQCD in MeV}
	\label{tab:J240_Tc}
\end{table}

In this subsection we study the results for condensate average at both nonzero
$T$ and $eB$. The main assumption here is that the parameters $G_0$, $\Lambda$,
$m$ and $c$ are independent of $eB$ and $T$ and we use the values fitted at
$T=0$ for all of these (Tables~\ref{tab:ParametersJ240},~\ref{tab:J240ChiSquared}), and compare the results
with the LQCD data~\cite{Bali:2012zg}. The value of the average condensate is very insensitive to $c$ if we vary it within the 
range $[0,1]$. Though we mention here the full range of $c$ as from $0$ to $1$ but for our analysis we never go beyond $c=1/2$ and the reason for this along with the implications have been mentioned in detail in the previous subsection before Eq.~\ref{eq:cPhysicalConstraints} and also in the next subsection after Eq.~\ref{eq:flavormixing}.
\begin{figure}[!htbp]
	\includegraphics[scale=0.62]{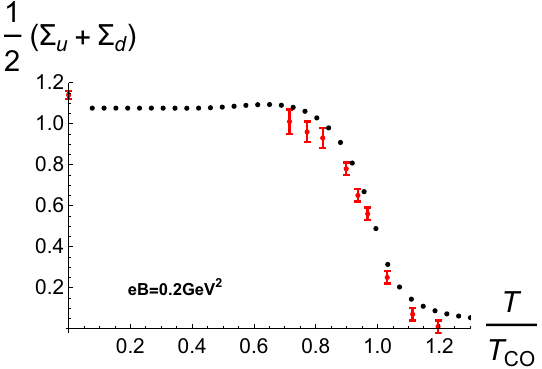}
	\includegraphics[scale=0.62]{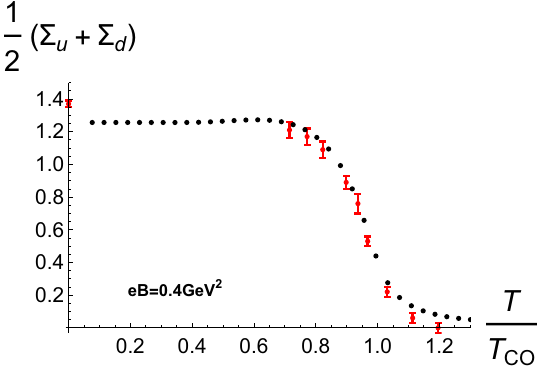}
	\includegraphics[scale=0.62]{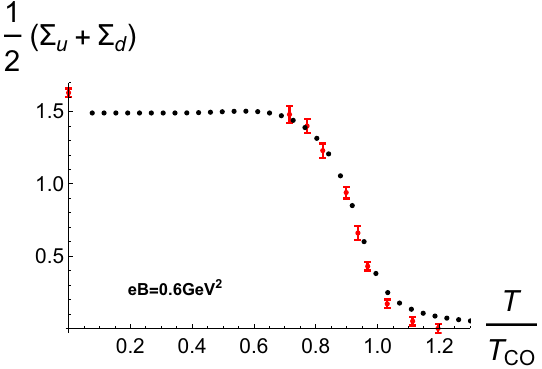}
	\includegraphics[scale=0.62]{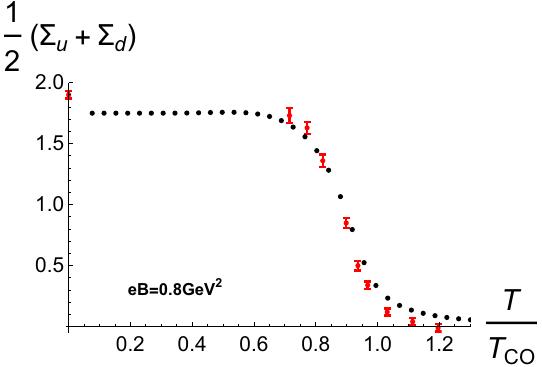}
	\includegraphics[scale=0.64]{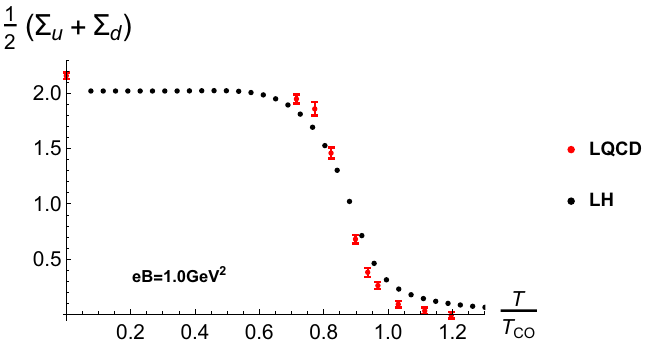}
	\caption{Plot of condensate average for different values of magnetic field as a function of temperature for parameter set LH of JLQCD, along with the comparison with LQCD data~\cite{Bali:2012zg}.}
	\label{fig:ASJ240TLH}
\end{figure}
In principle, we can start the discussion in this section from any parameter set
from Table~\ref{tab:ParametersJ240} and we design our analysis in the following
manner. First, we explore all the five available parameter sets and look for the
ones for which the IMC effect is obtained satisfactorily as compared to the
LQCD data~\cite{Bali:2012zg}. This part of the analysis is almost independent
of the values of $c$ and will leave us with a fewer number of parameter sets.
then we see the effect of $c$. 

\begin{figure}[!htbp]
	\includegraphics[scale=0.9]{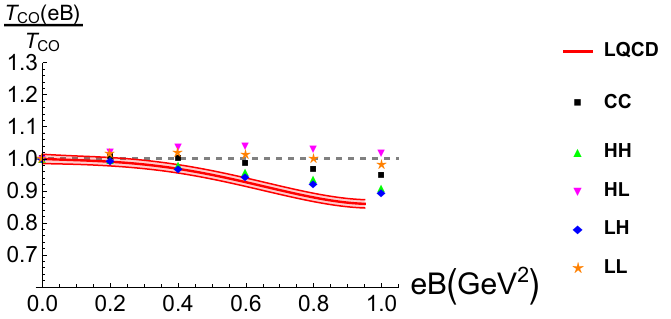}
    \caption{The phase diagram in $T-eB$ plane for parameter sets obtained to
    reproduce JLQCD vacuum observables, compared with that from LQCD
    study~\cite{Bali:2011qj}.}
	\label{fig:pd_jlqcd}
\end{figure}

Following the preceding argument, initially, we have two parameter sets to deal
with \textemdash\, one is the LH parameter set ($221.1,\,92.9$) and the other
is the HH ($228.6,\,92.9$) one, both of which reproduce the IMC effect
reasonably well (Fig.~\ref{fig:pd_jlqcd}) as compared to the LQCD
data~\cite{Bali:2012zg}. Then we refer back to the fitted $c$ values
(Table~\ref{tab:J240ChiSquared}) for these two parameter sets and chose LH
($0.149\pm0.103$) over HH ($0.044\pm0.079$) by the following arguments. Further
imposing the physical constraint in Eq.~\ref{eq:cPhysicalConstraints} eliminates
the allowed region for HH and hence from now on we focus on the LH data set
only. The allowed $c$ value for LH parameter set, after taking into consideration the constraint, becomes,
\begin{equation}
c=0.149^{+0.103}_{-0.029}~\label{eq:c_LH_pheno_cons}.
\end{equation}

As we have argued that the LH parameter set is the best possible parameter set in the model we display the plot of the temperature dependence of average condensate in Fig.~\ref{fig:ASJ240TLH} for the corresponding parameter set. In the figure we have used five different values of $eB$, the top left panel is for $0.2$ $\mathrm{GeV^2}$ and we increase $eB$ in steps of $0.2$ $\mathrm{GeV^2}$ from left to right. We observe that the results from the model (Fig.~\ref{fig:ASJ240TLH}) agree
reasonably well with the LQCD data. 

The crossover temperatures from the LQCD calculation~\cite{Bali:2012zg} for
$eB=0$ and in the present model are different. For a clear notation, we call
this $T_{CO}$. The crossover temperature at finite $eB$ we will call $T_{CO}(eB)$.
The value of $T_{CO}$ in Ref.~\cite{Bali:2012zg} is given to be $158$ MeV. The values for the model are shown in Table~\ref{tab:J240_Tc}. The
crossover temperature in both is defined as the inflection point of the average
condensate. 

It is to be noted here that the overall scale of the thermal transition is not
captured in the model, as the predicted critical temperatures
(Table~\ref{tab:J240_Tc}) are found to be relatively low as compared to the
known standard LQCD results. This is usually the case with the non-local NJL
model~\cite{Pagura:2016pwr}. This can be improved with the standard technique
of including Polyakov loop (PL)~\cite{GomezDumm:2017iex}. In spite of an
underprediction of the transition temperature in the model, to see if the model
describes how the condensate changes as a function of the ratio of the
temperature to the crossover scale, we follow previous literature and compare
our calculations with lattice calculations as a function of $T/T_{CO}$. We keep
the inclusion of PL for the future endeavour, particularly it will be really
interesting to observe the behaviour of the $U(1)_A$ symmetry breaking
parameter $c$ in the presence of background gauge field.

It is expected from the figure above that the crossover temperature in the model for the LH parameter set will be very close to that of LQCD as $eB$ is increased. This is what is observed from Fig.~\ref{fig:pd_jlqcd} where the predicted phase diagrams for all five parameter sets are displayed. As already mentioned, it is
noted from the phase diagram that the decrease of $T_{CO}(eB)$ as a function of the magnetic field as obtained in LQCD simulation is not satisfactorily reproduced with all the parameter sets. HL ($228.6,\,81.7$) and LL ($221.1,\,81.7$) parameter sets fail to produce the IMC effect. Though the CC ($224.8,\,87.3$) parameter set can reproduce the IMC effect but it is not very satisfactory as compared to LQCD data. Only for the LH and HH parameter sets we see a comparable reproduction of the IMC effect.

From the phase diagram, we find this important message that we get a better agreement with LQCD results as we decrease the condensate and/or increase $F_\pi$. The necessity of smaller values of the condensate has already been demonstrated by Scoccola et. al. (see the left panel of Fig. $3$ in Ref.~\cite{Pagura:2016pwr} for the Gaussian
form factor). But our analysis indicates that whether the non-local effective QCD model shows the IMC effect also depends on the values of $F_\pi$. This
observation becomes apparent when the analysis is performed with sets of
self-consistent parameters fitted to LQCD results~\footnote{We find that there is a difference in the value of $m_\pi$ between Ref.~\cite{Pagura:2016pwr} and us. Their quoted value is $139$ MeV.}. This observation in the effective model scenarios, to the extent of our knowledge, has not been reported before.

As discussed in Sec~\ref{ssec:para_set}, this observation can be related to the constants of the model, namely $\Lambda$ and the coupling constant ($G_0$). We observe that lower $\Lambda$ and higher $G_0$ gives a better agreement with LQCD results. One of the important features of the non-local NJL model is that it captures an important property of QCD, the running of coupling constant which gives rise to the asymptotic freedom. For higher $\Lambda$ the effective coupling constant decreases with a slower rate as compared to smaller $\Lambda$ and thus getting more contribution from higher momentum modes. It leads to delay in achieving asymptotic freedom, which is crucial for obtaining IMC
effect~\cite{Miransky:2002rp,Farias:2014eca,Ferreira:2014kpa}.

\subsubsection{Condensate difference~\label{sssec:condiff}}
In this subsection, we describe the condensate difference obtained by using the same LH parameter sets and compare them with the LQCD findings for
different strengths of $eB$.    

To understand the results it is helpful to review the idea of ``flavour mixing''~\cite{Frank:2003ve} in the presence of backgrounds that break iso-spin symmetry. The idea is most cleanly displayed for the local NJL
model where the constituent masses (the analog of Eq.~\ref{eq:cons_mass_gaus}) are given by,
\begin{eqnarray}
M_u&=&m-4G_1\langle \bar{u} u\rangle-4G_2\langle \bar{d} d\rangle \nonumber\\
M_d&=&m-4G_1\langle \bar{d} d\rangle-4G_2\langle \bar{u} u\rangle\;,
~\label{eq:flavormixing}
\end{eqnarray}
where $G_1, G_2$ are given in Eq.~\ref{eq:cdef}. In particular, for $c=0$, $G_2=0$ and $M_u$ decouples from $\langle \bar{d} d\rangle$ and vice versa and the equations for the two condensates are independent. For $c=1$, $G_1=0$ and the gap equations for the two condensates are maximally coupled. In terms of the Lagrangian, $c=0$ implies ${\cal L}_2$ (Eq.~\ref{eq:lag_2}) is set to zero which further signifies complete flavour decoupling and maximal ``flavour mixing'' corresponds to $c=1$ which implies that ${\cal L}_1$ (Eq.~\ref{eq:lag_1}) is set to zero. For the non-local model, the relation between $\sigma_s+\pi_s$ ($\sigma_s-\pi_s$) and $\langle\bar{u}u\rangle$ ($\langle\bar{d}d\rangle$) is more complicated but the intuition that the two gap equations decouple for $c=0$ still holds.  In our analysis we explore only up to the $c=1/2$ as the reference point, since beyond that ’t Hooft mass becomes imaginary, i.e., $\eta^*$ becomes a tachyon~\cite{Dmitrasinovic:1996fi}. 

\begin{figure}[!htbp]
	\includegraphics[scale=0.62]{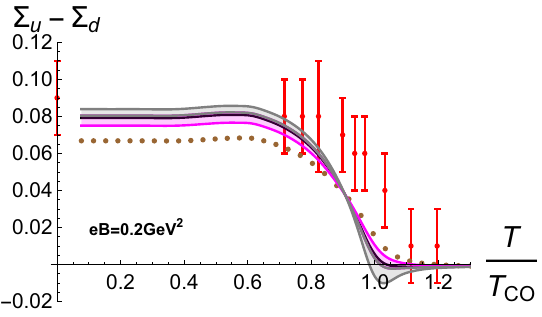}
	\includegraphics[scale=0.62]{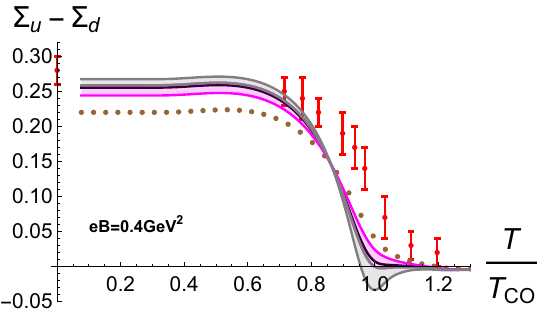}
	\includegraphics[scale=0.62]{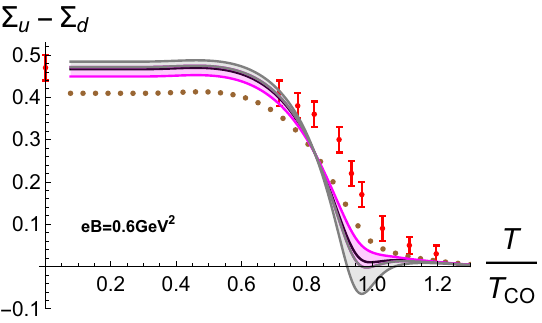}
	\includegraphics[scale=0.62]{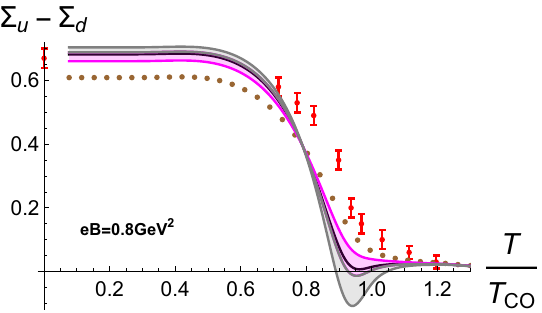}
	\includegraphics[scale=0.64]{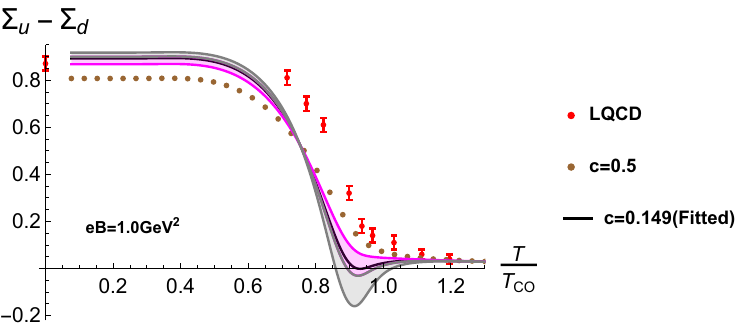}
	\caption{Plot of condensate difference for different values of magnetic field as a function of temperature for parameter set LH of JLQCD.}
	\label{fig:DSJ240TLH}
\end{figure}

With these intuitions in mind, we show the Fig.~\ref{fig:DSJ240TLH} displaying the plot for the condensate difference for the LH parameter set for different values of $eB$. The band shows the uncertainty in the fitted $c$ value. 

More concretely, the uncertainty marked in the figure by a grey band corresponds to the fitted value $0.149\pm0.103$ at $T=0$ (Table~\ref{tab:J240ChiSquared}). Further imposing the condition $c>0.12$ (Sec.~\ref{ssec:c_fit_eB0_T0}) a part of the gray band (from $c=0.046$ to $0.12$) gets excluded and the magenta band remains as a prediction from the model as given by our final fitted $c$ value ($0.149^{+0.103}_{-0.029}$) in Eq.~\ref{eq:c_LH_pheno_cons}. In the figure we have also shown the behaviour for $c=1/2$ as a reference line; at that value of $c$, the model reduces back to the usual NJL model. This figure sheds some light on how the ``flavour mixing" effects impact the behaviour of condensate difference in presence of $eB$ at finite $T$. We note that the condensate difference calculated in the model matches well with the LQCD data up to the temperature $0.8\,T/T_{CO}$ and then it falls at a bit faster rate than the LQCD data. 

One interesting thing to notice in Fig.~\ref{fig:DSJ240TLH} is that around the crossover temperature, for a part of the grey band corresponding to the lowest $c$ values, the condensate difference ($\Sigma_u-\Sigma_d$) becomes
negative and then increases and gradually merges with
the LQCD data at higher values of $T$. It is quite heartening that for the magenta band for which the lowest values of $c$ are excluded based on physical arguments, this peculiar behaviour is clipped and the results track closer to lattice results. This oscillatory behaviour about
$\Sigma_u-\Sigma_d=0$ at small $c$ in the model can be explained in the following manner.

For any arbitrary strength of the magnetic field at $T=0$, the $u$ condensate is always greater than the $d$ condensate because its coupling with the magnetic field is twice as strong as that of the $d$ quark.  For $c=0$, as discussed below Eq.~\ref{eq:flavormixing}, the $u$ and $d$ condensates decouple and can vary independently. For the $c$ values which are small but not $0$, the partial decoupling of the $u$ and $d$ condensates leads to an independent drop of $\Sigma_u$ at a faster rate as $T$ approaches $T_{CO}$ before $\Sigma_d$, i.e., the IMC effect is stronger for $\Sigma_u$ than $\Sigma_d$ due to a larger $|q_u|$. Hence $\Sigma_u-\Sigma_d$ becomes negative just before $T_{CO}$. Eventually, both the condensates catch up and drop asymptotically to $0$. At zero magnetic field, $\pi_s=0$, and $c$ does not have any influence on the condensates. Introduction of a magnetic field separates the two, hence the stronger the magnetic field greater the effect. 

This kind of behaviour just below $T_{CO}$ is not a peculiarity of this
particular model for small $c$. In fact, in the local NJL (3 flavour) model a
different kind of scenario is also observed, where the $d$ condensate decreases
with a faster rate as compared to that of $u$ one resulting in a `bump' like
behaviour in the condensate difference around the crossover temperature before
finally going to zero (\cite{Ferreira:2013tba,Ferreira:2014kpa}). The authors
explained this using a higher coupling constant of $u$ quark with the magnetic field
as compared to $d$, which relatively delays the decrease of $u$ condensate
around the crossover temperature since they consider a constant coupling
constant which does not reproduce IMC. This argument is indeed an important way in which such behaviour can arise, but as we have found, ``flavour mixing'' effects could possibly play an important part too.

\subsection{Topological susceptibility}
\label{ssec:res_top_sus}
In this section, we describe our model predictions for $\chi_t$. The field $a$ is connected to $U(1)_A$ transformations of the quark field and
hence the $\chi_t$ is related to the extent of the $U(1)_A$
symmetry breaking. ${\cal{L}}_2$ explicitly breaks $U(1)_A$ symmetry. (The
small quark mass $m$ also breaks $U(1)_A$ weakly.) Furthermore, the chiral
condensate spontaneously breaks $U(1)_A$. Hence we expect that the topological susceptibility is sensitive to both $c$ and the chiral condensate. In particular, in the chiral limit, if $c=0$, $\chi_t$ will be $0$.

\begin{figure}[!htbp]
	\includegraphics[scale=0.9]{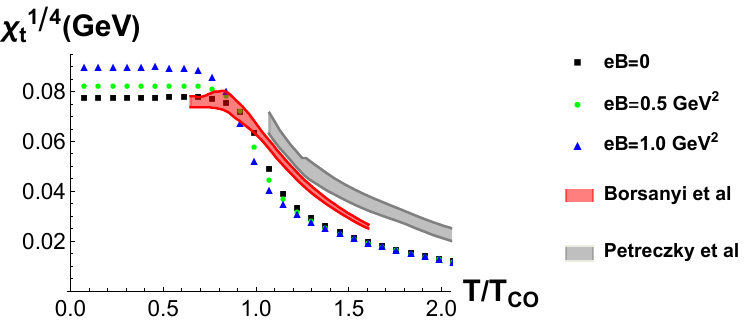}
	\caption{Topological susceptibility as a function of scaled temperature for the LH parameter set. The red and the gray bands represent lattice results from the Refs.~\cite{Borsanyi:2016ksw} and~\cite{Petreczky:2016vrs}, respectively.}
	\label{fig:top_sus}
\end{figure}

We use the fitted $c$ value for LH parameter set (the most suitable parameter set in the present model as established in the previous sections) from tables~\ref{tab:J240ChiSquared} to calculate the $\chi_t$ using Eq.~\ref{eq:top_sus}.  The plot is shown in Fig.~\ref{fig:top_sus} along with two different LQCD results~\cite{Borsanyi:2016ksw,Petreczky:2016vrs}. We show the central value of the fitted $c$, which is $0.149$, for three different values of the magnetic field, $eB=0,\,0.5\;{\mathrm{and}}\;1.0$ $\mathrm{GeV}^2$.

For comparison with the LQCD results, we have used two separate results from
LQCD calculations available in the literature. The red
band is obtained from the Ref.~\cite{Borsanyi:2016ksw}. In this reference, they
used a $2+1+1$ flavour LQCD and extend their analysis further to give the
equation of state in $2+1+1+1$ flavour QCD. But as charm quarks begin to
contribute to the equation of state above $300\,\mathrm{MeV}$, they used $2+1$
flavours dynamical quarks up to $250\,\mathrm{MeV}$. So this red band used in
the Fig.~\ref{fig:top_sus} is a $2+1$ flavour lattice result for
$\chi_t$. In that original paper~\cite{Borsanyi:2016ksw} the authors have given the plot as a function of
temperature. To scale it with the transition temperature we used
Ref.~\cite{Borsanyi:2010bp}, where the authors have given the range of the
$T_{CO}$ depending on the observables they use. Since in the model we are
calculating the $T_{CO}$ from the inflection point of the condensate, it makes
sense that we use the value calculated using the same observable in lattice
simulation, which is found to be $155\,\mathrm{MeV}$. There is another LQCD
result~\cite{Petreczky:2016vrs} shown by the grey band, which is also a $2+1$
flavour calculation. There the result is already provided as a function of
scaled temperature and is given starting from close to $T_{CO}$.

The black squares in the plot are the model predictions at zero $eB$ with the fitted $c$ value. Below the transition temperature, the model prediction is within the red band given by the only available LQCD data at that range~\cite{Borsanyi:2010bp}. Although the trend is similar, as the temperature increases further the model prediction falls at a faster rate. We understand that the model is simple and is deprived of some of the important features of full QCD as that in lattice QCD, but in our opinion, this numerical mismatch could possibly be arising from different flavour number considerations. This hunch can be tested by incorporating another flavour, which is beyond the scope of the present article and we plan to address it elsewhere. 

\begin{figure}[!htbp]
	\includegraphics[scale=0.9]{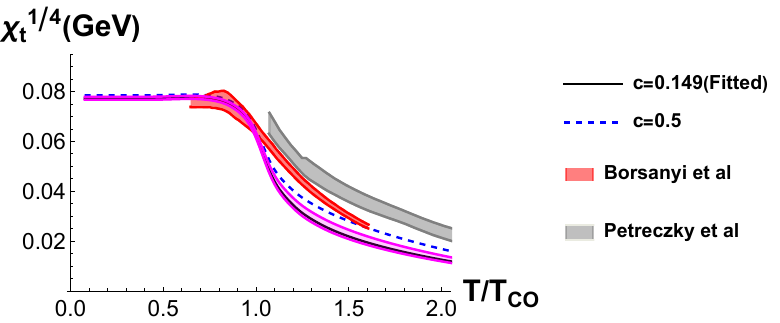}
	\caption{Topological susceptibility as a function of scaled temperature for different $c$ values for the LH parameter set. The blue dashed line is for $c=0.5$ and the black solid line is the model prediction with the magenta band representing the uncertainty in $c$. The red and the gray bands represent lattice results from the Refs.~\cite{Borsanyi:2016ksw} and~\cite{Petreczky:2016vrs}, respectively.}
	\label{fig:top_sus_j_cd}
\end{figure}

Then we further explore the impact of the magnetic field on $\chi_t$ and learn
that it increases with the increase of $eB$ below the transition temperature.
This is easily understood as $eB$ increases the condensate at low $T$.  This
finding has already been reported in Ref.~\cite{Bandyopadhyay:2019pml} in the
local NJL model.  There are two main differences between our calculation and
Ref.~\cite{Bandyopadhyay:2019pml}. The first is that we use a value of $c$
determined by matching to $T=0$ LQCD results at $T=0$.  Second, we observe that
after $1.3\,T_{CO}$ or so we don't see any effect of magnetic field on
$\chi_t$ and they all fall on top of each other, whereas
Ref.~\cite{Bandyopadhyay:2019pml} reported a considerable impact from the
magnetic field even after the transition temperature. It is also to be noted from the figure that the IMC effect is well reflected in $\chi_t$ around the $T_{CO}$.

In Fig.~\ref{fig:top_sus_j_cd} we show the sensitivity of the result to the
value of $c$ for $eB=0$. It is drawn for two different values of $c$: $c=0.149$ (the fitted one with solid black line), $c=1/2$ (standard NJL model with blue dashed line). For the fitted value, we have shown the uncertainty in $\chi_t$ as well by the magenta band. This uncertainty arises from the uncertainty in $c$, given by Eq.~\ref{eq:c_LH_pheno_cons}. At low temperature the $\chi_t$ for these three values are almost the same and as the temperature is enhanced they start diverging from one another, particular above the $T_{CO}$.  We see that currently the LQCD results for $\chi_t$ cannot distinguish among these values. 

\section{Conclusions and outlook}
\label{sec:con}
In this paper, we have studied the two flavour non-local NJL model in the
presence of a magnetic field and explored the chiral crossover. Our investigation
builds upon the non-local NJL model calculations of Refs.~\cite{GomezDumm:2006vz,Pagura:2016pwr} with some
important additions.

The first is that we add to the non-local form of the standard four Fermi NJL
interaction, the non-local form of the ’t Hooft determinant term with an
arbitrary coupling constant which is governed by a dimensionless parameter $c$.
In the Refs.~\cite{GomezDumm:2006vz,Pagura:2016pwr} $c$ is taken to be $1/2$,
which is generally the case in the usual NJL model. In absence of any isospin
symmetry breaking (we assume equal $u$ and $d$ current quark masses) the value of
$c$ does not play any role because only the sum of the $u$ and $d$ condensates
is non-trivial.  But in the presence of isospin symmetry breaking agents like
an isospin chemical potential~\cite{Frank:2003ve} or/and magnetic
field~\cite{Boomsma:2009yk}, the two condensates become different because of
their different couplings to the agents. The difference between the $u$ and $d$
condensates is particularly sensitive to $c$. This is evident from the gap equation 
associated with $\pi_{s}^3$ (Eq.~\ref{eq:gap_two}).  

Our second major addition is that we have attempted a more systematic analysis
of the parameters of the model by fitting them to a self-consistent set of lattice
results. For $eB=0$, the parameters $G_0$, $\Lambda$ and $m$ were fixed to
$m_\pi$, $F_\pi$, and the condensate (at $T=0$). We used the lattice result~\cite{Fukaya:2007pn} for which we performed the detailed analysis in the main text.  We also use another lattice result~\cite{Brandt:2013dua}, the central values of the condensates being different for these two references, and discuss briefly the results of our analysis in the appendix~\ref{sec:analysis_B13}.

Then considering $T=0$ results for finite $eB$, we found that an $eB$ independent $c$
describes the lattice results for the $u$, $d$ condensate difference at $T=0$
quite well, which allows us to extract the value of $c$ using lattice results
on the $u$, $d$ condensate difference. We estimate this observation to be
significant, as to our knowledge, for the first time $c$ has been constrained
using lattice results. 

In the past efforts have been made to constrain $c$ but from different
perspectives than us, particularly by these two
Refs.~\cite{Frank:2003ve,Boomsma:2009yk} which we summarize here to emphasize
the difference with our approach. Ref.~\cite{Frank:2003ve} discusses
the effect of $c$ on the phase diagram for small $T$ and high chemical
potential ($\mu$) region, in presence of isospin chemical potential($\mu_I$).
For zero instanton interactions, the quarks decouple, hence giving rise to
different transition lines in the $T-\mu$ plane.  Though the authors argued that in
their respective $T-\mu_I$ plane they will be identical. They obtained a
critical value for $c$ above which these two transition lines merge with each
other in the $T-\mu$ plane. They also drew an analogy from the 3-flavour NJL model to
estimate the value of $c$, which turned out to be close to the one obtained from
the phase diagram. Ref.~\cite{Boomsma:2009yk} has done similar analysis 
to Ref.~\cite{Frank:2003ve} with a non-zero magnetic field instead of isospin chemical potential. With
zero instanton effects, one obtains two different phase transitions. In this
paper, they showed that as one introduces the ``flavour mixing effects'' through
$c$, the transitions come closer to each other and beyond some critical value
of $c$ they merge to become a single phase transition. In these works, the value of $c$ is found to be approximately within the range of $0.1-0.2$.

In our calculation, we analysed the parameter sets accessible for a given LQCD study and picked the ones which are reproducing the IMC effect suitably as compared to the LQCD result~\cite{Bali:2011qj}. We found that for the JLQCD data~\cite{Fukaya:2007pn} LH ($221.1,\,92.9$) and HH ($228.6,\,92.9$) parameter sets replicate the IMC effect reasonably well. Further looking at the fitted $c$ values for these two parameter sets, which are $0.149\pm0.103$ and $0.044\pm0.079$, respectively and evoking some basic properties of $\eta^*$ we could constrain the plausible range of $c$. We argued that, as at $c=0$, $\eta^*$ becomes degenerate with $\pi^0$ and also expecting the mass of $\eta^*$ to be a few times that of $\pi^0$ (considering $\eta^*$ as a mixture of $\eta$ and $\eta'$), which sets the lower bound in $c$ as $c>0.12$, thus excluding the HH parameter as the fitted $c$ is smaller than the above mentioned lower bound. Using this $\eta^*$ motivated lower bound on $c$, our final fitted $c$ range in the LH parameter set becomes $0.149^{+0.103}_{-0.029}$. This range is also compatible with the other existing ones in the literature~\cite{Frank:2003ve,Boomsma:2009yk}. We would also like to comment here that the error in our fitted $c$ value can be reduced further once we have narrower error bands for the LQCD data on $\Sigma_u-\Sigma_d$ and thus further improving our estimation of $c$.

After fitting $c$ to $T=0$ results for the splitting
between the $u$ and $d$ condensate values, we use the model to analyze the
average condensate and the splitting as a function of $T$ and $eB$. These
results are summarized in Sec.~\ref{sssec:cond_ave_pd} and  Sec.~\ref{sssec:condiff}.
We found, like Ref.~\cite{Pagura:2016pwr}, that within the error band of
$\langle\bar{\psi}\psi\rangle^{1/3}$ IMC is obtained for the condensates near
the lower edge of the range. Furthermore, we observed that within the error band
of $F_\pi$ to get a better match with the phase diagram given by
LQCD~\cite{Bali:2011qj} one needs to consider $F_\pi$ towards the upper edge of
the range. 

We further test our model by calculating the topological susceptibility
($\chi_t$) and comparing that with the available LQCD results. We observed that with the fitted $eB$ independent $c$ value the model's prediction for $\chi_t$ at zero $eB$ can produce the trend well as found in the lattice results~\cite{Borsanyi:2016ksw,Petreczky:2016vrs}. Beyond $T_{CO}$, the quantitative mismatch with the lattice data may arise because of differences in the number of flavours in the two methods. For non-zero $eB$ (for
which, to our knowledge, there is no lattice study available for $\chi_t$) we
found that the $\chi_t$ increases as one increases the strength of the magnetic
field up to the crossover temperature. This conclusion is similar to what is
found in Ref.~\cite{Bandyopadhyay:2019pml}. We also observed that, in the
present model, $\chi_t$'s for different values of $eB$ fall on top of that at
zero $eB$ once we go beyond the crossover temperature and the IMC effect is reflected. All these
observations can be understood following the correlation between topological
susceptibility and condensate average, as we know that the condensate is
responsible for spontaneously breaking $U(1)_A$ symmetry along with the chiral
symmetry. This connection is well reflected in the present study. All these
predictions for nonzero $eB$ could be further tested in future when lattice
data becomes available for the same.

One natural extension of our study is the analysis of $2+1$ flavour QCD. The
$U(1)_A$ breaking term, in this case, is of dimension $9$ and the strength of the
interaction is well known to be related to the $\eta-\eta'$ mass
splitting~\cite{Klevansky:1992qe,Hatsuda:1994pi}. It will be interesting to see whether the results for
$\Sigma_u-\Sigma_d$ for finite $eB$ at both $0$ and finite $T$ can be
adequately described by the $2+1$ flavour model, or other terms in the effective
models are necessary. 

More recently, a magnetic field dependent 't Hooft
interaction strength for three flavour~\cite{Moreira:2020wau} has been
considered. All these facts validate our choice of considering arbitrary
strength of ’t Hooft interaction in presence of a magnetic field. In principle,
$c$ can also depend on $eB$ as well as $T$, though we do not see evidence of a
strong dependence on these variables in the range we consider. However, a
closer analysis of these effects will be interesting.

\vspace{1cm}
{\bf Acknowledgments:} The authors would like to thank Gunnar Bali and Gergely
Endrödi for their help with the LQCD data and comments. MSA and RS would like to
acknowledge discussions with Sourendu Gupta and Subrata Pal. CAI wish to thank
Aritra Bandyopadhyay for many fruitful discussions and helpful communications.
He would also like to acknowledge the facilities provided by the Tata Institute of
Fundamental Research, India where he was a visiting fellow when the preliminary
parts of the work were completed and his present institute, University of
Chinese Academy of Sciences, China. 

\clearpage
\begin{appendices}
\section{$M_{\eta^*}$ vs $c$~\label{sec:Metaplot}}
\begin{figure}[!htbp]
    \centering
    \includegraphics[scale=0.62]{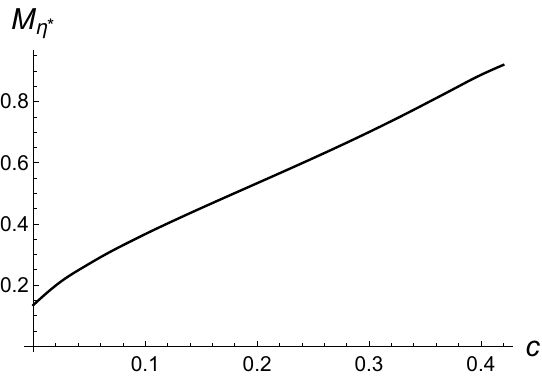}
    \caption{$\eta^*$ mass as a function of $c$ for the CC parameter set.}
    \label{fig:etamass}
\end{figure}
In Fig.~\ref{fig:etamass} we show the plot of $\eta^*$ mass as a function of $c$ for the CC parameter set. The plot is not very sensitive to the choice of parameter set (Table~\ref{tab:ParametersJ240}) in the range shown. It is also to be noted that $\eta^*$ and $\pi^0$ become degenerate at $c=0$.

\section{Analysis with another set of LQCD data}
\label{sec:analysis_B13}
\begin{table}[!htbp]
	\centering
	\begin{tabular}{|m{2.8cm} | m{2.8cm} | m{1.8cm} | m{1.8cm} || m{1.8cm} | m{1.6cm}| m{1.8cm} | m{1.6cm}| }
		\hline
		& $\langle\bar{\psi_f}\psi_f\rangle^{1/3}$(MeV) & $m_\pi$(MeV) & $F_\pi$(MeV)& $F_{\pi,0}$(MeV)& $m$(MeV) & $G_0({\rm GeV^{-2}})$ & $\Lambda$(MeV) \\  
		\hline
		Parameter Set CC  & 244.7    & 135  & 90.0     & 86.28 & 4.89  & 28.03  & 814.75  \\ 
		\hline
		Parameter Set HH  & 256.9  & 135   & 98.25  & 95.40 & 5.04  & 31.35   & 809.04  \\ 
		\hline
		Parameter Set HL  & 256.9   & 135   & 81.75  & 74.90 & 3.54  & 15.01   & 1012.84  \\ 
		\hline
		Parameter Set LH  & 232.5   & 135   & 98.25  & 96.52 & 6.71  & 78.19   & 627.60  \\ 
		\hline
		Parameter Set LL  & 232.5    & 135   & 81.75  & 76.80 & 4.71  & 25.34   & 821.43   \\ 
		\hline
	\end{tabular}
    \caption{Central and the four corner parameter sets for LQCD
	data~\cite{Brandt:2013dua}. The quantities on the right of the
	double bars pertain to the non-local NJL model. As in the case of JLQCD, the condensate values are estimated at $\mu=1$ GeV following the RG running shown in~\cite{Giusti:1998wy}.}
	\label{tab:ParametersB13}
\end{table}
We tried to perform the whole analysis with another set of LQCD data~\cite{Brandt:2013dua} (referred to as Brandt13 in the text), which has a higher central value of condensate ($\langle\bar{\psi}_f\psi_f\rangle^{1/3}=261\ {\rm MeV}$)\footnote{It is to be noted here that the central value of the condensate in Brandt13 is roughly $20$ MeV higher than that of JLQCD. This difference could be attributed to the systematic controls for LQCD discretisation artefacts by continuum extrapolation and controls for finite-size effects in the former data set.} This value is at renormalisation scale $\mu=2$ GeV. To be able to use it in the present effective model scenario, we have evaluated it at $\mu=1$ GeV following the procedures given in~\cite{Giusti:1998wy} as given in Table~\ref{tab:ParametersB13} along with the corner parameter sets, which are obtained using the errors in the values of both the condensate
($\langle\bar{\psi}\psi\rangle^{1/3}=261(13)(1)\rm MeV$) and the decay constant ($F_{\pi}=90(8)(2)\rm MeV$).  Apart from the condensate we also have the values for $m_\pi$ and $F_\pi$, which are used to estimate the model parameters $m$, $G_0$ and $\Lambda$.
\begin{table}[t]
	\centering
	\begin{tabular}{|m{2.8cm} | m{1.42cm} | m{2cm} |  }
		\hline
		                 &   c        &  $\rchi^2$ per DoF  \\  
		\hline
	    Parameter Set CC &   0.0749    &   0.334    \\ 
		\hline
		Parameter Set HH &  -0.0198   &   6.076   \\ 
		\hline
		Parameter Set HL &   0.2200    &   0.461   \\ 
		\hline
		Parameter Set LH &  -0.0255   &   2.753   \\ 
		\hline
		Parameter Set LL &   0.338    &   0.157   \\ 
		\hline
	\end{tabular}
	\caption{$\rchi^2$ fitting of condensate difference in $c$ for all the five parameter sets from Brandt13~\cite{Brandt:2013dua}.}
	\label{tab:B13ChiSquared}
\end{table}

We follow here the same procedures as we did for the JLQCD data set. We explore all the data set including the corner ones and look for the data set which could reproduce the IMC effect satisfactorily as compared to the LQCD data~\cite{Bali:2012zg}.

\begin{figure}[!h]
	\includegraphics[scale=0.9]{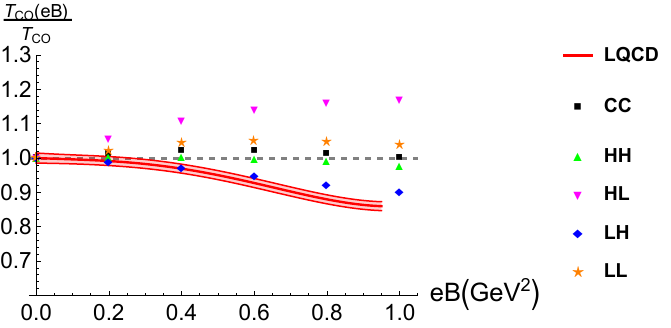}
	\caption{The phase diagram in $T-eB$ plane for all parameter sets of Table~\ref{tab:ParametersB13} (Brandt13~\cite{Brandt:2013dua}) along with that as given by LQCD~\cite{Bali:2011qj}.}
	\label{fig:PDB13Com}
\end{figure}

Looking at the Fig.~\ref{fig:PDB13Com}, we observe that for Brandt13 LH ($232.5,\,98.25$) is the lone parameter set that can replicate the IMC effect, quite satisfactorily. Then we move to the second part of our analysis, in which we look at the fitted $c$ values (Table~\ref{tab:B13ChiSquared}) corresponding to the IMC reproducing parameter set(s). In this case the fitted $c$ value is $-0.0255$ for LH. 

Now evoking the $\eta^*$ related phenomenological argument as given in the main text before Eq.~\ref{eq:cPhysicalConstraints}, where the plausible range of $c$ is given as $c\in (0.12,0.5]$, we can exclude the LH parameter set of Brandt13 data as a favourable one for the present model. Thus we conclude that from Brandt13 data set~\cite{Brandt:2013dua} we do not find any suitable parameter set for the present model, which can be reliably explored.

\section{$c$ dependence of $\chi_t$ in the LH parameter set of JLQCD}
\label{Chi_t_JLH_AsC}
\begin{figure}[!htbp]
	\includegraphics[scale=0.9]{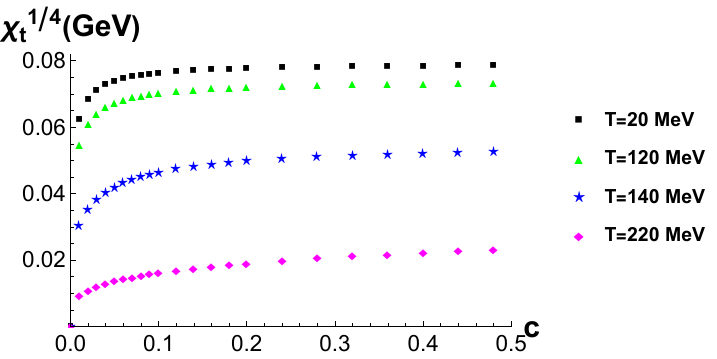}
	\caption{$\chi_t$ as a function of the explicit $U(1)_A$ symmetric breaking parameter $c$ at different temperatures for the LH parameter set of JLQCD~\cite{Fukaya:2007pn}.}
	\label{fig:top_sus_jlqcdlh}
\end{figure}

In Fig.~\ref{fig:top_sus_jlqcdlh} we show the topological susceptibility as a function of the parameter $c$ for different values of $T$ at zero $eB$ in the LH parameter set of JLQCD~\cite{Fukaya:2007pn}. For a fixed $eB$, the topological susceptibility is $0$ for $c=0$ and rises as
we increase $c$ from $0$. This rise is very sharp for smaller values of $c$ and saturates very fast, particularly for the low-temperature values. Thus a smaller $c$ value will not be able to reproduce the expected result for $\chi_t$. This gives us another reason, although in the hindsight, to exclude any parameter set which offers a smaller fitted $c$ value and corroborates our choice of excluding the HH parameter set (for which the fitted central $c$ value is $0.044$) from $\eta^*$ phenomenology. 

\end{appendices}

\bibliography{ref}

\begin{thebibliography}{78}%
\makeatletter
\providecommand \@ifxundefined [1]{%
 \@ifx{#1\undefined}
}%
\providecommand \@ifnum [1]{%
 \ifnum #1\expandafter \@firstoftwo
 \else \expandafter \@secondoftwo
 \fi
}%
\providecommand \@ifx [1]{%
 \ifx #1\expandafter \@firstoftwo
 \else \expandafter \@secondoftwo
 \fi
}%
\providecommand \natexlab [1]{#1}%
\providecommand \enquote  [1]{``#1''}%
\providecommand \bibnamefont  [1]{#1}%
\providecommand \bibfnamefont [1]{#1}%
\providecommand \citenamefont [1]{#1}%
\providecommand \href@noop [0]{\@secondoftwo}%
\providecommand \href [0]{\begingroup \@sanitize@url \@href}%
\providecommand \@href[1]{\@@startlink{#1}\@@href}%
\providecommand \@@href[1]{\endgroup#1\@@endlink}%
\providecommand \@sanitize@url [0]{\catcode `\\12\catcode `\$12\catcode
  `\&12\catcode `\#12\catcode `\^12\catcode `\_12\catcode `\%12\relax}%
\providecommand \@@startlink[1]{}%
\providecommand \@@endlink[0]{}%
\providecommand \url  [0]{\begingroup\@sanitize@url \@url }%
\providecommand \@url [1]{\endgroup\@href {#1}{\urlprefix }}%
\providecommand \urlprefix  [0]{URL }%
\providecommand \Eprint [0]{\href }%
\providecommand \doibase [0]{http://dx.doi.org/}%
\providecommand \selectlanguage [0]{\@gobble}%
\providecommand \bibinfo  [0]{\@secondoftwo}%
\providecommand \bibfield  [0]{\@secondoftwo}%
\providecommand \translation [1]{[#1]}%
\providecommand \BibitemOpen [0]{}%
\providecommand \bibitemStop [0]{}%
\providecommand \bibitemNoStop [0]{.\EOS\space}%
\providecommand \EOS [0]{\spacefactor3000\relax}%
\providecommand \BibitemShut  [1]{\csname bibitem#1\endcsname}%
\let\auto@bib@innerbib\@empty
%</preamble>
\bibitem [{\citenamefont {Pagura}\ \emph {et~al.}(2017)\citenamefont {Pagura},
  \citenamefont {Gomez~Dumm}, \citenamefont {Noguera},\ and\ \citenamefont
  {Scoccola}}]{Pagura:2016pwr}%
  \BibitemOpen
  \bibfield  {author} {\bibinfo {author} {\bibfnamefont {V.}~\bibnamefont
  {Pagura}}, \bibinfo {author} {\bibfnamefont {D.}~\bibnamefont {Gomez~Dumm}},
  \bibinfo {author} {\bibfnamefont {S.}~\bibnamefont {Noguera}}, \ and\
  \bibinfo {author} {\bibfnamefont {N.}~\bibnamefont {Scoccola}},\ }\href
  {\doibase 10.1103/PhysRevD.95.034013} {\bibfield  {journal} {\bibinfo
  {journal} {Phys. Rev. D}\ }\textbf {\bibinfo {volume} {95}},\ \bibinfo
  {pages} {034013} (\bibinfo {year} {2017})},\ \Eprint
  {http://arxiv.org/abs/1609.02025} {arXiv:1609.02025 [hep-ph]} \BibitemShut
  {NoStop}%
\bibitem [{\citenamefont {Kharzeev}\ \emph {et~al.}(2013)\citenamefont
  {Kharzeev}, \citenamefont {Landsteiner}, \citenamefont {Schmitt},\ and\
  \citenamefont {Yee}}]{Kharzeev:2013jha}%
  \BibitemOpen
  \bibinfo {editor} {\bibfnamefont {D.}~\bibnamefont {Kharzeev}}, \bibinfo
  {editor} {\bibfnamefont {K.}~\bibnamefont {Landsteiner}}, \bibinfo {editor}
  {\bibfnamefont {A.}~\bibnamefont {Schmitt}}, \ and\ \bibinfo {editor}
  {\bibfnamefont {H.-U.}\ \bibnamefont {Yee}},\ eds.,\ \href {\doibase
  10.1007/978-3-642-37305-3} {\emph {\bibinfo {title} {{Strongly Interacting
  Matter in Magnetic Fields}}}},\ Vol.\ \bibinfo {volume} {871}\ (\bibinfo
  {year} {2013})\BibitemShut {NoStop}%
\bibitem [{\citenamefont {Skokov}\ \emph {et~al.}(2009)\citenamefont {Skokov},
  \citenamefont {Illarionov},\ and\ \citenamefont {Toneev}}]{Skokov:2009qp}%
  \BibitemOpen
  \bibfield  {author} {\bibinfo {author} {\bibfnamefont {V.}~\bibnamefont
  {Skokov}}, \bibinfo {author} {\bibfnamefont {A.}~\bibnamefont {Illarionov}},
  \ and\ \bibinfo {author} {\bibfnamefont {V.}~\bibnamefont {Toneev}},\ }\href
  {\doibase 10.1142/S0217751X09047570} {\bibfield  {journal} {\bibinfo
  {journal} {Int. J. Mod. Phys. A}\ }\textbf {\bibinfo {volume} {24}},\
  \bibinfo {pages} {5925} (\bibinfo {year} {2009})},\ \Eprint
  {http://arxiv.org/abs/0907.1396} {arXiv:0907.1396 [nucl-th]} \BibitemShut
  {NoStop}%
\bibitem [{\citenamefont {Tuchin}(2016)}]{Tuchin:2015oka}%
  \BibitemOpen
  \bibfield  {author} {\bibinfo {author} {\bibfnamefont {K.}~\bibnamefont
  {Tuchin}},\ }\href {\doibase 10.1103/PhysRevC.93.014905} {\bibfield
  {journal} {\bibinfo  {journal} {Phys. Rev. C}\ }\textbf {\bibinfo {volume}
  {93}},\ \bibinfo {pages} {014905} (\bibinfo {year} {2016})},\ \Eprint
  {http://arxiv.org/abs/1508.06925} {arXiv:1508.06925 [hep-ph]} \BibitemShut
  {NoStop}%
\bibitem [{\citenamefont {Lattimer}\ and\ \citenamefont
  {Prakash}(2007)}]{Lattimer:2006xb}%
  \BibitemOpen
  \bibfield  {author} {\bibinfo {author} {\bibfnamefont {J.~M.}\ \bibnamefont
  {Lattimer}}\ and\ \bibinfo {author} {\bibfnamefont {M.}~\bibnamefont
  {Prakash}},\ }\href {\doibase 10.1016/j.physrep.2007.02.003} {\bibfield
  {journal} {\bibinfo  {journal} {Phys. Rept.}\ }\textbf {\bibinfo {volume}
  {442}},\ \bibinfo {pages} {109} (\bibinfo {year} {2007})},\ \Eprint
  {http://arxiv.org/abs/astro-ph/0612440} {arXiv:astro-ph/0612440} \BibitemShut
  {NoStop}%
\bibitem [{\citenamefont {Schramm}\ \emph {et~al.}(1992)\citenamefont
  {Schramm}, \citenamefont {Muller},\ and\ \citenamefont
  {Schramm}}]{Schramm:1991ex}%
  \BibitemOpen
  \bibfield  {author} {\bibinfo {author} {\bibfnamefont {S.}~\bibnamefont
  {Schramm}}, \bibinfo {author} {\bibfnamefont {B.}~\bibnamefont {Muller}}, \
  and\ \bibinfo {author} {\bibfnamefont {A.~J.}\ \bibnamefont {Schramm}},\
  }\href {\doibase 10.1142/S0217732392000860} {\bibfield  {journal} {\bibinfo
  {journal} {Mod. Phys. Lett. A}\ }\textbf {\bibinfo {volume} {7}},\ \bibinfo
  {pages} {973} (\bibinfo {year} {1992})}\BibitemShut {NoStop}%
\bibitem [{\citenamefont {Gusynin}\ \emph {et~al.}(1996)\citenamefont
  {Gusynin}, \citenamefont {Miransky},\ and\ \citenamefont
  {Shovkovy}}]{Gusynin:1995nb}%
  \BibitemOpen
  \bibfield  {author} {\bibinfo {author} {\bibfnamefont {V.}~\bibnamefont
  {Gusynin}}, \bibinfo {author} {\bibfnamefont {V.}~\bibnamefont {Miransky}}, \
  and\ \bibinfo {author} {\bibfnamefont {I.}~\bibnamefont {Shovkovy}},\ }\href
  {\doibase 10.1016/0550-3213(96)00021-1} {\bibfield  {journal} {\bibinfo
  {journal} {Nucl. Phys. B}\ }\textbf {\bibinfo {volume} {462}},\ \bibinfo
  {pages} {249} (\bibinfo {year} {1996})},\ \Eprint
  {http://arxiv.org/abs/hep-ph/9509320} {arXiv:hep-ph/9509320} \BibitemShut
  {NoStop}%
\bibitem [{\citenamefont {Lee}\ \emph {et~al.}(1997)\citenamefont {Lee},
  \citenamefont {Leung},\ and\ \citenamefont {Ng}}]{Lee:1997zj}%
  \BibitemOpen
  \bibfield  {author} {\bibinfo {author} {\bibfnamefont {D.}~\bibnamefont
  {Lee}}, \bibinfo {author} {\bibfnamefont {C.~N.}\ \bibnamefont {Leung}}, \
  and\ \bibinfo {author} {\bibfnamefont {Y.}~\bibnamefont {Ng}},\ }\href
  {\doibase 10.1103/PhysRevD.55.6504} {\bibfield  {journal} {\bibinfo
  {journal} {Phys. Rev. D}\ }\textbf {\bibinfo {volume} {55}},\ \bibinfo
  {pages} {6504} (\bibinfo {year} {1997})},\ \Eprint
  {http://arxiv.org/abs/hep-th/9701172} {arXiv:hep-th/9701172} \BibitemShut
  {NoStop}%
\bibitem [{\citenamefont {Fraga}\ and\ \citenamefont
  {Mizher}(2008)}]{Fraga:2008qn}%
  \BibitemOpen
  \bibfield  {author} {\bibinfo {author} {\bibfnamefont {E.~S.}\ \bibnamefont
  {Fraga}}\ and\ \bibinfo {author} {\bibfnamefont {A.~J.}\ \bibnamefont
  {Mizher}},\ }\href {\doibase 10.1103/PhysRevD.78.025016} {\bibfield
  {journal} {\bibinfo  {journal} {Phys. Rev. D}\ }\textbf {\bibinfo {volume}
  {78}},\ \bibinfo {pages} {025016} (\bibinfo {year} {2008})},\ \Eprint
  {http://arxiv.org/abs/0804.1452} {arXiv:0804.1452 [hep-ph]} \BibitemShut
  {NoStop}%
\bibitem [{\citenamefont {Boomsma}\ and\ \citenamefont
  {Boer}(2010)}]{Boomsma:2009yk}%
  \BibitemOpen
  \bibfield  {author} {\bibinfo {author} {\bibfnamefont {J.~K.}\ \bibnamefont
  {Boomsma}}\ and\ \bibinfo {author} {\bibfnamefont {D.}~\bibnamefont {Boer}},\
  }\href {\doibase 10.1103/PhysRevD.81.074005} {\bibfield  {journal} {\bibinfo
  {journal} {Phys. Rev. D}\ }\textbf {\bibinfo {volume} {81}},\ \bibinfo
  {pages} {074005} (\bibinfo {year} {2010})},\ \Eprint
  {http://arxiv.org/abs/0911.2164} {arXiv:0911.2164 [hep-ph]} \BibitemShut
  {NoStop}%
\bibitem [{\citenamefont {Gatto}\ and\ \citenamefont
  {Ruggieri}(2011)}]{Gatto:2010pt}%
  \BibitemOpen
  \bibfield  {author} {\bibinfo {author} {\bibfnamefont {R.}~\bibnamefont
  {Gatto}}\ and\ \bibinfo {author} {\bibfnamefont {M.}~\bibnamefont
  {Ruggieri}},\ }\href {\doibase 10.1103/PhysRevD.83.034016} {\bibfield
  {journal} {\bibinfo  {journal} {Phys. Rev. D}\ }\textbf {\bibinfo {volume}
  {83}},\ \bibinfo {pages} {034016} (\bibinfo {year} {2011})},\ \Eprint
  {http://arxiv.org/abs/1012.1291} {arXiv:1012.1291 [hep-ph]} \BibitemShut
  {NoStop}%
\bibitem [{\citenamefont {Chatterjee}\ \emph {et~al.}(2011)\citenamefont
  {Chatterjee}, \citenamefont {Mishra},\ and\ \citenamefont
  {Mishra}}]{Chatterjee:2011ry}%
  \BibitemOpen
  \bibfield  {author} {\bibinfo {author} {\bibfnamefont {B.}~\bibnamefont
  {Chatterjee}}, \bibinfo {author} {\bibfnamefont {H.}~\bibnamefont {Mishra}},
  \ and\ \bibinfo {author} {\bibfnamefont {A.}~\bibnamefont {Mishra}},\ }\href
  {\doibase 10.1103/PhysRevD.84.014016} {\bibfield  {journal} {\bibinfo
  {journal} {Phys. Rev. D}\ }\textbf {\bibinfo {volume} {84}},\ \bibinfo
  {pages} {014016} (\bibinfo {year} {2011})},\ \Eprint
  {http://arxiv.org/abs/1101.0498} {arXiv:1101.0498 [hep-ph]} \BibitemShut
  {NoStop}%
\bibitem [{\citenamefont {Ferrari}\ \emph {et~al.}(2012)\citenamefont
  {Ferrari}, \citenamefont {Garcia},\ and\ \citenamefont
  {Pinto}}]{Ferrari:2012yw}%
  \BibitemOpen
  \bibfield  {author} {\bibinfo {author} {\bibfnamefont {G.~N.}\ \bibnamefont
  {Ferrari}}, \bibinfo {author} {\bibfnamefont {A.~F.}\ \bibnamefont {Garcia}},
  \ and\ \bibinfo {author} {\bibfnamefont {M.~B.}\ \bibnamefont {Pinto}},\
  }\href {\doibase 10.1103/PhysRevD.86.096005} {\bibfield  {journal} {\bibinfo
  {journal} {Phys. Rev. D}\ }\textbf {\bibinfo {volume} {86}},\ \bibinfo
  {pages} {096005} (\bibinfo {year} {2012})},\ \Eprint
  {http://arxiv.org/abs/1207.3714} {arXiv:1207.3714 [hep-ph]} \BibitemShut
  {NoStop}%
\bibitem [{\citenamefont {Ferreira}\ \emph
  {et~al.}(2014{\natexlab{a}})\citenamefont {Ferreira}, \citenamefont {Costa},
  \citenamefont {Menezes}, \citenamefont {Providência},\ and\ \citenamefont
  {Scoccola}}]{Ferreira:2013tba}%
  \BibitemOpen
  \bibfield  {author} {\bibinfo {author} {\bibfnamefont {M.}~\bibnamefont
  {Ferreira}}, \bibinfo {author} {\bibfnamefont {P.}~\bibnamefont {Costa}},
  \bibinfo {author} {\bibfnamefont {D.~P.}\ \bibnamefont {Menezes}}, \bibinfo
  {author} {\bibfnamefont {C.}~\bibnamefont {Providência}}, \ and\ \bibinfo
  {author} {\bibfnamefont {N.}~\bibnamefont {Scoccola}},\ }\href {\doibase
  10.1103/PhysRevD.89.016002} {\bibfield  {journal} {\bibinfo  {journal} {Phys.
  Rev. D}\ }\textbf {\bibinfo {volume} {89}},\ \bibinfo {pages} {016002}
  (\bibinfo {year} {2014}{\natexlab{a}})},\ \bibinfo {note} {[Addendum:
  Phys.Rev.D 89, 019902 (2014)]},\ \Eprint {http://arxiv.org/abs/1305.4751}
  {arXiv:1305.4751 [hep-ph]} \BibitemShut {NoStop}%
\bibitem [{\citenamefont {D'Elia}\ \emph {et~al.}(2010)\citenamefont {D'Elia},
  \citenamefont {Mukherjee},\ and\ \citenamefont {Sanfilippo}}]{DElia:2010abb}%
  \BibitemOpen
  \bibfield  {author} {\bibinfo {author} {\bibfnamefont {M.}~\bibnamefont
  {D'Elia}}, \bibinfo {author} {\bibfnamefont {S.}~\bibnamefont {Mukherjee}}, \
  and\ \bibinfo {author} {\bibfnamefont {F.}~\bibnamefont {Sanfilippo}},\
  }\href {\doibase 10.1103/PhysRevD.82.051501} {\bibfield  {journal} {\bibinfo
  {journal} {Phys. Rev. D}\ }\textbf {\bibinfo {volume} {82}},\ \bibinfo
  {pages} {051501} (\bibinfo {year} {2010})},\ \Eprint
  {http://arxiv.org/abs/1005.5365} {arXiv:1005.5365 [hep-lat]} \BibitemShut
  {NoStop}%
\bibitem [{\citenamefont {Bali}\ \emph
  {et~al.}(2012{\natexlab{a}})\citenamefont {Bali}, \citenamefont {Bruckmann},
  \citenamefont {Endrodi}, \citenamefont {Fodor}, \citenamefont {Katz},
  \citenamefont {Krieg}, \citenamefont {Schafer},\ and\ \citenamefont
  {Szabo}}]{Bali:2011qj}%
  \BibitemOpen
  \bibfield  {author} {\bibinfo {author} {\bibfnamefont {G.}~\bibnamefont
  {Bali}}, \bibinfo {author} {\bibfnamefont {F.}~\bibnamefont {Bruckmann}},
  \bibinfo {author} {\bibfnamefont {G.}~\bibnamefont {Endrodi}}, \bibinfo
  {author} {\bibfnamefont {Z.}~\bibnamefont {Fodor}}, \bibinfo {author}
  {\bibfnamefont {S.}~\bibnamefont {Katz}}, \bibinfo {author} {\bibfnamefont
  {S.}~\bibnamefont {Krieg}}, \bibinfo {author} {\bibfnamefont
  {A.}~\bibnamefont {Schafer}}, \ and\ \bibinfo {author} {\bibfnamefont
  {K.}~\bibnamefont {Szabo}},\ }\href {\doibase 10.1007/JHEP02(2012)044}
  {\bibfield  {journal} {\bibinfo  {journal} {JHEP}\ }\textbf {\bibinfo
  {volume} {02}},\ \bibinfo {pages} {044} (\bibinfo {year}
  {2012}{\natexlab{a}})},\ \Eprint {http://arxiv.org/abs/1111.4956}
  {arXiv:1111.4956 [hep-lat]} \BibitemShut {NoStop}%
\bibitem [{\citenamefont {Bali}\ \emph
  {et~al.}(2012{\natexlab{b}})\citenamefont {Bali}, \citenamefont {Bruckmann},
  \citenamefont {Endrodi}, \citenamefont {Fodor}, \citenamefont {Katz},\ and\
  \citenamefont {Schafer}}]{Bali:2012zg}%
  \BibitemOpen
  \bibfield  {author} {\bibinfo {author} {\bibfnamefont {G.}~\bibnamefont
  {Bali}}, \bibinfo {author} {\bibfnamefont {F.}~\bibnamefont {Bruckmann}},
  \bibinfo {author} {\bibfnamefont {G.}~\bibnamefont {Endrodi}}, \bibinfo
  {author} {\bibfnamefont {Z.}~\bibnamefont {Fodor}}, \bibinfo {author}
  {\bibfnamefont {S.}~\bibnamefont {Katz}}, \ and\ \bibinfo {author}
  {\bibfnamefont {A.}~\bibnamefont {Schafer}},\ }\href {\doibase
  10.1103/PhysRevD.86.071502} {\bibfield  {journal} {\bibinfo  {journal} {Phys.
  Rev. D}\ }\textbf {\bibinfo {volume} {86}},\ \bibinfo {pages} {071502}
  (\bibinfo {year} {2012}{\natexlab{b}})},\ \Eprint
  {http://arxiv.org/abs/1206.4205} {arXiv:1206.4205 [hep-lat]} \BibitemShut
  {NoStop}%
\bibitem [{\citenamefont {Endrodi}(2015)}]{Endrodi:2015oba}%
  \BibitemOpen
  \bibfield  {author} {\bibinfo {author} {\bibfnamefont {G.}~\bibnamefont
  {Endrodi}},\ }\href {\doibase 10.1007/JHEP07(2015)173} {\bibfield  {journal}
  {\bibinfo  {journal} {JHEP}\ }\textbf {\bibinfo {volume} {07}},\ \bibinfo
  {pages} {173} (\bibinfo {year} {2015})},\ \Eprint
  {http://arxiv.org/abs/1504.08280} {arXiv:1504.08280 [hep-lat]} \BibitemShut
  {NoStop}%
\bibitem [{\citenamefont {D'Elia}\ \emph {et~al.}(2018)\citenamefont {D'Elia},
  \citenamefont {Manigrasso}, \citenamefont {Negro},\ and\ \citenamefont
  {Sanfilippo}}]{DElia:2018xwo}%
  \BibitemOpen
  \bibfield  {author} {\bibinfo {author} {\bibfnamefont {M.}~\bibnamefont
  {D'Elia}}, \bibinfo {author} {\bibfnamefont {F.}~\bibnamefont {Manigrasso}},
  \bibinfo {author} {\bibfnamefont {F.}~\bibnamefont {Negro}}, \ and\ \bibinfo
  {author} {\bibfnamefont {F.}~\bibnamefont {Sanfilippo}},\ }\href {\doibase
  10.1103/PhysRevD.98.054509} {\bibfield  {journal} {\bibinfo  {journal} {Phys.
  Rev. D}\ }\textbf {\bibinfo {volume} {98}},\ \bibinfo {pages} {054509}
  (\bibinfo {year} {2018})},\ \Eprint {http://arxiv.org/abs/1808.07008}
  {arXiv:1808.07008 [hep-lat]} \BibitemShut {NoStop}%
\bibitem [{\citenamefont {Endrodi}\ \emph {et~al.}(2019)\citenamefont
  {Endrodi}, \citenamefont {Giordano}, \citenamefont {Katz}, \citenamefont
  {Kovács},\ and\ \citenamefont {Pittler}}]{Endrodi:2019zrl}%
  \BibitemOpen
  \bibfield  {author} {\bibinfo {author} {\bibfnamefont {G.}~\bibnamefont
  {Endrodi}}, \bibinfo {author} {\bibfnamefont {M.}~\bibnamefont {Giordano}},
  \bibinfo {author} {\bibfnamefont {S.~D.}\ \bibnamefont {Katz}}, \bibinfo
  {author} {\bibfnamefont {T.}~\bibnamefont {Kovács}}, \ and\ \bibinfo
  {author} {\bibfnamefont {F.}~\bibnamefont {Pittler}},\ }\href {\doibase
  10.1007/JHEP07(2019)007} {\bibfield  {journal} {\bibinfo  {journal} {JHEP}\
  }\textbf {\bibinfo {volume} {07}},\ \bibinfo {pages} {007} (\bibinfo {year}
  {2019})},\ \Eprint {http://arxiv.org/abs/1904.10296} {arXiv:1904.10296
  [hep-lat]} \BibitemShut {NoStop}%
\bibitem [{\citenamefont {Chao}\ \emph {et~al.}(2013)\citenamefont {Chao},
  \citenamefont {Chu},\ and\ \citenamefont {Huang}}]{Chao:2013qpa}%
  \BibitemOpen
  \bibfield  {author} {\bibinfo {author} {\bibfnamefont {J.}~\bibnamefont
  {Chao}}, \bibinfo {author} {\bibfnamefont {P.}~\bibnamefont {Chu}}, \ and\
  \bibinfo {author} {\bibfnamefont {M.}~\bibnamefont {Huang}},\ }\href
  {\doibase 10.1103/PhysRevD.88.054009} {\bibfield  {journal} {\bibinfo
  {journal} {Phys. Rev. D}\ }\textbf {\bibinfo {volume} {88}},\ \bibinfo
  {pages} {054009} (\bibinfo {year} {2013})},\ \Eprint
  {http://arxiv.org/abs/1305.1100} {arXiv:1305.1100 [hep-ph]} \BibitemShut
  {NoStop}%
\bibitem [{\citenamefont {Farias}\ \emph {et~al.}(2014)\citenamefont {Farias},
  \citenamefont {Gomes}, \citenamefont {Krein},\ and\ \citenamefont
  {Pinto}}]{Farias:2014eca}%
  \BibitemOpen
  \bibfield  {author} {\bibinfo {author} {\bibfnamefont {R.}~\bibnamefont
  {Farias}}, \bibinfo {author} {\bibfnamefont {K.}~\bibnamefont {Gomes}},
  \bibinfo {author} {\bibfnamefont {G.}~\bibnamefont {Krein}}, \ and\ \bibinfo
  {author} {\bibfnamefont {M.}~\bibnamefont {Pinto}},\ }\href {\doibase
  10.1103/PhysRevC.90.025203} {\bibfield  {journal} {\bibinfo  {journal} {Phys.
  Rev. C}\ }\textbf {\bibinfo {volume} {90}},\ \bibinfo {pages} {025203}
  (\bibinfo {year} {2014})},\ \Eprint {http://arxiv.org/abs/1404.3931}
  {arXiv:1404.3931 [hep-ph]} \BibitemShut {NoStop}%
\bibitem [{\citenamefont {Ferreira}\ \emph
  {et~al.}(2014{\natexlab{b}})\citenamefont {Ferreira}, \citenamefont {Costa},
  \citenamefont {Louren{\c c}o}, \citenamefont {Frederico},\ and\ \citenamefont
  {Providência}}]{Ferreira:2014kpa}%
  \BibitemOpen
  \bibfield  {author} {\bibinfo {author} {\bibfnamefont {M.}~\bibnamefont
  {Ferreira}}, \bibinfo {author} {\bibfnamefont {P.}~\bibnamefont {Costa}},
  \bibinfo {author} {\bibfnamefont {O.}~\bibnamefont {Louren{\c c}o}}, \bibinfo
  {author} {\bibfnamefont {T.}~\bibnamefont {Frederico}}, \ and\ \bibinfo
  {author} {\bibfnamefont {C.}~\bibnamefont {Providência}},\ }\href {\doibase
  10.1103/PhysRevD.89.116011} {\bibfield  {journal} {\bibinfo  {journal} {Phys.
  Rev. D}\ }\textbf {\bibinfo {volume} {89}},\ \bibinfo {pages} {116011}
  (\bibinfo {year} {2014}{\natexlab{b}})},\ \Eprint
  {http://arxiv.org/abs/1404.5577} {arXiv:1404.5577 [hep-ph]} \BibitemShut
  {NoStop}%
\bibitem [{\citenamefont {Ayala}\ \emph {et~al.}(2014)\citenamefont {Ayala},
  \citenamefont {Loewe}, \citenamefont {Mizher},\ and\ \citenamefont
  {Zamora}}]{Ayala:2014iba}%
  \BibitemOpen
  \bibfield  {author} {\bibinfo {author} {\bibfnamefont {A.}~\bibnamefont
  {Ayala}}, \bibinfo {author} {\bibfnamefont {M.}~\bibnamefont {Loewe}},
  \bibinfo {author} {\bibfnamefont {A.~J.}\ \bibnamefont {Mizher}}, \ and\
  \bibinfo {author} {\bibfnamefont {R.}~\bibnamefont {Zamora}},\ }\href
  {\doibase 10.1103/PhysRevD.90.036001} {\bibfield  {journal} {\bibinfo
  {journal} {Phys. Rev. D}\ }\textbf {\bibinfo {volume} {90}},\ \bibinfo
  {pages} {036001} (\bibinfo {year} {2014})},\ \Eprint
  {http://arxiv.org/abs/1406.3885} {arXiv:1406.3885 [hep-ph]} \BibitemShut
  {NoStop}%
\bibitem [{\citenamefont {Ayala}\ \emph {et~al.}(2015)\citenamefont {Ayala},
  \citenamefont {Loewe},\ and\ \citenamefont {Zamora}}]{Ayala:2014gwa}%
  \BibitemOpen
  \bibfield  {author} {\bibinfo {author} {\bibfnamefont {A.}~\bibnamefont
  {Ayala}}, \bibinfo {author} {\bibfnamefont {M.}~\bibnamefont {Loewe}}, \ and\
  \bibinfo {author} {\bibfnamefont {R.}~\bibnamefont {Zamora}},\ }\href
  {\doibase 10.1103/PhysRevD.91.016002} {\bibfield  {journal} {\bibinfo
  {journal} {Phys. Rev. D}\ }\textbf {\bibinfo {volume} {91}},\ \bibinfo
  {pages} {016002} (\bibinfo {year} {2015})},\ \Eprint
  {http://arxiv.org/abs/1406.7408} {arXiv:1406.7408 [hep-ph]} \BibitemShut
  {NoStop}%
\bibitem [{\citenamefont {Ferrer}\ \emph {et~al.}(2015)\citenamefont {Ferrer},
  \citenamefont {de~la Incera},\ and\ \citenamefont {Wen}}]{Ferrer:2014qka}%
  \BibitemOpen
  \bibfield  {author} {\bibinfo {author} {\bibfnamefont {E.}~\bibnamefont
  {Ferrer}}, \bibinfo {author} {\bibfnamefont {V.}~\bibnamefont {de~la
  Incera}}, \ and\ \bibinfo {author} {\bibfnamefont {X.}~\bibnamefont {Wen}},\
  }\href {\doibase 10.1103/PhysRevD.91.054006} {\bibfield  {journal} {\bibinfo
  {journal} {Phys. Rev. D}\ }\textbf {\bibinfo {volume} {91}},\ \bibinfo
  {pages} {054006} (\bibinfo {year} {2015})},\ \Eprint
  {http://arxiv.org/abs/1407.3503} {arXiv:1407.3503 [nucl-th]} \BibitemShut
  {NoStop}%
\bibitem [{\citenamefont {Andersen}\ \emph {et~al.}(2015)\citenamefont
  {Andersen}, \citenamefont {Naylor},\ and\ \citenamefont
  {Tranberg}}]{Andersen:2014oaa}%
  \BibitemOpen
  \bibfield  {author} {\bibinfo {author} {\bibfnamefont {J.~O.}\ \bibnamefont
  {Andersen}}, \bibinfo {author} {\bibfnamefont {W.~R.}\ \bibnamefont
  {Naylor}}, \ and\ \bibinfo {author} {\bibfnamefont {A.}~\bibnamefont
  {Tranberg}},\ }\href {\doibase 10.1007/JHEP02(2015)042} {\bibfield  {journal}
  {\bibinfo  {journal} {JHEP}\ }\textbf {\bibinfo {volume} {02}},\ \bibinfo
  {pages} {042} (\bibinfo {year} {2015})},\ \Eprint
  {http://arxiv.org/abs/1410.5247} {arXiv:1410.5247 [hep-ph]} \BibitemShut
  {NoStop}%
\bibitem [{\citenamefont {Yu}\ \emph {et~al.}(2015)\citenamefont {Yu},
  \citenamefont {Van~Doorsselaere},\ and\ \citenamefont {Huang}}]{Yu:2014xoa}%
  \BibitemOpen
  \bibfield  {author} {\bibinfo {author} {\bibfnamefont {L.}~\bibnamefont
  {Yu}}, \bibinfo {author} {\bibfnamefont {J.}~\bibnamefont
  {Van~Doorsselaere}}, \ and\ \bibinfo {author} {\bibfnamefont
  {M.}~\bibnamefont {Huang}},\ }\href {\doibase 10.1103/PhysRevD.91.074011}
  {\bibfield  {journal} {\bibinfo  {journal} {Phys. Rev. D}\ }\textbf {\bibinfo
  {volume} {91}},\ \bibinfo {pages} {074011} (\bibinfo {year} {2015})},\
  \Eprint {http://arxiv.org/abs/1411.7552} {arXiv:1411.7552 [hep-ph]}
  \BibitemShut {NoStop}%
\bibitem [{\citenamefont {Providência}\ \emph {et~al.}(2015)\citenamefont
  {Providência}, \citenamefont {Ferreira},\ and\ \citenamefont
  {Costa}}]{Providencia:2014txa}%
  \BibitemOpen
  \bibfield  {author} {\bibinfo {author} {\bibfnamefont {C.}~\bibnamefont
  {Providência}}, \bibinfo {author} {\bibfnamefont {M.}~\bibnamefont
  {Ferreira}}, \ and\ \bibinfo {author} {\bibfnamefont {P.}~\bibnamefont
  {Costa}},\ }\href {\doibase 10.5506/APhysPolBSupp.8.207} {\bibfield
  {journal} {\bibinfo  {journal} {Acta Phys. Polon. Supp.}\ }\textbf {\bibinfo
  {volume} {8}},\ \bibinfo {pages} {207} (\bibinfo {year} {2015})},\ \Eprint
  {http://arxiv.org/abs/1412.8308} {arXiv:1412.8308 [hep-ph]} \BibitemShut
  {NoStop}%
\bibitem [{\citenamefont {Mueller}\ and\ \citenamefont
  {Pawlowski}(2015)}]{Mueller:2015fka}%
  \BibitemOpen
  \bibfield  {author} {\bibinfo {author} {\bibfnamefont {N.}~\bibnamefont
  {Mueller}}\ and\ \bibinfo {author} {\bibfnamefont {J.~M.}\ \bibnamefont
  {Pawlowski}},\ }\href {\doibase 10.1103/PhysRevD.91.116010} {\bibfield
  {journal} {\bibinfo  {journal} {Phys. Rev. D}\ }\textbf {\bibinfo {volume}
  {91}},\ \bibinfo {pages} {116010} (\bibinfo {year} {2015})},\ \Eprint
  {http://arxiv.org/abs/1502.08011} {arXiv:1502.08011 [hep-ph]} \BibitemShut
  {NoStop}%
\bibitem [{\citenamefont {Mao}(2016)}]{Mao:2016fha}%
  \BibitemOpen
  \bibfield  {author} {\bibinfo {author} {\bibfnamefont {S.}~\bibnamefont
  {Mao}},\ }\href {\doibase 10.1016/j.physletb.2016.05.018} {\bibfield
  {journal} {\bibinfo  {journal} {Phys. Lett. B}\ }\textbf {\bibinfo {volume}
  {758}},\ \bibinfo {pages} {195} (\bibinfo {year} {2016})},\ \Eprint
  {http://arxiv.org/abs/1602.06503} {arXiv:1602.06503 [hep-ph]} \BibitemShut
  {NoStop}%
\bibitem [{\citenamefont {Farias}\ \emph {et~al.}(2017)\citenamefont {Farias},
  \citenamefont {Timoteo}, \citenamefont {Avancini}, \citenamefont {Pinto},\
  and\ \citenamefont {Krein}}]{Farias:2016gmy}%
  \BibitemOpen
  \bibfield  {author} {\bibinfo {author} {\bibfnamefont {R.}~\bibnamefont
  {Farias}}, \bibinfo {author} {\bibfnamefont {V.}~\bibnamefont {Timoteo}},
  \bibinfo {author} {\bibfnamefont {S.}~\bibnamefont {Avancini}}, \bibinfo
  {author} {\bibfnamefont {M.}~\bibnamefont {Pinto}}, \ and\ \bibinfo {author}
  {\bibfnamefont {G.}~\bibnamefont {Krein}},\ }\href {\doibase
  10.1140/epja/i2017-12320-8} {\bibfield  {journal} {\bibinfo  {journal} {Eur.
  Phys. J. A}\ }\textbf {\bibinfo {volume} {53}},\ \bibinfo {pages} {101}
  (\bibinfo {year} {2017})},\ \Eprint {http://arxiv.org/abs/1603.03847}
  {arXiv:1603.03847 [hep-ph]} \BibitemShut {NoStop}%
\bibitem [{\citenamefont {Gómez~Dumm}\ \emph {et~al.}(2017)\citenamefont
  {Gómez~Dumm}, \citenamefont {Izzo~Villafañe}, \citenamefont {Noguera},
  \citenamefont {Pagura},\ and\ \citenamefont {Scoccola}}]{GomezDumm:2017iex}%
  \BibitemOpen
  \bibfield  {author} {\bibinfo {author} {\bibfnamefont {D.}~\bibnamefont
  {Gómez~Dumm}}, \bibinfo {author} {\bibfnamefont {M.}~\bibnamefont
  {Izzo~Villafañe}}, \bibinfo {author} {\bibfnamefont {S.}~\bibnamefont
  {Noguera}}, \bibinfo {author} {\bibfnamefont {V.}~\bibnamefont {Pagura}}, \
  and\ \bibinfo {author} {\bibfnamefont {N.}~\bibnamefont {Scoccola}},\ }\href
  {\doibase 10.1103/PhysRevD.96.114012} {\bibfield  {journal} {\bibinfo
  {journal} {Phys. Rev. D}\ }\textbf {\bibinfo {volume} {96}},\ \bibinfo
  {pages} {114012} (\bibinfo {year} {2017})},\ \Eprint
  {http://arxiv.org/abs/1709.04742} {arXiv:1709.04742 [hep-ph]} \BibitemShut
  {NoStop}%
\bibitem [{\citenamefont {Nambu}\ and\ \citenamefont
  {Jona-Lasinio}(1961{\natexlab{a}})}]{Nambu:1961tp}%
  \BibitemOpen
  \bibfield  {author} {\bibinfo {author} {\bibfnamefont {Y.}~\bibnamefont
  {Nambu}}\ and\ \bibinfo {author} {\bibfnamefont {G.}~\bibnamefont
  {Jona-Lasinio}},\ }\href {\doibase 10.1103/PhysRev.122.345} {\bibfield
  {journal} {\bibinfo  {journal} {Phys. Rev.}\ }\textbf {\bibinfo {volume}
  {122}},\ \bibinfo {pages} {345} (\bibinfo {year}
  {1961}{\natexlab{a}})}\BibitemShut {NoStop}%
\bibitem [{\citenamefont {Nambu}\ and\ \citenamefont
  {Jona-Lasinio}(1961{\natexlab{b}})}]{Nambu:1961fr}%
  \BibitemOpen
  \bibfield  {author} {\bibinfo {author} {\bibfnamefont {Y.}~\bibnamefont
  {Nambu}}\ and\ \bibinfo {author} {\bibfnamefont {G.}~\bibnamefont
  {Jona-Lasinio}},\ }\href {\doibase 10.1103/PhysRev.124.246} {\bibfield
  {journal} {\bibinfo  {journal} {Phys. Rev.}\ }\textbf {\bibinfo {volume}
  {124}},\ \bibinfo {pages} {246} (\bibinfo {year}
  {1961}{\natexlab{b}})}\BibitemShut {NoStop}%
\bibitem [{\citenamefont {Bowler}\ and\ \citenamefont
  {Birse}(1995)}]{Bowler:1994ir}%
  \BibitemOpen
  \bibfield  {author} {\bibinfo {author} {\bibfnamefont {R.}~\bibnamefont
  {Bowler}}\ and\ \bibinfo {author} {\bibfnamefont {M.}~\bibnamefont {Birse}},\
  }\href {\doibase 10.1016/0375-9474(94)00481-2} {\bibfield  {journal}
  {\bibinfo  {journal} {Nucl. Phys. A}\ }\textbf {\bibinfo {volume} {582}},\
  \bibinfo {pages} {655} (\bibinfo {year} {1995})},\ \Eprint
  {http://arxiv.org/abs/hep-ph/9407336} {arXiv:hep-ph/9407336} \BibitemShut
  {NoStop}%
\bibitem [{\citenamefont {Plant}\ and\ \citenamefont
  {Birse}(1998)}]{Plant:1997jr}%
  \BibitemOpen
  \bibfield  {author} {\bibinfo {author} {\bibfnamefont {R.~S.}\ \bibnamefont
  {Plant}}\ and\ \bibinfo {author} {\bibfnamefont {M.~C.}\ \bibnamefont
  {Birse}},\ }\href {\doibase 10.1016/S0375-9474(97)00635-0} {\bibfield
  {journal} {\bibinfo  {journal} {Nucl. Phys. A}\ }\textbf {\bibinfo {volume}
  {628}},\ \bibinfo {pages} {607} (\bibinfo {year} {1998})},\ \Eprint
  {http://arxiv.org/abs/hep-ph/9705372} {arXiv:hep-ph/9705372} \BibitemShut
  {NoStop}%
\bibitem [{\citenamefont {General}\ \emph {et~al.}(2001)\citenamefont
  {General}, \citenamefont {Gomez~Dumm},\ and\ \citenamefont
  {Scoccola}}]{General:2000zx}%
  \BibitemOpen
  \bibfield  {author} {\bibinfo {author} {\bibfnamefont {I.}~\bibnamefont
  {General}}, \bibinfo {author} {\bibfnamefont {D.}~\bibnamefont {Gomez~Dumm}},
  \ and\ \bibinfo {author} {\bibfnamefont {N.}~\bibnamefont {Scoccola}},\
  }\href {\doibase 10.1016/S0370-2693(01)00240-4} {\bibfield  {journal}
  {\bibinfo  {journal} {Phys. Lett. B}\ }\textbf {\bibinfo {volume} {506}},\
  \bibinfo {pages} {267} (\bibinfo {year} {2001})},\ \Eprint
  {http://arxiv.org/abs/hep-ph/0010034} {arXiv:hep-ph/0010034} \BibitemShut
  {NoStop}%
\bibitem [{\citenamefont {Praszalowicz}\ and\ \citenamefont
  {Rostworowski}(2001)}]{Praszalowicz:2001wy}%
  \BibitemOpen
  \bibfield  {author} {\bibinfo {author} {\bibfnamefont {M.}~\bibnamefont
  {Praszalowicz}}\ and\ \bibinfo {author} {\bibfnamefont {A.}~\bibnamefont
  {Rostworowski}},\ }\href {\doibase 10.1103/PhysRevD.64.074003} {\bibfield
  {journal} {\bibinfo  {journal} {Phys. Rev. D}\ }\textbf {\bibinfo {volume}
  {64}},\ \bibinfo {pages} {074003} (\bibinfo {year} {2001})},\ \Eprint
  {http://arxiv.org/abs/hep-ph/0105188} {arXiv:hep-ph/0105188} \BibitemShut
  {NoStop}%
\bibitem [{\citenamefont {Gomez~Dumm}\ and\ \citenamefont
  {Scoccola}(2002)}]{GomezDumm:2001fz}%
  \BibitemOpen
  \bibfield  {author} {\bibinfo {author} {\bibfnamefont {D.}~\bibnamefont
  {Gomez~Dumm}}\ and\ \bibinfo {author} {\bibfnamefont {N.~N.}\ \bibnamefont
  {Scoccola}},\ }\href {\doibase 10.1103/PhysRevD.65.074021} {\bibfield
  {journal} {\bibinfo  {journal} {Phys. Rev. D}\ }\textbf {\bibinfo {volume}
  {65}},\ \bibinfo {pages} {074021} (\bibinfo {year} {2002})},\ \Eprint
  {http://arxiv.org/abs/hep-ph/0107251} {arXiv:hep-ph/0107251} \BibitemShut
  {NoStop}%
\bibitem [{\citenamefont {Kashiwa}(2011)}]{Kashiwa:2011js}%
  \BibitemOpen
  \bibfield  {author} {\bibinfo {author} {\bibfnamefont {K.}~\bibnamefont
  {Kashiwa}},\ }\href {\doibase 10.1103/PhysRevD.83.117901} {\bibfield
  {journal} {\bibinfo  {journal} {Phys. Rev. D}\ }\textbf {\bibinfo {volume}
  {83}},\ \bibinfo {pages} {117901} (\bibinfo {year} {2011})},\ \Eprint
  {http://arxiv.org/abs/1104.5167} {arXiv:1104.5167 [hep-ph]} \BibitemShut
  {NoStop}%
\bibitem [{\citenamefont {Ritus}(1978)}]{Ritus:1978cj}%
  \BibitemOpen
  \bibfield  {author} {\bibinfo {author} {\bibfnamefont {V.}~\bibnamefont
  {Ritus}},\ }\href@noop {} {\bibfield  {journal} {\bibinfo  {journal} {Sov.
  Phys. JETP}\ }\textbf {\bibinfo {volume} {48}},\ \bibinfo {pages} {788}
  (\bibinfo {year} {1978})}\BibitemShut {NoStop}%
\bibitem [{\citenamefont {Gomez~Dumm}\ \emph {et~al.}(2006)\citenamefont
  {Gomez~Dumm}, \citenamefont {Grunfeld},\ and\ \citenamefont
  {Scoccola}}]{GomezDumm:2006vz}%
  \BibitemOpen
  \bibfield  {author} {\bibinfo {author} {\bibfnamefont {D.}~\bibnamefont
  {Gomez~Dumm}}, \bibinfo {author} {\bibfnamefont {A.}~\bibnamefont
  {Grunfeld}}, \ and\ \bibinfo {author} {\bibfnamefont {N.}~\bibnamefont
  {Scoccola}},\ }\href {\doibase 10.1103/PhysRevD.74.054026} {\bibfield
  {journal} {\bibinfo  {journal} {Phys. Rev. D}\ }\textbf {\bibinfo {volume}
  {74}},\ \bibinfo {pages} {054026} (\bibinfo {year} {2006})},\ \Eprint
  {http://arxiv.org/abs/hep-ph/0607023} {arXiv:hep-ph/0607023} \BibitemShut
  {NoStop}%
\bibitem [{\citenamefont {Frank}\ \emph {et~al.}(2003)\citenamefont {Frank},
  \citenamefont {Buballa},\ and\ \citenamefont {Oertel}}]{Frank:2003ve}%
  \BibitemOpen
  \bibfield  {author} {\bibinfo {author} {\bibfnamefont {M.}~\bibnamefont
  {Frank}}, \bibinfo {author} {\bibfnamefont {M.}~\bibnamefont {Buballa}}, \
  and\ \bibinfo {author} {\bibfnamefont {M.}~\bibnamefont {Oertel}},\ }\href
  {\doibase 10.1016/S0370-2693(03)00607-5} {\bibfield  {journal} {\bibinfo
  {journal} {Phys. Lett. B}\ }\textbf {\bibinfo {volume} {562}},\ \bibinfo
  {pages} {221} (\bibinfo {year} {2003})},\ \Eprint
  {http://arxiv.org/abs/hep-ph/0303109} {arXiv:hep-ph/0303109} \BibitemShut
  {NoStop}%
\bibitem [{\citenamefont {Klevansky}(1992)}]{Klevansky:1992qe}%
  \BibitemOpen
  \bibfield  {author} {\bibinfo {author} {\bibfnamefont {S.}~\bibnamefont
  {Klevansky}},\ }\href {\doibase 10.1103/RevModPhys.64.649} {\bibfield
  {journal} {\bibinfo  {journal} {Rev. Mod. Phys.}\ }\textbf {\bibinfo {volume}
  {64}},\ \bibinfo {pages} {649} (\bibinfo {year} {1992})}\BibitemShut
  {NoStop}%
\bibitem [{\citenamefont {Hatsuda}\ and\ \citenamefont
  {Kunihiro}(1994)}]{Hatsuda:1994pi}%
  \BibitemOpen
  \bibfield  {author} {\bibinfo {author} {\bibfnamefont {T.}~\bibnamefont
  {Hatsuda}}\ and\ \bibinfo {author} {\bibfnamefont {T.}~\bibnamefont
  {Kunihiro}},\ }\href {\doibase 10.1016/0370-1573(94)90022-1} {\bibfield
  {journal} {\bibinfo  {journal} {Phys. Rept.}\ }\textbf {\bibinfo {volume}
  {247}},\ \bibinfo {pages} {221} (\bibinfo {year} {1994})},\ \Eprint
  {http://arxiv.org/abs/hep-ph/9401310} {arXiv:hep-ph/9401310} \BibitemShut
  {NoStop}%
\bibitem [{\citenamefont {Aoki}\ \emph {et~al.}(2020)\citenamefont {Aoki} \emph
  {et~al.}}]{Aoki:2019cca}%
  \BibitemOpen
  \bibfield  {author} {\bibinfo {author} {\bibfnamefont {S.}~\bibnamefont
  {Aoki}} \emph {et~al.} (\bibinfo {collaboration} {Flavour Lattice Averaging
  Group}),\ }\href {\doibase 10.1140/epjc/s10052-019-7354-7} {\bibfield
  {journal} {\bibinfo  {journal} {Eur. Phys. J. C}\ }\textbf {\bibinfo {volume}
  {80}},\ \bibinfo {pages} {113} (\bibinfo {year} {2020})},\ \Eprint
  {http://arxiv.org/abs/1902.08191} {arXiv:1902.08191 [hep-lat]} \BibitemShut
  {NoStop}%
\bibitem [{\citenamefont {Fukaya}\ \emph {et~al.}(2008)\citenamefont {Fukaya},
  \citenamefont {Aoki}, \citenamefont {Hashimoto}, \citenamefont {Kaneko},
  \citenamefont {Matsufuru}, \citenamefont {Noaki}, \citenamefont {Ogawa},
  \citenamefont {Onogi},\ and\ \citenamefont {Yamada}}]{Fukaya:2007pn}%
  \BibitemOpen
  \bibfield  {author} {\bibinfo {author} {\bibfnamefont {H.}~\bibnamefont
  {Fukaya}}, \bibinfo {author} {\bibfnamefont {S.}~\bibnamefont {Aoki}},
  \bibinfo {author} {\bibfnamefont {S.}~\bibnamefont {Hashimoto}}, \bibinfo
  {author} {\bibfnamefont {T.}~\bibnamefont {Kaneko}}, \bibinfo {author}
  {\bibfnamefont {H.}~\bibnamefont {Matsufuru}}, \bibinfo {author}
  {\bibfnamefont {J.}~\bibnamefont {Noaki}}, \bibinfo {author} {\bibfnamefont
  {K.}~\bibnamefont {Ogawa}}, \bibinfo {author} {\bibfnamefont
  {T.}~\bibnamefont {Onogi}}, \ and\ \bibinfo {author} {\bibfnamefont
  {N.}~\bibnamefont {Yamada}} (\bibinfo {collaboration} {JLQCD}),\ }\href
  {\doibase 10.1103/PhysRevD.77.074503} {\bibfield  {journal} {\bibinfo
  {journal} {Phys. Rev. D}\ }\textbf {\bibinfo {volume} {77}},\ \bibinfo
  {pages} {074503} (\bibinfo {year} {2008})},\ \Eprint
  {http://arxiv.org/abs/0711.4965} {arXiv:0711.4965 [hep-lat]} \BibitemShut
  {NoStop}%
\bibitem [{\citenamefont {Brandt}\ \emph {et~al.}(2013)\citenamefont {Brandt},
  \citenamefont {Jüttner},\ and\ \citenamefont {Wittig}}]{Brandt:2013dua}%
  \BibitemOpen
  \bibfield  {author} {\bibinfo {author} {\bibfnamefont {B.~B.}\ \bibnamefont
  {Brandt}}, \bibinfo {author} {\bibfnamefont {A.}~\bibnamefont {Jüttner}}, \
  and\ \bibinfo {author} {\bibfnamefont {H.}~\bibnamefont {Wittig}},\ }\href
  {\doibase 10.1007/JHEP11(2013)034} {\bibfield  {journal} {\bibinfo  {journal}
  {JHEP}\ }\textbf {\bibinfo {volume} {11}},\ \bibinfo {pages} {034} (\bibinfo
  {year} {2013})},\ \Eprint {http://arxiv.org/abs/1306.2916} {arXiv:1306.2916
  [hep-lat]} \BibitemShut {NoStop}%
\bibitem [{\citenamefont {Bazavov}\ \emph {et~al.}(2012)\citenamefont {Bazavov}
  \emph {et~al.}}]{Bazavov:2012qja}%
  \BibitemOpen
  \bibfield  {author} {\bibinfo {author} {\bibfnamefont {A.}~\bibnamefont
  {Bazavov}} \emph {et~al.} (\bibinfo {collaboration} {HotQCD}),\ }\href
  {\doibase 10.1103/PhysRevD.86.094503} {\bibfield  {journal} {\bibinfo
  {journal} {Phys. Rev. D}\ }\textbf {\bibinfo {volume} {86}},\ \bibinfo
  {pages} {094503} (\bibinfo {year} {2012})},\ \Eprint
  {http://arxiv.org/abs/1205.3535} {arXiv:1205.3535 [hep-lat]} \BibitemShut
  {NoStop}%
\bibitem [{\citenamefont {Lu}\ and\ \citenamefont
  {Ruggieri}(2019)}]{Lu:2018ukl}%
  \BibitemOpen
  \bibfield  {author} {\bibinfo {author} {\bibfnamefont {Z.-Y.}\ \bibnamefont
  {Lu}}\ and\ \bibinfo {author} {\bibfnamefont {M.}~\bibnamefont {Ruggieri}},\
  }\href {\doibase 10.1103/PhysRevD.100.014013} {\bibfield  {journal} {\bibinfo
   {journal} {Phys. Rev. D}\ }\textbf {\bibinfo {volume} {100}},\ \bibinfo
  {pages} {014013} (\bibinfo {year} {2019})},\ \Eprint
  {http://arxiv.org/abs/1811.05102} {arXiv:1811.05102 [hep-ph]} \BibitemShut
  {NoStop}%
\bibitem [{\citenamefont {Bandyopadhyay}\ \emph {et~al.}(2019)\citenamefont
  {Bandyopadhyay}, \citenamefont {Farias}, \citenamefont {Lopes},\ and\
  \citenamefont {Ramos}}]{Bandyopadhyay:2019pml}%
  \BibitemOpen
  \bibfield  {author} {\bibinfo {author} {\bibfnamefont {A.}~\bibnamefont
  {Bandyopadhyay}}, \bibinfo {author} {\bibfnamefont {R.~L.}\ \bibnamefont
  {Farias}}, \bibinfo {author} {\bibfnamefont {B.~S.}\ \bibnamefont {Lopes}}, \
  and\ \bibinfo {author} {\bibfnamefont {R.~O.}\ \bibnamefont {Ramos}},\ }\href
  {\doibase 10.1103/PhysRevD.100.076021} {\bibfield  {journal} {\bibinfo
  {journal} {Phys. Rev. D}\ }\textbf {\bibinfo {volume} {100}},\ \bibinfo
  {pages} {076021} (\bibinfo {year} {2019})},\ \Eprint
  {http://arxiv.org/abs/1906.09250} {arXiv:1906.09250 [hep-ph]} \BibitemShut
  {NoStop}%
\bibitem [{\citenamefont {Hell}\ \emph {et~al.}(2009)\citenamefont {Hell},
  \citenamefont {Roessner}, \citenamefont {Cristoforetti},\ and\ \citenamefont
  {Weise}}]{Hell:2008cc}%
  \BibitemOpen
  \bibfield  {author} {\bibinfo {author} {\bibfnamefont {T.}~\bibnamefont
  {Hell}}, \bibinfo {author} {\bibfnamefont {S.}~\bibnamefont {Roessner}},
  \bibinfo {author} {\bibfnamefont {M.}~\bibnamefont {Cristoforetti}}, \ and\
  \bibinfo {author} {\bibfnamefont {W.}~\bibnamefont {Weise}},\ }\href
  {\doibase 10.1103/PhysRevD.79.014022} {\bibfield  {journal} {\bibinfo
  {journal} {Phys. Rev. D}\ }\textbf {\bibinfo {volume} {79}},\ \bibinfo
  {pages} {014022} (\bibinfo {year} {2009})},\ \Eprint
  {http://arxiv.org/abs/0810.1099} {arXiv:0810.1099 [hep-ph]} \BibitemShut
  {NoStop}%
\bibitem [{\citenamefont {'t~Hooft}(1976)}]{tHooft:1976snw}%
  \BibitemOpen
  \bibfield  {author} {\bibinfo {author} {\bibfnamefont {G.}~\bibnamefont
  {'t~Hooft}},\ }\href {\doibase 10.1103/PhysRevD.14.3432} {\bibfield
  {journal} {\bibinfo  {journal} {Phys. Rev. D}\ }\textbf {\bibinfo {volume}
  {14}},\ \bibinfo {pages} {3432} (\bibinfo {year} {1976})},\ \bibinfo {note}
  {[Erratum: Phys.Rev.D 18, 2199 (1978)]}\BibitemShut {NoStop}%
\bibitem [{\citenamefont {'t~Hooft}(1986)}]{tHooft:1986ooh}%
  \BibitemOpen
  \bibfield  {author} {\bibinfo {author} {\bibfnamefont {G.}~\bibnamefont
  {'t~Hooft}},\ }\href {\doibase 10.1016/0370-1573(86)90117-1} {\bibfield
  {journal} {\bibinfo  {journal} {Phys. Rept.}\ }\textbf {\bibinfo {volume}
  {142}},\ \bibinfo {pages} {357} (\bibinfo {year} {1986})}\BibitemShut
  {NoStop}%
\bibitem [{\citenamefont {Weinberg}(1996)}]{weinberg_1996}%
  \BibitemOpen
  \bibfield  {author} {\bibinfo {author} {\bibfnamefont {S.}~\bibnamefont
  {Weinberg}},\ }\href {\doibase 10.1017/CBO9781139644174} {\emph {\bibinfo
  {title} {The Quantum Theory of Fields}}},\ Vol.~\bibinfo {volume} {2}\
  (\bibinfo  {publisher} {Cambridge University Press},\ \bibinfo {year}
  {1996})\BibitemShut {NoStop}%
\bibitem [{\citenamefont {Del~Debbio}\ \emph {et~al.}(2005)\citenamefont
  {Del~Debbio}, \citenamefont {Giusti},\ and\ \citenamefont
  {Pica}}]{DelDebbio:2004ns}%
  \BibitemOpen
  \bibfield  {author} {\bibinfo {author} {\bibfnamefont {L.}~\bibnamefont
  {Del~Debbio}}, \bibinfo {author} {\bibfnamefont {L.}~\bibnamefont {Giusti}},
  \ and\ \bibinfo {author} {\bibfnamefont {C.}~\bibnamefont {Pica}},\ }\href
  {\doibase 10.1103/PhysRevLett.94.032003} {\bibfield  {journal} {\bibinfo
  {journal} {Phys. Rev. Lett.}\ }\textbf {\bibinfo {volume} {94}},\ \bibinfo
  {pages} {032003} (\bibinfo {year} {2005})},\ \Eprint
  {http://arxiv.org/abs/hep-th/0407052} {arXiv:hep-th/0407052} \BibitemShut
  {NoStop}%
\bibitem [{\citenamefont {Fukushima}\ \emph {et~al.}(2001)\citenamefont
  {Fukushima}, \citenamefont {Ohnishi},\ and\ \citenamefont
  {Ohta}}]{Fukushima:2001hr}%
  \BibitemOpen
  \bibfield  {author} {\bibinfo {author} {\bibfnamefont {K.}~\bibnamefont
  {Fukushima}}, \bibinfo {author} {\bibfnamefont {K.}~\bibnamefont {Ohnishi}},
  \ and\ \bibinfo {author} {\bibfnamefont {K.}~\bibnamefont {Ohta}},\ }\href
  {\doibase 10.1103/PhysRevC.63.045203} {\bibfield  {journal} {\bibinfo
  {journal} {Phys. Rev. C}\ }\textbf {\bibinfo {volume} {63}},\ \bibinfo
  {pages} {045203} (\bibinfo {year} {2001})},\ \Eprint
  {http://arxiv.org/abs/nucl-th/0101062} {arXiv:nucl-th/0101062} \BibitemShut
  {NoStop}%
\bibitem [{\citenamefont {Weinberg}(1978)}]{Weinberg:1977ma}%
  \BibitemOpen
  \bibfield  {author} {\bibinfo {author} {\bibfnamefont {S.}~\bibnamefont
  {Weinberg}},\ }\href {\doibase 10.1103/PhysRevLett.40.223} {\bibfield
  {journal} {\bibinfo  {journal} {Phys. Rev. Lett.}\ }\textbf {\bibinfo
  {volume} {40}},\ \bibinfo {pages} {223} (\bibinfo {year} {1978})}\BibitemShut
  {NoStop}%
\bibitem [{\citenamefont {Wantz}\ and\ \citenamefont
  {Shellard}(2010)}]{Wantz:2009mi}%
  \BibitemOpen
  \bibfield  {author} {\bibinfo {author} {\bibfnamefont {O.}~\bibnamefont
  {Wantz}}\ and\ \bibinfo {author} {\bibfnamefont {E.}~\bibnamefont
  {Shellard}},\ }\href {\doibase 10.1016/j.nuclphysb.2009.12.005} {\bibfield
  {journal} {\bibinfo  {journal} {Nucl. Phys. B}\ }\textbf {\bibinfo {volume}
  {829}},\ \bibinfo {pages} {110} (\bibinfo {year} {2010})},\ \Eprint
  {http://arxiv.org/abs/0908.0324} {arXiv:0908.0324 [hep-ph]} \BibitemShut
  {NoStop}%
\bibitem [{\citenamefont {Fujikawa}(1980)}]{Fujikawa:1980eg}%
  \BibitemOpen
  \bibfield  {author} {\bibinfo {author} {\bibfnamefont {K.}~\bibnamefont
  {Fujikawa}},\ }\href {\doibase 10.1103/PhysRevD.21.2848} {\bibfield
  {journal} {\bibinfo  {journal} {Phys. Rev. D}\ }\textbf {\bibinfo {volume}
  {21}},\ \bibinfo {pages} {2848} (\bibinfo {year} {1980})},\ \bibinfo {note}
  {[Erratum: Phys.Rev.D 22, 1499 (1980)]}\BibitemShut {NoStop}%
\bibitem [{\citenamefont {Boer}\ and\ \citenamefont
  {Boomsma}(2008)}]{Boer:2008ct}%
  \BibitemOpen
  \bibfield  {author} {\bibinfo {author} {\bibfnamefont {D.}~\bibnamefont
  {Boer}}\ and\ \bibinfo {author} {\bibfnamefont {J.~K.}\ \bibnamefont
  {Boomsma}},\ }\href {\doibase 10.1103/PhysRevD.78.054027} {\bibfield
  {journal} {\bibinfo  {journal} {Phys. Rev. D}\ }\textbf {\bibinfo {volume}
  {78}},\ \bibinfo {pages} {054027} (\bibinfo {year} {2008})},\ \Eprint
  {http://arxiv.org/abs/0806.1669} {arXiv:0806.1669 [hep-ph]} \BibitemShut
  {NoStop}%
\bibitem [{\citenamefont {Boomsma}\ and\ \citenamefont
  {Boer}(2009)}]{Boomsma:2009eh}%
  \BibitemOpen
  \bibfield  {author} {\bibinfo {author} {\bibfnamefont {J.~K.}\ \bibnamefont
  {Boomsma}}\ and\ \bibinfo {author} {\bibfnamefont {D.}~\bibnamefont {Boer}},\
  }\href {\doibase 10.1103/PhysRevD.80.034019} {\bibfield  {journal} {\bibinfo
  {journal} {Phys. Rev. D}\ }\textbf {\bibinfo {volume} {80}},\ \bibinfo
  {pages} {034019} (\bibinfo {year} {2009})},\ \Eprint
  {http://arxiv.org/abs/0905.4660} {arXiv:0905.4660 [hep-ph]} \BibitemShut
  {NoStop}%
\bibitem [{\citenamefont {Schwartz}(2014)}]{Schwartz:2014sze}%
  \BibitemOpen
  \bibfield  {author} {\bibinfo {author} {\bibfnamefont {M.~D.}\ \bibnamefont
  {Schwartz}},\ }\href@noop {} {\emph {\bibinfo {title} {{Quantum Field Theory
  and the Standard Model}}}}\ (\bibinfo  {publisher} {Cambridge University
  Press},\ \bibinfo {year} {2014})\BibitemShut {NoStop}%
\bibitem [{\citenamefont {Buballa}(2005)}]{Buballa:2003qv}%
  \BibitemOpen
  \bibfield  {author} {\bibinfo {author} {\bibfnamefont {M.}~\bibnamefont
  {Buballa}},\ }\emph {\bibinfo {title} {{NJL model analysis of quark matter at
  large density}}},\ \href {\doibase 10.1016/j.physrep.2004.11.004} {\bibinfo
  {type} {Other thesis}} (\bibinfo {year} {2005}),\ \Eprint
  {http://arxiv.org/abs/hep-ph/0402234} {arXiv:hep-ph/0402234} \BibitemShut
  {NoStop}%
\bibitem [{\citenamefont {Vermaseren}\ \emph {et~al.}(1997)\citenamefont
  {Vermaseren}, \citenamefont {Larin},\ and\ \citenamefont {van
  Ritbergen}}]{Vermaseren:1997fq}%
  \BibitemOpen
  \bibfield  {author} {\bibinfo {author} {\bibfnamefont {J.~A.~M.}\
  \bibnamefont {Vermaseren}}, \bibinfo {author} {\bibfnamefont {S.~A.}\
  \bibnamefont {Larin}}, \ and\ \bibinfo {author} {\bibfnamefont
  {T.}~\bibnamefont {van Ritbergen}},\ }\href {\doibase
  10.1016/S0370-2693(97)00660-6} {\bibfield  {journal} {\bibinfo  {journal}
  {Phys. Lett. B}\ }\textbf {\bibinfo {volume} {405}},\ \bibinfo {pages} {327}
  (\bibinfo {year} {1997})},\ \Eprint {http://arxiv.org/abs/hep-ph/9703284}
  {arXiv:hep-ph/9703284} \BibitemShut {NoStop}%
\bibitem [{\citenamefont {van Ritbergen}\ \emph {et~al.}(1997)\citenamefont
  {van Ritbergen}, \citenamefont {Vermaseren},\ and\ \citenamefont
  {Larin}}]{vanRitbergen:1997va}%
  \BibitemOpen
  \bibfield  {author} {\bibinfo {author} {\bibfnamefont {T.}~\bibnamefont {van
  Ritbergen}}, \bibinfo {author} {\bibfnamefont {J.~A.~M.}\ \bibnamefont
  {Vermaseren}}, \ and\ \bibinfo {author} {\bibfnamefont {S.~A.}\ \bibnamefont
  {Larin}},\ }\href {\doibase 10.1016/S0370-2693(97)00370-5} {\bibfield
  {journal} {\bibinfo  {journal} {Phys. Lett. B}\ }\textbf {\bibinfo {volume}
  {400}},\ \bibinfo {pages} {379} (\bibinfo {year} {1997})},\ \Eprint
  {http://arxiv.org/abs/hep-ph/9701390} {arXiv:hep-ph/9701390} \BibitemShut
  {NoStop}%
\bibitem [{\citenamefont {Chetyrkin}(1997)}]{Chetyrkin:1997dh}%
  \BibitemOpen
  \bibfield  {author} {\bibinfo {author} {\bibfnamefont {K.~G.}\ \bibnamefont
  {Chetyrkin}},\ }\href {\doibase 10.1016/S0370-2693(97)00535-2} {\bibfield
  {journal} {\bibinfo  {journal} {Phys. Lett. B}\ }\textbf {\bibinfo {volume}
  {404}},\ \bibinfo {pages} {161} (\bibinfo {year} {1997})},\ \Eprint
  {http://arxiv.org/abs/hep-ph/9703278} {arXiv:hep-ph/9703278} \BibitemShut
  {NoStop}%
\bibitem [{\citenamefont {Giusti}\ \emph {et~al.}(1999)\citenamefont {Giusti},
  \citenamefont {Rapuano}, \citenamefont {Talevi},\ and\ \citenamefont
  {Vladikas}}]{Giusti:1998wy}%
  \BibitemOpen
  \bibfield  {author} {\bibinfo {author} {\bibfnamefont {L.}~\bibnamefont
  {Giusti}}, \bibinfo {author} {\bibfnamefont {F.}~\bibnamefont {Rapuano}},
  \bibinfo {author} {\bibfnamefont {M.}~\bibnamefont {Talevi}}, \ and\ \bibinfo
  {author} {\bibfnamefont {A.}~\bibnamefont {Vladikas}},\ }\href {\doibase
  10.1016/S0550-3213(98)00659-2} {\bibfield  {journal} {\bibinfo  {journal}
  {Nucl. Phys. B}\ }\textbf {\bibinfo {volume} {538}},\ \bibinfo {pages} {249}
  (\bibinfo {year} {1999})},\ \Eprint {http://arxiv.org/abs/hep-lat/9807014}
  {arXiv:hep-lat/9807014} \BibitemShut {NoStop}%
\bibitem [{\citenamefont {Dosch}\ and\ \citenamefont
  {Narison}(1998)}]{Dosch:1997wb}%
  \BibitemOpen
  \bibfield  {author} {\bibinfo {author} {\bibfnamefont {H.~G.}\ \bibnamefont
  {Dosch}}\ and\ \bibinfo {author} {\bibfnamefont {S.}~\bibnamefont
  {Narison}},\ }\href {\doibase 10.1016/S0370-2693(97)01370-1} {\bibfield
  {journal} {\bibinfo  {journal} {Phys. Lett. B}\ }\textbf {\bibinfo {volume}
  {417}},\ \bibinfo {pages} {173} (\bibinfo {year} {1998})},\ \Eprint
  {http://arxiv.org/abs/hep-ph/9709215} {arXiv:hep-ph/9709215} \BibitemShut
  {NoStop}%
\bibitem [{\citenamefont {Berges}\ and\ \citenamefont
  {Rajagopal}(1999)}]{Berges:1998rc}%
  \BibitemOpen
  \bibfield  {author} {\bibinfo {author} {\bibfnamefont {J.}~\bibnamefont
  {Berges}}\ and\ \bibinfo {author} {\bibfnamefont {K.}~\bibnamefont
  {Rajagopal}},\ }\href {\doibase 10.1016/S0550-3213(98)00620-8} {\bibfield
  {journal} {\bibinfo  {journal} {Nucl. Phys. B}\ }\textbf {\bibinfo {volume}
  {538}},\ \bibinfo {pages} {215} (\bibinfo {year} {1999})},\ \Eprint
  {http://arxiv.org/abs/hep-ph/9804233} {arXiv:hep-ph/9804233} \BibitemShut
  {NoStop}%
\bibitem [{\citenamefont {Dmitrasinovic}(1996)}]{Dmitrasinovic:1996fi}%
  \BibitemOpen
  \bibfield  {author} {\bibinfo {author} {\bibfnamefont {V.}~\bibnamefont
  {Dmitrasinovic}},\ }\href {\doibase 10.1103/PhysRevC.53.1383} {\bibfield
  {journal} {\bibinfo  {journal} {Phys. Rev. C}\ }\textbf {\bibinfo {volume}
  {53}},\ \bibinfo {pages} {1383} (\bibinfo {year} {1996})}\BibitemShut
  {NoStop}%
\bibitem [{\citenamefont {Sabir}\ \emph {et~al.}(2021)\citenamefont {Sabir},
  \citenamefont {Islam},\ and\ \citenamefont {Sharma}}]{Sabir:2021nws}%
  \BibitemOpen
  \bibfield  {author} {\bibinfo {author} {\bibfnamefont {M.}~\bibnamefont
  {Sabir}}, \bibinfo {author} {\bibfnamefont {C.~A.}\ \bibnamefont {Islam}}, \
  and\ \bibinfo {author} {\bibfnamefont {R.}~\bibnamefont {Sharma}},\
  }\href@noop {} {\  (\bibinfo {year} {2021})},\ \Eprint
  {http://arxiv.org/abs/2103.15849} {arXiv:2103.15849 [hep-ph]} \BibitemShut
  {NoStop}%
\bibitem [{\citenamefont {Miransky}\ and\ \citenamefont
  {Shovkovy}(2002)}]{Miransky:2002rp}%
  \BibitemOpen
  \bibfield  {author} {\bibinfo {author} {\bibfnamefont {V.}~\bibnamefont
  {Miransky}}\ and\ \bibinfo {author} {\bibfnamefont {I.}~\bibnamefont
  {Shovkovy}},\ }\href {\doibase 10.1103/PhysRevD.66.045006} {\bibfield
  {journal} {\bibinfo  {journal} {Phys. Rev. D}\ }\textbf {\bibinfo {volume}
  {66}},\ \bibinfo {pages} {045006} (\bibinfo {year} {2002})},\ \Eprint
  {http://arxiv.org/abs/hep-ph/0205348} {arXiv:hep-ph/0205348} \BibitemShut
  {NoStop}%
\bibitem [{\citenamefont {Borsanyi}\ \emph {et~al.}(2016)\citenamefont
  {Borsanyi} \emph {et~al.}}]{Borsanyi:2016ksw}%
  \BibitemOpen
  \bibfield  {author} {\bibinfo {author} {\bibfnamefont {S.}~\bibnamefont
  {Borsanyi}} \emph {et~al.},\ }\href {\doibase 10.1038/nature20115} {\bibfield
   {journal} {\bibinfo  {journal} {Nature}\ }\textbf {\bibinfo {volume}
  {539}},\ \bibinfo {pages} {69} (\bibinfo {year} {2016})},\ \Eprint
  {http://arxiv.org/abs/1606.07494} {arXiv:1606.07494 [hep-lat]} \BibitemShut
  {NoStop}%
\bibitem [{\citenamefont {Petreczky}\ \emph {et~al.}(2016)\citenamefont
  {Petreczky}, \citenamefont {Schadler},\ and\ \citenamefont
  {Sharma}}]{Petreczky:2016vrs}%
  \BibitemOpen
  \bibfield  {author} {\bibinfo {author} {\bibfnamefont {P.}~\bibnamefont
  {Petreczky}}, \bibinfo {author} {\bibfnamefont {H.-P.}\ \bibnamefont
  {Schadler}}, \ and\ \bibinfo {author} {\bibfnamefont {S.}~\bibnamefont
  {Sharma}},\ }\href {\doibase 10.1016/j.physletb.2016.09.063} {\bibfield
  {journal} {\bibinfo  {journal} {Phys. Lett. B}\ }\textbf {\bibinfo {volume}
  {762}},\ \bibinfo {pages} {498} (\bibinfo {year} {2016})},\ \Eprint
  {http://arxiv.org/abs/1606.03145} {arXiv:1606.03145 [hep-lat]} \BibitemShut
  {NoStop}%
\bibitem [{\citenamefont {Borsanyi}\ \emph {et~al.}(2010)\citenamefont
  {Borsanyi}, \citenamefont {Fodor}, \citenamefont {Hoelbling}, \citenamefont
  {Katz}, \citenamefont {Krieg}, \citenamefont {Ratti},\ and\ \citenamefont
  {Szabo}}]{Borsanyi:2010bp}%
  \BibitemOpen
  \bibfield  {author} {\bibinfo {author} {\bibfnamefont {S.}~\bibnamefont
  {Borsanyi}}, \bibinfo {author} {\bibfnamefont {Z.}~\bibnamefont {Fodor}},
  \bibinfo {author} {\bibfnamefont {C.}~\bibnamefont {Hoelbling}}, \bibinfo
  {author} {\bibfnamefont {S.~D.}\ \bibnamefont {Katz}}, \bibinfo {author}
  {\bibfnamefont {S.}~\bibnamefont {Krieg}}, \bibinfo {author} {\bibfnamefont
  {C.}~\bibnamefont {Ratti}}, \ and\ \bibinfo {author} {\bibfnamefont {K.~K.}\
  \bibnamefont {Szabo}} (\bibinfo {collaboration} {Wuppertal-Budapest}),\
  }\href {\doibase 10.1007/JHEP09(2010)073} {\bibfield  {journal} {\bibinfo
  {journal} {JHEP}\ }\textbf {\bibinfo {volume} {09}},\ \bibinfo {pages} {073}
  (\bibinfo {year} {2010})},\ \Eprint {http://arxiv.org/abs/1005.3508}
  {arXiv:1005.3508 [hep-lat]} \BibitemShut {NoStop}%
\bibitem [{\citenamefont {Moreira}\ \emph {et~al.}(2020)\citenamefont
  {Moreira}, \citenamefont {Costa},\ and\ \citenamefont
  {Restrepo}}]{Moreira:2020wau}%
  \BibitemOpen
  \bibfield  {author} {\bibinfo {author} {\bibfnamefont {J.}~\bibnamefont
  {Moreira}}, \bibinfo {author} {\bibfnamefont {P.}~\bibnamefont {Costa}}, \
  and\ \bibinfo {author} {\bibfnamefont {T.~E.}\ \bibnamefont {Restrepo}},\
  }\href {\doibase 10.1103/PhysRevD.102.014032} {\bibfield  {journal} {\bibinfo
   {journal} {Phys. Rev. D}\ }\textbf {\bibinfo {volume} {102}},\ \bibinfo
  {pages} {014032} (\bibinfo {year} {2020})},\ \Eprint
  {http://arxiv.org/abs/2005.07049} {arXiv:2005.07049 [hep-ph]} \BibitemShut
  {NoStop}%
\end{thebibliography}%

\end{document}